\documentclass[aps,prd,amsmath,amssymb,showpacs,twocolumn,floatfix]{revtex4}
\usepackage{epsfig,graphics,color}
\usepackage{graphicx}% Include figure files
\usepackage{dcolumn}% Align table columns on decimal point
\usepackage{bm}% bold math
\usepackage{amsmath}

\newcommand \tie {{\it i.e.}}
\newcommand \ie {{\it i.e.} }
\newcommand \f {\not\!}

\newcommand \kd  {\delta}
\newcommand \ra  {\rightarrow}

\newcommand \fp {{\bf p}}
\newcommand \fP {{\bf P}}
\newcommand \fn {{\bf n}}

\newcommand \fk {{\bf k}}
\newcommand \fq {{\bf q}}
\newcommand \fx {{\bf x}}
\newcommand \fy {{\bf y}}
\newcommand \fl {{\bf l}}
\newcommand \fgma {{\bf \gamma}}
\newcommand \h {\theta}
\newcommand \im {\Rightarrow}
\newcommand \vk {\vec{k}}
\newcommand \vl {\vec{l}}
\newcommand \vq {\vec{q}}

\newcommand \vp {\vec{p}}

\newcommand \mat {{\mathcal M}}
\newcommand \mal {{\mathcal L}}

\newcommand \g {\gamma}
\newcommand \ro {\rho}
\newcommand \si {\sigma}
\newcommand \e {\epsilon}
\newcommand \ve {\varepsilon}

\newcommand \x {\cdot}

\newcommand \A {\alpha}
\newcommand \B {\beta}

\newcommand \lc {\langle}
\newcommand \rc {\rangle}
\newcommand \prt {\partial}

\newcommand \nt {\noindent}
\newcommand \T {\tilde}
\newcommand \al {\alpha}

\newcommand \mh {\mathcal{H}}
\newcommand \mhp {\mathcal{H'}}
\newcommand \ml {\mathcal{L}}

\newcommand \bvec{\left( \begin{array}{c} }
\newcommand \evec{\end{array} \right)}
\newcommand \tr {\mbox{{\bf Tr}}}
\newcommand \eg {{\it e.g.}}  
\newcommand \bea{\begin{eqnarray} }
\newcommand \eea{\end{eqnarray} }
\newcommand \nn {\nonumber}
\newcommand {\be} {\begin{equation}}
\newcommand {\ee} {\end{equation}}
\newcommand {\epem} {$e^+ e^-$}
\newcommand {\mbx} {\mbox{}}

\newcommand {\gev} {\mbox{GeV}}

\begin{document}

\title{ The dihadron fragmentation function and its evolution  }

\author{A. Majumder} 

\affiliation{Nuclear Science Division, 
Lawrence Berkeley National Laboratory\\
1 Cyclotron road, Berkeley, CA 94720}

\author{Xin-Nian Wang}
\affiliation{Nuclear Science Division, 
Lawrence Berkeley National Laboratory\\
1 Cyclotron road, Berkeley, CA 94720}

\date{ \today}

\begin{abstract} 
Dihadron fragmentation functions and their evolution are
studied in the process of $e^+e^-$ annihilation. Under the 
collinear factorization approximation and facilitated by the
cut-vertex technique, the two hadron inclusive
cross section at leading order (LO) is shown to factorize
into a short distance parton cross section and a long distance
dihadron fragmentation function. We provide the definition of
such a dihadron fragmentation function in terms of parton matrix 
elements and derive its DGLAP evolution equation at leading log. 
The evolution equation for the 
non-singlet quark fragmentation function is solved numerically
with a simple ansatz for the initial condition and results are
presented for cases of physical interest.
\end{abstract}


\preprint{LBNL}

\maketitle

%%%%%%%%%%%%%%%%%%%%%%%%%%%%%%
%%%%%%%%%%%%%%%%%%%%%%%%%%%%%%
%%%%%%%%%%%%%%%%%%%%%%%%%%%%%%

\section{INTRODUCTION}

%%%%%%%%%%%%%%%%%%%%%%%%%%%%%%
%%%%%%%%%%%%%%%%%%%%%%%%%%%%%%
%%%%%%%%%%%%%%%%%%%%%%%%%%%%%%

%need a few more words here about why our thing rules!  

Lattice QCD calculations \cite{lattice} predict a phase 
transition  from a hadronic gas to a quark gluon plasma (QGP) at very high 
energy densities in which quarks and gluons are no longer confined 
to the size of individual hadrons. To create such a dense and hot
matter, heavy ions are accelerated to extremely high energies to 
collide with each other. If formed in such heavy-ion collisions, 
the QGP is rather short lived and hadronizes quickly into a plethora of 
mesons and baryons. Hence, the existence of such a state in the 
history of a given collision must be surmised through a 
variety of indirect probes. One of the most promising signatures 
has been that of jet quenching \cite{quenching}, which leads to the suppression 
of high $p_T$ particles emanating from such collisions. Such jet 
quenching phenomena have been among the most striking experimental 
discoveries from the Relativistic Heavy Ion Collider (RHIC) at 
Brookhaven National Laboratory. Not only has the suppression of
single inclusive high $p_T$ hadron spectra been observed \cite{highpt},
but also the disappearance of 
back-to-back correlations of high $p_T$
hadrons has been noted \cite{adl03}. 
Both phenomena and the observed azimuthal
anisotropy of high $p_T$ hadron spectra are qualitatively consistent
with the picture of parton energy loss that leads to jet quenching.
This is indicative of the formation of a hot medium which is 
opaque to energetic
partons and has a parton density about 30 times higher than in a
cold heavy nucleus.

In the investigation of jet suppression, correlations between two high $p_T$ 
hadrons in azimuthal angle are used to study the change of jet 
structure \cite{adl03}. 
While the back-to-back correlations are suppressed in central $Au+Au$ 
collisions, indicating parton energy loss, the same-side 
correlations remain approximately the same as in $p+p$ and $d+Au$ collisions.
Given the experimental kinematics, this is considered as an indication
of parton hadronization outside the medium. However, since the
same-side correlation corresponds to two-hadron distribution
within a single jet, the observed phenomenon is highly nontrivial.
To answer the question as to why 
a parton with a reduced energy would give the same two-hadron
distribution, one has to take a closer
look at the single and double hadron fragmentation functions and
their modification in medium. In this paper we take the first
step by studying the dihadron fragmentation functions in the
process of $e^+e^-$ annihilations.

Inclusive hadron production cross sections in $e^+e^-$ collisions 
have turned out to be one of the many successful predictions of 
perturbative QCD \cite{ste77,dok77,fie78}.
%These are not, as yet, computable directly  within QCD. 
For reactions at an energy scale much above $\Lambda_{QCD}$ 
one can factorize the cross section into a short-distance 
parton cross section which is computable order by order 
as a series in $\al_s(Q^2)$; and a long-distance phenomenological 
object (the single hadron inclusive fragmentation function)
which contains the non-perturbative information of parton 
hadronization \cite{col89}. These fragmentation functions can be  
defined in an operator formalism \cite{mue78} and hence are valid 
beyond the perturbative theory. They, however, cannot 
be calculated perturbatively and have to be, instead, inferred from 
experiments. The definition of these functions affords them the 
mantle of being universal or process-independent. Once
measured in one process, {\it e.g.} $e^+e^-$ annihilation, they
can be applied to another, {\it e.g.} deep inelastic scattering
or $p+p$ collisions, and therein lies the predictive power.
Another contribution of pQCD rests in the fact that once these 
functions are measured 
%in any experiment for  all values of species and momentum fraction 
at a given energy scale, they can be predicted for all other 
energy scales via  the Dokshitzer-Gribov-Lipatov-Altarelli-Parisi (DGLAP) 
evolution equations \cite{gri72,dok77b,alt77}.

% need to say something about evolution. 
%The two properties of fragmentation functions mentioned in the 
%previous paragraph 
%allow for an extension of 

The single inclusive fragmentation function $D_q^h(z,Q^2)$   
can be interpreted as a multiplicity distribution for hadrons 
of  type $h$ with a longitudinal momentum fraction 
$z$ that materialize from a fragmenting parton of flavour $q$.
One can have, in principle, an $n$-hadron fragmentation function 
$D_q^{h_1,h_2,...h_n}(z_1,z_2,...z_n,Q^2)$ which counts the 
number of hadrons of type $h_1,h_2,...h_n$ with momentum 
fractions $z_1,z_2,...z_n$ materializing from a fragmenting 
parton $q$. In this article, we will be concerned with the 
double inclusive fragmentation function $D_q^{h_1,h_2}(z_1,z_2,Q^2)$
or the dihadron fragmentation function.
The operator definition of this function is not merely a 
trivial extension of the single hadron case; 
there are no straightforward sum rules connecting it
to single inclusive fragmentation functions.
%the reader will immediately note that the partonic 
%analogue of this function 
%vanishes at leading order. 
Similar functions have been 
considered previously \cite{kon78,kon79a,kon79b}.
However, such formulations considered the double 
fragmentation function with a fixed angle
$\kd$ (or a fixed tranverse momentum difference $q_T$) 
between the two hadrons. For a large enough choice 
of $\kd$ (or $q_T$) the dominant contribution to 
$D_q^{h_1,h_2}(z_1,z_2,\kd,Q^2)$ was postulated 
to emanate from a process where the fragmenting parton had 
undergone a split into two partons which then fragmented 
independently. This formulation, however, requires $\kd$ 
(or $q_T$) to be large enough that the splitting process may 
be calculated in perturbation theory. In our formulation, 
the internal angle $\kd$ (or relative
momentum $q_T$) will be explicitly integrated over. In this 
regard, our calculations are similar in spirit to 
those of Ref.~\cite{suk80}. In that effort, 
general evolution equations for $n$-hadron 
fragmentation functions were motivated, and 
an algebraic solution of the moments of 
the fragmentation functions was obtained.
In none of the previous studies, however, 
has an attempt been made to generalize the operator 
definition of the fragmentation function from one hadron 
to many hadrons. Evaluation of the $n$-hadron production 
cross section at leading order (LO) and leading twist 
allows one to \emph{analytically define} such a function. 
Evaluation of corrections to the same cross section 
at leading log (LL) and leading twist allows one to 
\emph{derive} the evolution equations for such functions. 
%at the LL level. 
This is the objective of the 
current study. To the knowledge of the 
authors, no such calculation has hitherto been attempted.    

Our eventual goal is to derive the medium modification to the
dihadron fragmentation function as the fragmenting parton propagates
through the medium. As shown in the case of single 
hadron fragmentation functions \cite{guowang}, the medium 
modification of the fragmentation function due to multiple
scattering and induced gluon radiation resembles very much
that of radiative corrections due to gluon bremsstrahlung in 
vacuum. Therefore, the study of DGLAP evolution of dihadron
fragmentation functions in this paper can already provide
us hints of what one may expect for medium modifications.

%A recent effort has measured 
%the two particle correlation between two high $p_T$ hadrons with a 
%variety of angles between them \cite{adl03}. While it may be 
%argued that for large 
%enough angles the process essentially consists of independent partons 
%traversing the medium, escaping and eventually fragmenting; the case at 
%very small angles (also called the same-side correlation) is the result of 
%two hadrons fragmenting from the same parton. The complete 
%calculation of this quantity will require the inclusion of medium effects 
%leading to energy loss of the high $p_T$ parton preceding its fragmentation. 
%This complication will be incorporated in a subsequent calculation. 
%In this effort we 
%will concentrate on the definition of the double fragmentation function and 
%derive the equations that describe its evolution with the energy scale of the 
%fragmenting parton. 

The remaining sections are organized as follows: in Sec. II we present a 
general discussion of the  double fragmentation function. We outline how 
such a function may be isolated in the expression for 
a double differential inclusive cross section and discuss the 
possible nature of its evolution equations. In Sec. III we begin with the 
S-matrix expression for the double differential cross section for the production 
of two hadrons in $e^+ e^-$ collisions at leading order (LO); 
we then factorize the expression 
at leading twist into the conventional hard part and the double 
fragmentation function. We also derive the rules for such an object in the 
cut-vertex technique of Mueller. In Sec. IV we write down the 
double differential cross section at next-to-leading order (NLO) and once
again factorize it at leading twist into the conventional hard part and 
NLO correction to the fragmentation function. Using these we derive the 
DGLAP \cite{gri72,dok77b,alt77} evolution equations for the 
double fragmentation functions. 
In Sec. V we focus on the evolution equation for the non-singlet (NS) 
quark fragmentation function and solve its evolution equation numerically. 
Finally in Sec. VI we discuss the results of our calculation and 
present our conclusions.

%%%%%%%%%%%%%%%%%%%%%%%%%%%%%%%%%%%%%%%
%%%%%%%%%%%%%%%%%%%%%%%%%%%%%%%%%%%%%%%
%%%%%%%%%%%%%%%%%%%%%%%%%%%%%%%%%%%%%%%
%%%%%%%%%%%%%%%%%%%%%%%%%%%%%%%%%%%%%%%

%\section{The double inclusive differential cross section, factorization 
%of divergences and evolution}

\section{The parton model}

%%%%%%%%%%%%%%%%%%%%%%%%%%%%%%%%%%%%%%%
%%%%%%%%%%%%%%%%%%%%%%%%%%%%%%%%%%%%%%%
%%%%%%%%%%%%%%%%%%%%%%%%%%%%%%%%%%%%%%%
%%%%%%%%%%%%%%%%%%%%%%%%%%%%%%%%%%%%%%%

In this section, we present a general discussion of the 
properties of a dihadron fragmentation function within
a parton-model-like picture with collinear factorization.
% \ie factorization into a pertubative 
%hard piece and a phenomenalogical soft part will be assumed. 
However, all our assumptions will be demonstrated to hold 
explicitly in an operator formalism at leading log and leading 
twist in the subsequent sections.
 
%In this section we will concentrate on the factorization of the 
%infrared divergences into the dihadron fragmentation function 
%followed by the resummation of higher order diagrams that lead 
%to the DGLAP evolution of this function. Various sum rules that 
%such a function will satisfy will be discussed, they will be 
%derived in the operator formalism in subsequent sections. 
%In this article we will be explicitly concerned with 

We consider the following semi-inclusive process 
\bea
e^+ + e^- \ra \gamma^* \ra h_1 + h_2 + X \nonumber
\eea
of $e^+ e^-$ annihilation.
%goes to a virtual photon which then 
%undergoes a transition to jets of hadrons,
%where we identify two particular hadrons.
%We should also point out that in this article we will explicitly 
%be concerned with the scenario where there are 
We consider two-jet events where both the 
identified hadrons $h_1$ and $h_2$ emanate from the same jet. 
At leading order in the strong coupling this occurs 
from the conversion of the virtual photon into a 
back-to-back quark and antiquark pair which 
fragment into two jets of hadrons. 

In this scenario, at a large $Q^2$ of the reaction, 
one may factorize the cross section of single inclusive
hadron production as \cite{fie95}
\bea
\frac{d \si}{dz} &=& \sum_{q} \sigma_0^{q\bar{q}}  
\left[ D_{q}^{h} (z) + D_{\bar{q}}^{h} (z) 
\right] . 
\label{LO_Dz}
\eea
with  $D_{q}^{h}(z)$ and $D_{\bar{q}}^{h}(z)$ as the 
single inclusive quark fragmentation functions.
The total cross section for the annihilation of an electron 
positron pair to a quark and an anti-quark, $\si_0^{q\bar{q}}$ 
at leading order is
\bea
\si_0^{q\bar{q}} = e_q^2 N_c\frac{4 \pi \al^2}{3 s}. \label{si_0}
\eea 
Here, $e_q$ is the fractional charge of the quark in units of
an electron charge, $s$ is the square of center of mass energy 
of the $e^+e^-$ pair, and $N_c=3$ is the number of colors in the
fundamental representation of QCD. The fractional momentum
$z$ represents the the light-cone momentum fraction 
of the hadrons to the parent partons, \tie,

\[
z = \frac{\fp_{h} \x \fn}{\fp \x \fn}, 
%
%z_2 = \frac{\fp_{h_2} \x \fn}{\fp \x \fn},
\]
where we use the notation of bold face letters representing 
four-vectors. The lightlike four-vector 
$\fn \equiv [n^+, n^-, n_\perp] = [0,1,0]$
is taken conjugate to a given momentum, as yet unspecified.

Similarly, one can expect to obtain the two-hadron inclusive cross
section from Eq.~(\ref{LO_Dz}) by replacing the single 
inclusive fragmentation functions $D^h_{q}(z_1)$ 
with the double inclusive functions $D^{h_1 h_2}_{q}(z_1,z_2)$,
\bea
\frac{d^2 \si}{dz_1 dz_2} = \sum_{q} \si_0^{q\bar{q}} 
\left[ D_{q}^{h_1 h_2} (z_1,z_2) + 
D_{\bar{q}}^{h_1 h_2} (z_1,z_2) 
\right].
\label{LO_Dz1z2}
\eea
We will not discuss cases in which each of the two 
hadrons emerges independently from each of the back-to-back 
quark and antiquark jets.

%\[
%3 \si(\mu \mu) \sum_{f=1}^{n_f} Q^2_f 
%
%\left[ D_{q_i}(z_1) D_{\bar{q}_i} (z_2) 
%
%+ D_{q_i}(z_2)D_{\bar{q}_i} (z_1) \right] .
%\]
%
%\nt
%Such cases will be removed by hand from the 
%later evaluation of the dihadron inclusive cross section. 

%Even without the exact operator definitions of the fragmentation 
%functions one may still proceed to the NLO double inclusive 
%cross sections with 
%the mere notion of a dihadron fragmentation function. Unlike the 
%well defined overlap integral of Eqs.~(\ref{dz1},\ref{dz1_op}), one 
%may imagine the product of a scaling function and a double infinitesimal 
%interval 
%$D_q^{h_1,h_2}(z_1,z_2) dz_1 dz_2$
%representing the probability of pairs of hadrons of type 
%$h_1$ with momentum fraction 
%$z_1$ within $dz_1$ of the parent parton $q$ and of type $h_2$ with 
%momentum fraction $z_2$ within $dz_2$, materializing from the parton.

In the parton model \cite{fie95}, at NLO with a single gluon 
radiative correction, the two hadron inclusive cross 
section in $e^+ e^-$ annihilation can be expressed as a
convolution of the fragmentation functions with the 
differential partonic cross sections (see Sec. 3.3 of 
Ref.~\cite{fie95}),
\bea 
d^2 \si (z_1, z_2, Q^2) &=& \left( \frac{d \si_q}{ dy} \right) dy 
D_q^{h_1,h_2}(x_1,x_2) dx_1 dx_2 \nn \\
&+& \left( \frac{d \si_{\bar{q}} }{ dy} \right) dy 
D_{\bar{q}}^{h_1,h_2}(x_1,x_2) dx_1 dx_2 \nn \\
&+& \left( \frac{d \si_{g} }{ dy} \right) dy 
D_g^{h_1,h_2}(x_1,x_2) dx_1 dx_2 \nn \\
&+& \left( \frac{d \si_{g q} }{ dy} \right) dy 
\left[ D_g^{h_1}(x_1)  D_q^{h_2}(x_2) + \right. \nn \\
& & \left. (h_1,z_1) \ra (h_2,z_2) \right]dx_1dx_2 \nn \\
&+& \left( \frac{d \si_{g \bar{q}} }{ dy} \right) dy 
\left[ D_g^{h_1}(x_1)  D_{\bar{q}}^{h_2}(x_2) + \right. \nn \\
& & \left. (h_1,z_1) \ra (h_2,z_2) \right]dx_1dx_2 .
\label{sigma_generic}
\eea 
%represent the pQCD differential cross sections 
%for the production of
%a quark, antiquark or a gluon 
%represent cross 
%sections for the production of a quark and a gluon 
%(antiquark and a gluon) in the same direction after splitting 
%off from the same parent parton, such that 
%These then undergo uncorrelated fragmentation such that 
There are two distinct types of contributions in the above equation.
The first one is determined by the two-hadron fragmentation functions
of single partons that have the pQCD differential cross
sections, $d\si_q/dy$, $d\si_{\bar{q}}/dy$, $d\si_{g}/dy$, 
and momentum fraction $y$. The second contribution corresponds to
two, almost collinear, partons (a gluon and a quark or antiquark with
pQCD cross sections $d\si_{g q}/dy$ and $d\si_{g\bar{q}}/dy$)
splitting from the same parent parton and then fragmenting 
independently into hadrons. In this case,
the quark (antiquark) carries momentum fraction $y$ and the 
gluon carries $1-y$. One of the identified hadrons comes 
from each of these partons. 

The two hadrons $h_1$ and $h_2$ have momentum fractions 
$x_1$ and $x_2$ of their 
immediate parent parton which itself is endowed with a 
momemtum fraction $y$ or $1-y$. 
Relating the inside parton variables $x_1,x_2,y$ to the outside hadron  
variables we obtain 
\[
x_1 = z_1/y , x_2 = z_2/y 
\]
for the first three terms and
\[
x_1 = z_1/y , x_2 = z_2/(1-y) 
\]
for the last two terms. Using the pQCD partonic cross 
sections in the massive gluon scheme \cite{fie95}, one can obtain the 
double differential cross section for the production of 
two hadrons with momentum  fractions $z_1,z_2$ as, 
\bea 
\frac{d^2 \si}{dz_1 dz_2} &=& \sum_q \si_0^{q\bar{q}} \Bigg[  
\int_{\!\!\!\!\!\!\mbox{}_{\mbox{}_{z_1+z_2}}}^1 
\frac{dy}{y^2} 
\Bigg\{ 
\Big( 1 + \frac{\A_s}{\pi} \Big) \kd(1-y) \nn \\
&+& \frac{\A_s}{2\pi} P_{q\ra qg} (y) \log(Q^2/m_g^2) 
+ \A_s f_q^{e^+e^-} (y) \Bigg\} \nn \\
& \times & \Big\{ D_q^{h_1,h_2}(z_1/y,z_2/y) 
+ D_{\bar{q}}^{h_1,h_2}(z_1/y,z_2/y) \Big\} \nn \\
&+& 2 \Bigg\{ \frac{\A_s}{2\pi} P_{q\ra g q} (y) \log(Q^2/m_g^2) 
+  \A_s f_g^{e^+e^-} (y) \Bigg\} \nn \\ 
& \times & D_{g}^{h_1,h_2}(z_1/y,z_2/y) 
%
%\Bigg] \nn \\
%
%&+& N_c \si(\mu \mu) 
+\int_{z_1}^{1-z_2} \frac{dy}{y(1-y)} \nn \\
%
%\Bigg[ \sum_i e_{q_1}^2 \nn \\
%
& \times & \frac{\A_s}{2\pi} \hat{P}_{q\ra q g} (y) \log(Q^2/m_g^2)  
\nn \\
&\times& 
\Big\{ D_q^{h_1}(z_1/y) D_{g}^{h_2} (z_2/(1-y)) \nn \\
&+& D_{\bar{q}}^{h_1} (z_1/y) D_{g}^{h_2} ( z_2/(1-y) ) \Big\}
\Bigg] + 1 \ra 2 .
\label{sigma_glue_mass}
\eea
\nt 
Here the switch between the indices $1\ra2$
is only meant for the last $y$ integration. 
The $P(y)$ functions are the regular 
splitting functions which contain both the real
and virtual contributions and thus have no 
infrared divergences. The $\hat{P} (y)$ functions 
contain no contributions from virtual diagrams. 
The $f(y)$'s are the 
scheme dependent functions obtained in the 
massive gluon scheme, where $m_g$ is the 
fictitious gluon mass introduced to 
regulate the collinear divergences. 
In the subsequent discussion we will 
focus on the leading log (LL) piece of the 
above expression. Thus one can drop the 
scheme dependent $f$ functions. 

Note that up to this point we have simply retraced the 
sequence of steps in the evaluation of the radiative corrections 
to the single inclusive fragmentation functions in the parton model. 
What is new in the case of two hadron inclusive cross section 
is the contribution from the splitting into a quark and gluon 
followed by independent fragmentation.
The $\log(Q^2/m_g^2)$ in this contribution originates from an 
integration over the transverse momentum $q_{\perp}$ 
of the quark and gluon emanating from the split. For 
very small values of $q_{\perp}$, other higher order
and nonperturbative processes become important that will 
invalidate the
picture of independent fragmentation of the two partons.
For $q_{\perp}  >> \Lambda_{QCD}$, however, the 
higher order corrections will be suppressed and the 
quark and gluon will fragment incoherently in LL
approximation (this was first pointed out in Ref.~\cite{kon79b}). 
A simple proof of this statement has been included in the Appendix. 
In this paper, we introduce a cut-off scale $\mu_\perp$
that separates two regimes of two-parton fragmentation
according to the value of $q_\perp$: independent fragmentation
for $q_\perp>\mu_\perp$ and coherent fragmentation 
for $q_\perp<\mu_\perp$. Unlike the factorization scales
that we will discuss shortly, $\mu_\perp$ is not introduced 
to renormalize the fragmentation functions but to define
the perturbative (or non-perturbative) part of the 
dihadron fragmentation functions. It is quite analogous to
the cone-size of jet definitions \cite{cone}.

%the  integration into two pieces, one below and one 
%above the physical scale $\mu_\perp$. The dependence of 
%this process on the value of $\mu_\perp$ introduces 
%a sort of scale dependence on the definition of these 
%fragmentation functions. This is quite unlike the 
%inevitable scale dependence which will be brought in
%by the factorization of infrared divergences. We will 
%return to this point shortly.  

To simplify the discussion in this paper, we will 
concentrate on the non-singlet fragmentation functions: 
\bea
D_{NS}^{h_1,h_2} (z_1,z_2) =  D_q^{h_1,h_2} (z_1,z_2) -  
D_{\bar{q}}^{h_1,h_2} (z_1,z_2). \label{non_singlet}
\eea
We also use the following convolution notations,
\[
A * B \big|_a^b= \int_a^b \frac{dy}{y^2} A(z_1/y,z_2/y) B(y)
\]

\[
A \bar{*} B \big|_a^b = \int_a^b \frac{dy}{y(1-y)} 
A(z_1/y,z_2/(1-y)) B(y).
\]

The \emph{bare} fragmentation functions 
$D_{q,\bar{q},g}(z_1,z_2)$ in 
Eqs.~(\ref{sigma_generic}) and (\ref{sigma_glue_mass})
are not as yet physical, measurable quantities 
and are scheme dependent, since the cross sections
expressed in terms of them contain collinear divergences.
%but instead are non-perturbative objects which only contain 
%the information of turning partons into hadrons.
One can however introduce renormalized
fragmentation functions such that the double inclusive
cross section can be factorized in the form of
Eq.~(\ref{LO_Dz1z2}) and is free of collinear divergences.

%we can define experimentally observable fragmentation 
%functions by absorbing all factors beyond 
%the hard cross section 
%($N_c \sum_{q_i} e_{q_i} \si (\mu \mu) $)
%into the definition of the experimentally 
%observed fragmentation functions. 

Factoring out the $e^+e^-$ annihilation cross section in
Eq.~(\ref{sigma_glue_mass}), we are left with 
the scale dependent \emph{physical} fragmentation 
functions which should be free of collinear divergences.
The ``non-singlet'' physical fragmentation functions are,

\bea 
D_{NS}^{h_1,h_2} (z_1,z_2,Q^2) &=& D_{NS}^{h_1,h_2}  \label{dqq_coll} \\ 
& &\hspace{-1.0in}+ D_{NS}^{h_1,h_2}*\frac{\A_s}{2\pi} 
P_{q\ra q g}\bigg|_{z_1+z_2}^1 \log(Q^2/m_g^2) \nn \\
%
%+ ... \Bigg)_{z_1+z_2}^1 \nn \\
%
& &\hspace{-1.0in} + \bigg(D_{NS}^{h_1} D_g^{h_2}\bigg) \bar{*} 
\frac{\A_s}{2\pi} \hat{P}_{q\ra q g} \bigg|_{z_1}^{1-z_2}
\log(Q^2/\mu_\perp^2)  \nn \\ 
& &\hspace{-1.0in} + \bigg(D_{NS}^{h_1} D_g^{h_2}\bigg) \bar{*} 
\frac{\A_s}{2\pi} \hat{P}_{q\ra q g} \bigg|_{z_1}^{1-z_2}
\log(\mu_\perp^2 / m_g^2) \nn \\
& &\hspace{-1.0in} + 1 \ra 2 \mbox{} \nn
\eea

\nt 
Here the switch $1 \ra 2$ is meant solely for the 
second and third term. Such exchange will be made
implicit in the rest of this paper.
We will also drop the limits of the convolutions in the
notation for brevity.
%We have further concentrated 
%solely on the leading log piece and dropped the 
%scheme dependent $f$ functions from the discussion.
In the case of independent fragmentation, we have also
split the expressions into a term that solely includes 
contributions with $q_\perp$ above the scale $\mu_\perp$. 
The second piece contains contributions below $\mu_\perp$ 
and thus receives large corrections from higher order
and non-perturbative processes. 
This piece will have to be absorbed into 
a redefinition of the bare dihadron fragmentation function.
%Alternatively, one may have begun (in Eq.~(\ref{LO_Dz1z2})) 
%with a scale dependent 
%fragmentation function $D(z_1,z_2,\mu_\perp)$, which only 
%recieves contributions 
%from incoherent fragmentation (2nd $dy$ integration in 
%Eq.~(\ref{sigma_glue_mass})),
%at NLO, from processes with a 
%$q_{\perp}$ greater than $\mu_\perp$.
%
%\bea
%D_{NS} (z_1,z_2,\mu_\perp) &=& D_{NS} (z_1,z_2) \nn \\ 
%
%\mbx + D_{NS}^{h_1} D_g^{h_2} \bar{*} && \!\!\!\!\!\!\!\!\!\!\! 
%
%\Bigg( \frac{\A_s}{2\pi} \hat{P}_{q\ra q g} \log(\mu_\perp^2 / m_g^2) 
%
%\Bigg)_{z_1}^{1-z_2} \!\!\!\!\!\!\!+ ...
%
%\eea
%The elipsis at the end of the above equation are meant to indicate 
%all higher order corrections to the above process. All such 
%processes essentially indicate cases where the the two 
%hadrons fragment coherently. Hence these must be included in 
%the phenomenological non-perturbative dihadron fragmentation 
%function. 

We now have to introduce the factorization scale $\mu$ and
redefine the bare dihadron fragmentation function
in terms of a renormalized one and the single fragmentation
functions,
\bea 
D_{NS}^{h_1,h_2} &=&  \label{dqq_bar} \\ 
\bar{D}_{NS}^{h_1,h_2} (\mu,\mu_\perp) 
* \Bigg( 1 
&+& \frac{\A_s}{2\pi} P_{q\ra q g} \log(m_g^2/\mu^2) 
+ ... \Bigg) \nn \\
\mbox{} + D_{NS}^{h_1} D_g^{h_2} \bar{*} 
\Bigg( 
&& \!\!\!\!\!\!\!\!\! \frac{\A_s}{2\pi} P_{q\ra q g} \log(m_g^2/\mu_\perp^2) 
+ ...   \Bigg). \nn
\eea
Note that the log in the second term can be separated into two
pieces

\bea
\log(m_g^2/\mu_\perp^2)=\log(m_g^2/\mu^2)+\log(\mu^2/\mu_\perp^2) ,  \label{split_log}
\eea

\nt
with the first one containing the collinear divergence and
the second piece defining the independent fragmentation
of two collinear partons.
%not contain an infinite divergence, as $\mu_\perp$ 
%cannot be set to zero. This is a finite term which is 
%being reabsorbed into the dihadron fragmentation
%function. The sole purpose of this term is to replace 
%the intermediate scale $\mu_\perp$ with 
%the double fragmentation factorization scale $\mu$. 
In the case where $\mu < \mu_\perp$, we essentially have a function that
depends on two scales: the collinear divergences that are extracted from the 
second term in Eq.~(\ref{dqq_bar}) cannot be factorized out at a scale below 
$\mu_\perp$. However, we may chose to have $\mu > \mu_\perp$, in which case the 
second piece in Eq.~(\ref{split_log}) is a finite constant that may be simply reabsorbed into 
the definition of the renormalized fragmentation function.  
With the factorization scale $\mu$ chosen above the physical scale $\mu_\perp$, 
we may now express Eq.~(\ref{dqq_bar}) as 

\bea 
D_{NS}^{h_1,h_2} &=&  \label{dqq_bar_2} \\ 
\bar{D}_{NS}^{h_1,h_2} (\mu) 
* \Bigg( 1 
&+& \frac{\A_s}{2\pi} P_{q\ra q g} \log(m_g^2/\mu^2) 
+ ... \Bigg) \nn \\
\mbox{} + D_{NS}^{h_1} D_g^{h_2} \bar{*} 
\Bigg( 
&& \!\!\!\!\!\!\!\!\! \frac{\A_s}{2\pi} P_{q\ra q g} \log(m_g^2/\mu^2) 
+ ...   \Bigg). \nn
\eea

\nt 
In what follows we will always compute in the region where $\mu > \mu_\perp $.

%It should be noted that this replacement 
%is only valid if $\mu > \mu_\perp$ \footnote{For 
%$\mu < \mu_\perp$ there will be no  
%independent single fragmentation piece in the NLO contribution. 
%The resulting DGLAP evolution equation will be 
%no different from that of the single 
%fragmentation function.}. 

We may now substitute  Eq.~(\ref{dqq_bar_2}) into 
Eq.~(\ref{dqq_coll}) and concentrate on the leading order and 
leading log (LL) sector of the
fragmentation functions (\ie we only keep terms to order 
$\A_s(Q^2) \log(Q^2) $) to get 

\bea
D_{NS}^{h_1,h_2} (z_1,z_2,Q^2) &=& \label{dqq_full} \\
\bar{D}_{NS}^{h_1,h_2} (\mu)
* \Bigg( 1 
&+& \frac{\A_s}{2\pi} P_{q\ra q g} \log(Q^2/\mu^2) 
+ ... \Bigg) \nn \\
\mbox{} + D_{NS}^{h_1}D_g^{h_2}\bar{*} 
\Bigg( 
&& \!\!\!\!\!\!\!\!\! \frac{\A_s}{2\pi} P_{q\ra q g} \log(Q^2/\mu^2) 
+ ...   \Bigg). \nn
\eea
The above dihadron fragmentation function in the NLO still
contains the bare single fragmentation functions in the
contribution from two independent parton fragmentation.

\begin{widetext}

\begin{figure}[!htb]
\begin{center}
%  \epsfxsize 80mm
%\hspace{0cm}
\resizebox{6in}{6in}{\includegraphics[1in,2in][8in,9in]{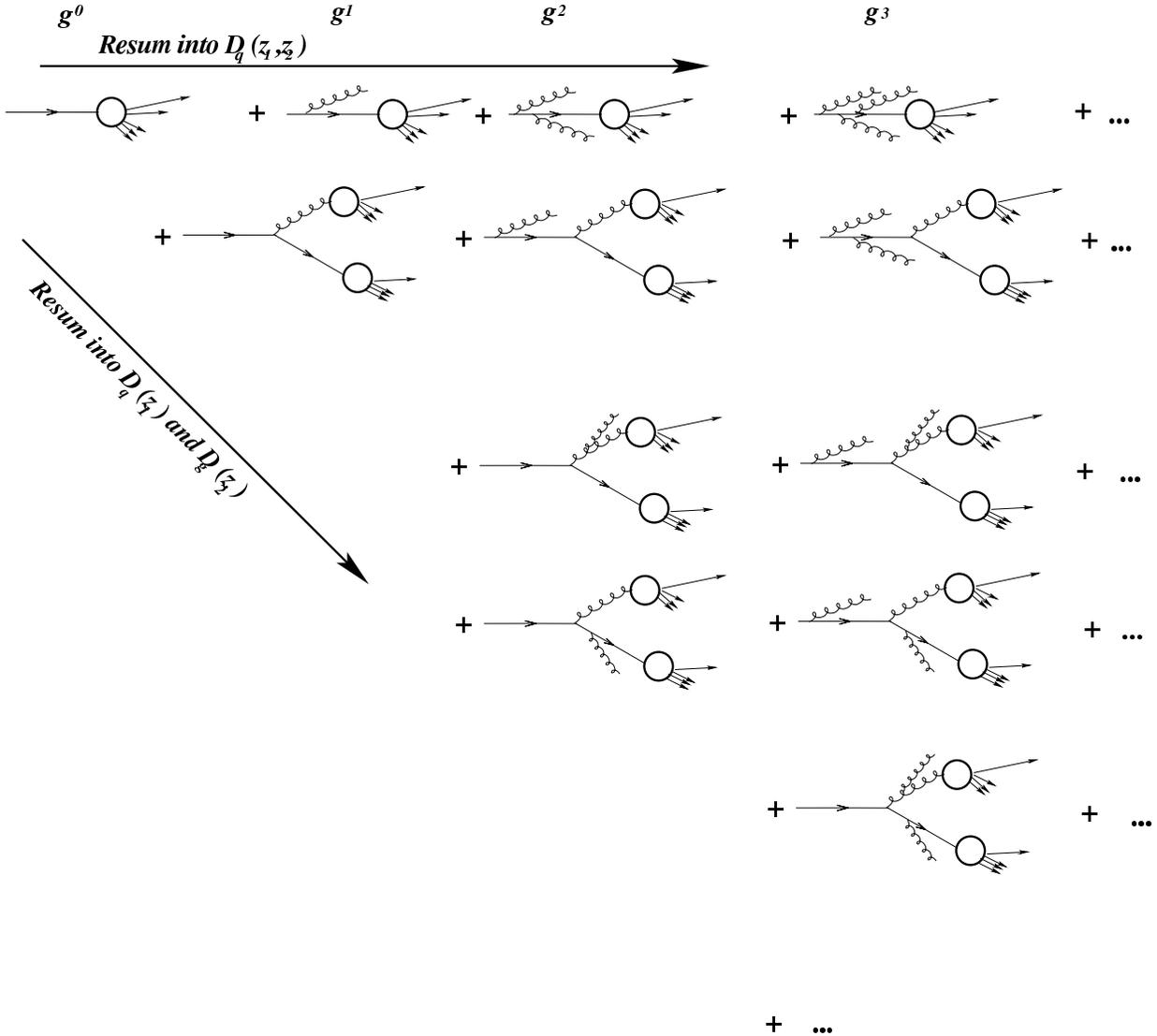}} 
%\vspace{0.25cm}
\caption{ Diagrams to be resummed iteratively to obtain the evolution of the
fragmentation functions.  }
\label{insanity}
\end{center}
\end{figure}

\end{widetext} 

If we consider higher order processes in which an additional
gluon radiation takes place after the split but before
independent fragmentation, as shown in Fig.~\ref{insanity},
another collinear divergence will arise. This is exactly
the same as in the NLO correction to the single inclusive
fragmentation functions.
%We note, as in the case of the single inclusive  
%cross sections, the expressions contain a collinear divergence as 
%$m_g \ra 0$. Recall that $m_g$ is not a realistic mass; it is 
%solely an artifact of the regulation scheme and must be set to 
%zero in the end. 
%In the case of the single inclusive functions 
%one identifies that the \emph{bare} single inclusive functions 
%must diverge in such a way that the physical measurable 
%functions are finite, this is done by introducing the 
%structure,
One has to introduce renormalized single hadron fragmentation
functions at a factorization scale $\mu_1$
\bea 
D_{NS}^{h_1} = \!\bar{D}_{NS}^{h_1}(\mu_1)\! \otimes \!\bigg( 1\! +\! 
\frac{\A_s}{2\pi} P_{q\ra q g}\log(m_g^2/\mu_1^2) + ... \bigg),
\label{dq_bar}
\eea
where, the $\otimes$ indicates the regular convolution notation, \ie 
$A\otimes B = \int \frac{dy}{y}  A (z/y) B(y)$. In addition, the
renormalized gluon fragmentation function is defined as,

\bea 
D_{g}^{h_1} &=& \bar{D}_{g}^{h_1}(\mu_1^2)  \otimes  \bigg( 1 + 
\frac{\A_s}{2\pi} P_{g\ra g g}\log(m_g^2/\mu_1^2) + ... \bigg) \nn \\
\mbx + & \sum\limits_q & \bar{D}_{q/\bar{q}}^{h_1} (\mu_1^2) \otimes
\frac{\A_s}{2\pi} \bigg( P_{q\ra q g}\log(m_g^2/\mu_1^2) + ...\bigg),
\label{dg_bar}
\eea

\nt
where $\bar{D}_{q/\bar{q}}^{h_1}$ represents the quark or antiquark 
fragmentation function and the sum includes all flavours.
%In the above equations the quantities $\bar{D}_{NS}^{h_1},\bar{D}_{g}^{h_1}$ 
%and $\bar{D}_{q_i/\bar{q}_i}^{h_1} $ are all finite in the limit of 
%vanishing $m_g$. The factorization scale $\mu$ is arbitratry and
%may be set equal to the renormalization scale.
The factorization scale $\mu_1$ for the single 
fragmentation functions needs not be the same as the 
factorization scale for double fragmentation functions.

%\bea 
%D_{NS}^{h_1,h_2} (z_1,z_2,Q^2) &=& \label{dqq_coll2} \\ 
%
%D_{NS}^{h_1,h_2}  
%
%* \Bigg( 1 
%
%&+& \frac{\A_s}{2\pi} P_{q\ra q g} \log(Q^2/m_g^2) 
%
%+ ... \Bigg) \nn \\
%
%\mbox{} + \bar{D}_{NS}^{h_1}(\mu_1) \bar{D}_g^{h_2}(\mu_1) \bar{*} 
%
%\Bigg( 
%
%&& \!\!\!\!\!\!\!\!\! \frac{\A_s}{2\pi} P_{q\ra q g} \log(Q^2/\mu_\perp^2) 
%
%+ ...   \Bigg). \nn
%
%\eea
%
%\nt
%We can now factorize the collinear 
%divergence into the Non-Singlet fragmentation
%function by introducing the following structure in the bare 
%dihadron fragmentation functions,

With both the renormalized single and double hadron fragmentation
functions, we obtain the leading log and NLO expressions of the 
double hadron fragmentation functions
%with the collinear divergences absorbed into the bare 
%fragmentation functions: 
\bea
D_{NS}^{h_1,h_2} (z_1,z_2,Q^2) &=& \label{dqq_full_2} \\
\bar{D}_{NS}^{h_1,h_2} (\mu)
* \Bigg( 1 
&+& \frac{\A_s}{2\pi} P_{q\ra q g} \log(Q^2/\mu^2) 
+ ... \Bigg) \nn \\
\mbox{} + \bar{D}_{NS}^{h_1}(\mu_1) \bar{D}_g^{h_2}(\mu_1) \bar{*} 
\Bigg( 
&& \!\!\!\!\!\!\!\!\! \frac{\A_s}{2\pi} P_{q\ra q g} \log(Q^2/\mu^2) 
+ ...   \Bigg). \nn
\eea

\nt
In the above discussion we have set $Q^2$ 
to be large such that there exists a hierarchy 
of scales $\Lambda_{QCD}^2 << \mu_{\perp}^2 << Q^2$. 
The above factorization is 
valid in the regime in which the fragmentation 
functions are measured at a scale $\mu$ such 
that $\mu_\perp < \mu << Q$. In this limit 
we may also set $\mu_1 = \mu$ to
define both single and double 
fragmentation functions at the same scale. Note 
that the single fragmentation functions 
at the new scale $\mu$ differ from those 
at $\mu_1$ at higher order in $\A_s$ and 
thus the correction due to these may be 
ignored in the leading log 
expressions for large enough $Q^2$.

The remaining task in our parton model evaluation of the 
double hadron fragmentation function is to iterate the 
radiative 
process as shown in Fig.~\ref{insanity}. 
%The figure represents the 
%expression for $D_{NS}^{h_1 h_2}(Q^2)$.  
Each of the circular blobs represents a fragmentation 
function at the scale $\mu$.
Differentiating the series with respect to $\log(Q^2)$ followed by a 
reorganization of  the various terms leads to the 
evolution equation:

\bea 
\frac{\prt D_{NS}^{h_1 h_2} (Q^2)}{\prt \log{Q^2}} &=& 
\frac{\A_s}{2\pi} \Bigg[ P_{q\ra q g} * D_{NS}^{h_1 h_2} (Q^2)  \nn \\
\mbx + && \!\!\!\!\!\!\!\!\!\! \hat{P}_{q \ra q g} 
\bar{*} D_{NS}^{h_1} (Q^2) D_g^{h_2} (Q^2)  + 1 \ra 2 \Bigg] \label{ns_dglap}
\eea

Within the framework of the parton model we can picture the process as 
the free propagation of a parton followed by its fragmentation into 
hadrons of which two are identified. Fragmentation may be preceded 
by the radiation of multiple soft gluons (this is the top 
line in Fig.~\ref{insanity}).
Occasionally the parent parton undergoes a semi-hard split into two 
offspring partons which then propagate freely of each other and then 
fragment independently into hadrons and one hadron from each of 
these offspring is identified. Prior to their fragmentation, the 
offspring may radiate multiple soft gluons as well.

%The existance of strong angular ordering between the radiation 
%from the parent parton and that from the offspring partons \cite{dok89}, 
%allows one to resum all the radiation from the offspring into separate 
%exponential series for each offspring parton.  
%In light-cone gauge the differential cross section for multiple 
%gluon emissions becomes just the product of the cross sections of 
%single emissions. One essentially resums all the gluon emissions 
%of the secondary partons after the hard split into the single 
%inclusive fragmentation functions. 

%In this section, explicit expressions for 
%the double differential cross sections at both leading and 
%next-to-leading order have been written down. The collinear divergences 
%were then factorized into the phenomenological fragmentation functions, and 
%their evolution with the energy scale of the process derived.
 
The excercise in this section is based on the validity of our 
assumptions about the nature of the fragmentation process, especially on 
the validity of Eqs.~(\ref{LO_Dz1z2},\ref{sigma_glue_mass}). 
It was assumed that in progressing from single inclusive 
to double inclusive cross sections the parton model dynamics 
would remain the single leading behaviour and more importantly 
would lead to Eq.~(\ref{sigma_glue_mass}). Such a proof exists 
for the single inclusive fragmentation functions that requires 
an exact operator definition of the single fragmentation functions. 
One can demonstrate both the factorized form of Eq.~(\ref{LO_Dz}) and 
the evolution equations of the single inclusive fragmentation functions, 
exactly, as the leading log behaviour at large $Q^2$ in an operator 
formalism (see Refs. \cite{col89,mue78,mue81} and Ref.~\cite{mut98}). 
Mounting such a proof for the dihadron fragmentation functions
will require us to provide a definition of the dihadron fragmentation 
function in the operator formalism. This will be the subject 
of the next section. We will extend the cut-vertex formalism of 
Mueller \cite{mue78,mue81} to dihadron fragmentation in this paper. 
The factorization of 
the NLO expressions at leading twist and leading log will be 
demonstrated in Sec. IV. 
%This will complete our proof of the evolution equations motivated 
%from parton model ideas in this section. 

%%%%%%%%%%%%%%%%%%%%%%%%%%%%%%
%%%%%%%%%%%%%%%%%%%%%%%%%%%%%%
%%%%%%%%%%%%%%%%%%%%%%%%%%%%%%
%%%%%%%%%%%%%%%%%%%%%%%%%%%%%%

\section{The single and double fragmention functions }

%%%%%%%%%%%%%%%%%%%%%%%%%%%%%%
%%%%%%%%%%%%%%%%%%%%%%%%%%%%%%
%%%%%%%%%%%%%%%%%%%%%%%%%%%%%%
%%%%%%%%%%%%%%%%%%%%%%%%%%%%%%

In the previous section, we made use of the  
parton model \cite{fie78} to motivate a 
double inclusive fragmentation function and 
assumed a factorized form as the leading behaviour 
of the two hadron inclusive cross sections. To prove 
the factorized behaviour we need to first obtain a 
consistent definition of the dihadron fragmentation 
function.

We begin with the matrix element 
for the electron positron annihilation into 
a given state of hadrons  in the single photon 
approximation,

\bea 
\mat_{e^+ e^- \ra S_{had}}  &=& e^2 
\int d^4 y \lc S_{had} | J^\nu (y) | 0 \rc \nn \\
\mbox{} \times \frac{-i g_{\nu \mu}}{( \fk_1 + \fk_2 )^2} & & \!\!\!\!\! 
e^{-i(\fk_1 + \fk_2) \x \fy} 
\,\,\,\bar{v} (\fk_2) \g^\mu u (\fk_1) .
\eea
In the above equation, $J^\nu(y)=\sum_qe_q\bar{\psi}_q(y)\gamma^\nu\psi_q(y)$ 
is the hadronic eletromagnetic current
%$eQ_f$ is the electric charge of this current
and $\fk_1,\fk_2$ are the momentum four-vectors 
of the electron and positron. Here the
sum over the number of colors in the fundamental
representation of QCD is implied.
Squaring the matrix element, summing over all final 
states of hadrons and averaging over all initial 
spins of hadrons, one obtains the total cross section for 
\epem annihilation into hadrons,

\bea 
\si &=& \frac{1}{2s} \sum_{S_{had}}
\int \frac{d^3 p_f}{2 E_f (2 \pi)^3}  (2 \pi)^4 
\kd(\fk_1 + \fk_2 - \fP_{S_{had}}) \nn \\
&\times& \frac{e^4}{4 (\fq^2)^2} \mal_{\mu \nu } 
\lc 0 | J^\mu(0) | S_{had} \rc \lc S_{had} | J^{\nu} (0) | 0\rc \nn \\
&=& \frac{e^4}{2s \fq^4} 
\frac{\mal_{\mu \nu} W^{\mu\nu}}{4}, \label{s_w}
\eea   

\nt
where $\mal_{\mu \nu}$ is the leptonic tensor and $W^{\mu \nu}$ 
is the hadronic tensor. The four-momentum of the virtual photon
is $\fq = \fk_1 + \fk_2 \equiv (Q,0,0,0)$ and the Mandelstam
variable $s = \fq^2 = Q^2$. 
The sum over $S_{had}$ includes both the sum of the complete set of
states and the phase space integration
$\prod_{f \in S_{had}} d^3p_f/2E_f (2\pi)^3$. 
%We also note that the coordinate variable $y$ has 
%been integrated out to obtain the $\kd-$function that 
%equates the momentum of the virtual photon to the 
%sum of the four-momenta of the hadrons.  
%%%%%%%%%%%%%%%%%%%%%%%%%%%%%%%%%%%%%%%%%%%%%%

One can evaluate the single inclusive cross 
section by summing over all possible hadronic final
states that contain the identified hadron $h$. In the
leading order and leading twist in a collinear
approximation, one can obtain Eq.~(\ref{LO_Dz}).
%in an exact operator formalism and obtain an expression 
%formally similar to that of Eq.~(\ref{LO_Dz}). 
%Factoring out the hard part represented by 
%$N_c \sum_f Q^2_f \si(\mu \mu)$, will leave us 
%with an operator expression for the fragmentation
%function. In such an evaluation performed 
In a light-cone gauge $(\fn \x {\bf A} = 0)$,
the operator expression for the single 
inclusive fragmentation function at leading twist 
is obtained as \cite{mue78,mue81,osb03},
\bea
D_{q(\bar{q})}^h (z_h) &=& \frac{z_h^3}{2} T_{q(\bar{q})} (z_h)  \nn \\
&=& \frac{z_h^3}{4} \int \frac{d^4 p}{(2\pi)^4}
\kd \bigg( z_h - \frac{\fp_h \x \fn}{\fp \x \fn} \bigg) \nn \\
&\times& \mbox{Tr} \bigg[  
\frac{\fgma \x \fn }{ \fp_h \x \fn } 
\hat{T}_{q(\bar{q})} (\fp,\fp_h)   \bigg],
\label{dz1}
\eea
where the the Dirac operators $\hat{T}_{q(\bar{q})} (\fp,\fp_h)$ 
are given by
\bea 
(\hat{T}_{q})_{\al,\B} (\fp,\fp_h) &=& \int d^4 x \sum_{S_{had}-1} 
\lc 0 | \psi_{\al} (0) | \fp_h, S_{had}-1 \rc \nn \\
& \times &\lc \fp_h, S-1 | \bar{\psi}_{\B} (x) | 0 \rc  e^{i\fp \x \fx} \\
(\hat{T}_{\bar{q}})_{\al,\B} (\fp,\fp_h) &=& \int d^4 x \sum_{S_{had}-1} 
\lc 0 | \bar{\psi}_{\B} (0) | \fp_h, S_{had}-1 \rc  \nn \\
& \times &\lc \fp_h, S_{had}-1 | \psi_{\A} (x) | 0 \rc  e^{i\fp \x \fx} .
\label{dz1_op}
\eea
Here, the sums are taken over all physical final 
states of hadrons, which always contain, at least, the 
single hadronic state with momentum $\fp_h$. In this 
case $\fn$ is chosen such that its spatial components
are antiparallel to the spatial components of the 
observed hadron. This implies $\fn \x \fp_h = p_h^+ = (p_h^0 + |\vp_h| )/2$.
In our choice of light-cone momenta $p^- = p^0 - p_z$. The gauge 
links required to make this expression gauge invariant have been suppressed
as they do not contribute to the leading twist fragmentation 
functions in light-cone gauge. 

\begin{figure}[ht!]
%\begin{center}
%  \epsfxsize 80mm
%\hspace{0cm}
  \resizebox{4in}{4in}{\includegraphics[0.5in,0.0in][5.5in,5in]{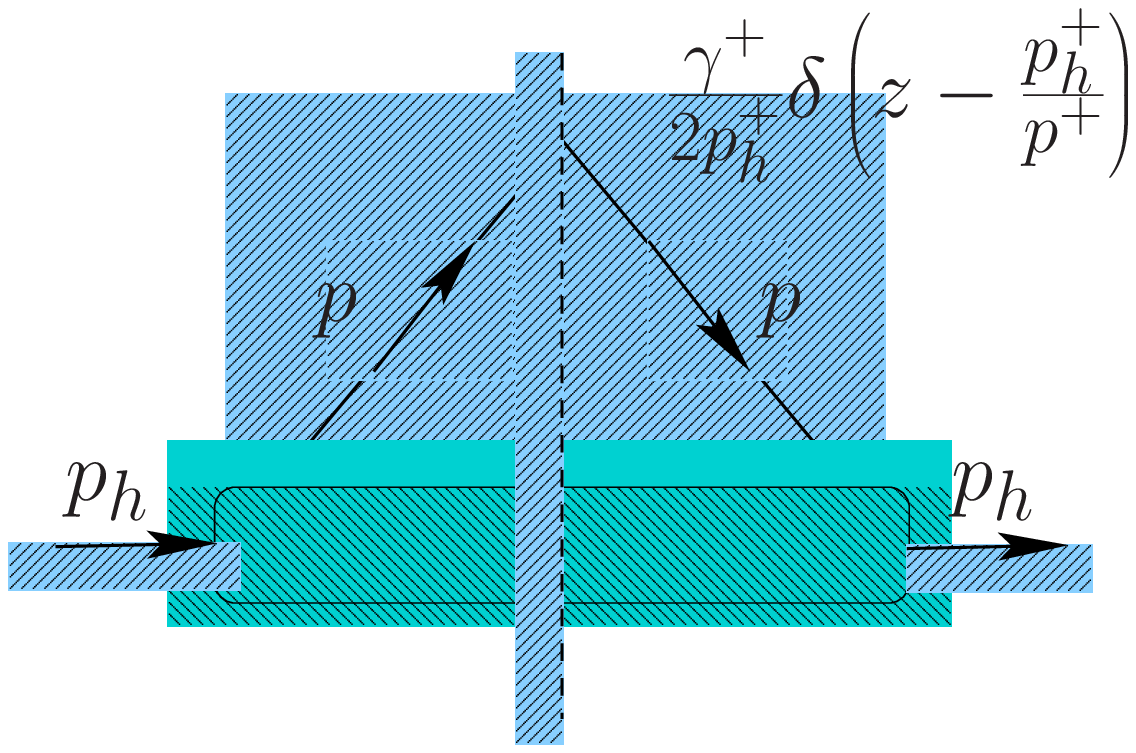}} 
%\vspace{0.25cm}
    \caption{cut-vertex for quark fragmention function at LO.}
    \label{cut_dz1}
%  \end{center}
\end{figure}

The fragmentation functions can also be reexpressed in the cut-vertex 
technique of Mueller. These represent a powerful computational tool 
that may be used to calculate inclusive cross sections and the scale 
dependence of these functions in perturbation theory in a diagramatic 
languange. The Feynman diagrams 
illustrating the leading order expressions for the single inclusive 
fragmentation functions are shown in Fig.~\ref{cut_dz1}. In this 
Feynman diagram the rule for the bare quark cut-vertex is
\bea
\frac{ \fgma \x \fn }{ 2 \fp_h \x \fn }  
\kd \bigg( z_h - \frac{\fp_h \x \fn}{\fp \x \fn} \bigg) .
\label{cut_ver_q}
\eea

\nt
The derivation of operator definitions for dihadron fragmetation functions
and the extension of the cut-vertex technique to incorporate these 
functions is the focus of this section.

%It is a fact of experiment, that an \epem 
%annihilation leads, at high energies, to the appearence of 
We will concentrate on the two-jet events in \epem annihilation 
and are interested only in two hadron production
off one single jet. The fate of the ``back-side'' jet will not be 
dwelled over here. We assume that the sum over all hadronic 
states in $W^{\mu \nu}$ can be simplified into
two complete sets of states and that each overlaps 
independently with the quark and antiquark jet.
This assumption neglects the interference between the two jets
and is valid in leading log and leading twist. 
We also assume the duality between the complete hadronic
states and partonic states. Thus the sum over hadronic states
in the ``back-side'' jet will be replaced by partonic states.
%consist of a sum over all hadron states in one jet and another
%sum over all possible states of a 
%single antiquark or quark in the opposite 
%direction. The justification behind this 
%procedure is the assumption of an equivalence between a sum over 
%hadronic states and over partonic states and the 
%leading order behaviour of the partonic states 
%to look like a quark and antiquark moving in the 
%opposite direction followed by almost independent 
%fragmentation. 
Under this assumption, we evaluate the 
hadronic current operator, order by order in $\A_s$, by 
expanding the QCD interaction Hamiltonian in the 
interaction picture. 
%At LO we have 
%$J^\mu = ig \bar{\psi} \g^\mu \psi$. In the next section 
%we will expand the interaction Lagrangian to higher order.
%Substituting this into the expression for the hadronic tensor, 
In the leading order (LO), one obtains,
\bea
W^{\mu \nu} &=& \sum_{S_{had} - 2} \sum_q e_q^2 
%\Bigg( \prod_{f \in S_{had}-2} 
%
%\frac{d^3p_f}{2E_f (2\pi)^3} \Bigg)
%
\int \frac{d^3p_1d^3p_2}{4E_1E_2 (2\pi)^6}   \nn \\
&& \!\!\!\!\!\!\!\!\!\!\!\! \int \frac{d^3k}{2E_k (2\pi)^3}     (2\pi)^4
\kd^4 (\fq - \fp_1 -\fp_2 - \fk - 
\!\!\!\!\!\!\!\! \sum_{f \in S_{had} -2}  \fp_f ) \nn \\
&\times& \lc 0 | \bar{\psi}_q (0) \g^\mu \psi_q(0) 
| k,p_1,p_2,S_{had} -2  \rc \nn \\
&\times& \lc k,p_1,p_2,S_{had} -2 | \bar{\psi_q} (0) \g^\nu \psi_q(0) |0 \rc . 
\label{w_mu_nu_1}
\eea

\begin{figure}[htb!]
%\begin{center}
%  \epsfxsize 80mm
%\hspace{0cm}
  \resizebox{4in}{4in}{\includegraphics[2in,1in][10in,9in]{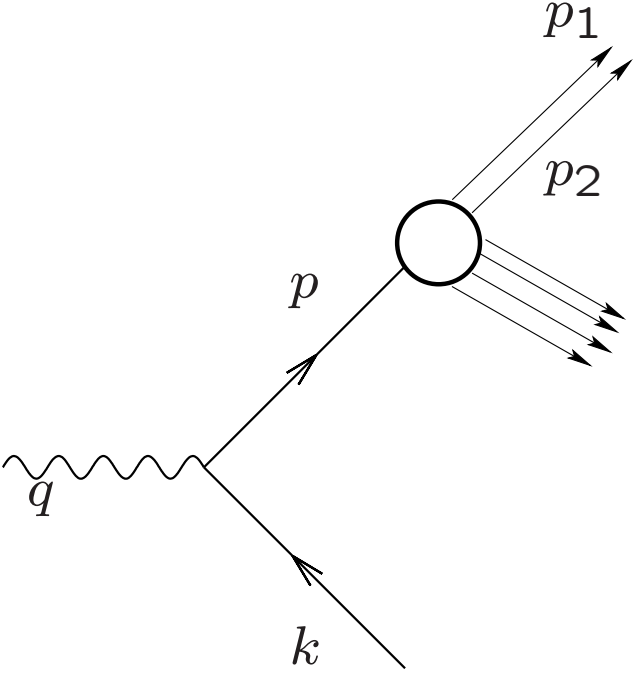}} 
%\vspace{0.25cm}
    \caption{The leading order Feynman diagram contributing to the 
    double inclusive fragmentation function.}
    \label{fig3}
%  \end{center}
\end{figure}

\nt 
In the above equation, $S_{had}$ is a complete set of hadronic states.
%on only one \emph{side} of the reaction. 
In the remaining discussion we will drop the subscript ($\mbx_{had}$).
We have extracted two particular hadronic state sums, 
labeled as $p_1,p_2$, from the full sum over states $S$. The 
hadronic tensor may be represented by the Feynman diagram 
in Fig.~\ref{fig3}.

On Fourier decomposition of one of the quark or antiquark operators, followed 
by a sum over all spins of the outgoing antiquark (quark) state $\fk$, 
we obtain the hadronic tensor as,
\bea 
W^{\mu \nu} &=&  \sum_{q,S - 2} e_q^2 
%\Bigg( \prod_{f \in S-2} 
%\frac{d^3p_f}{2E_f (2\pi)^3} \Bigg)
%
\int \frac{d^3p_1d^3 p_2}{4E_1E_2 (2\pi)^6}    \label{w_mu_nu_2}\\
&& \hspace{-0.5in} \int \frac{d^4k}{(2\pi)^4} 
2\pi \kd^+ (\fk^2) (2\pi)^4 \kd^4 (\fq - \fp_1 -\fp_2 - \fk - \fp_{S-2}) \nn \\
& & \hspace{-0.5in}
\Bigg[ \Big\{ \lc 0 | \bar{\psi}_q (0) | p_1,p_2,S -2  \rc 
\g^\mu \f k \g^\nu  \lc p_1,p_2,S-2 |  \psi_q(0) |0 \rc  \Big\}\nn \\
&& \hspace{-0.5in}
+ \Big\{  \g^\mu \lc 0 | \psi(0) | p_1,p_2,S -2  \rc 
\lc p_1,p_2,S -2 | \bar{\psi} (0) |0 \rc \g^\nu \f k \Big\} \Bigg]. \nn
\eea

\nt 
One may rewrite the $\kd-$function as a four-space integration 
of an exponent $\exp(-i \fp \x \fx)$ that in turn can be used 
to transform the quark wavefunction operator $\psi(0) \ra \psi(x)$. 
A shift in the $d^4 k$ integration can be performed 
\ie $\fk \equiv \fq - \fp$. Summing over the spins (and colors)
of all the final (parton) states we obtain the general form of the 
hadronic tensor as 
\bea 
W^{\mu \nu} &=& N_c \sum_q e_q^2 
\int \frac{d^3p_1d^3 p_2}{4E_1E_2 (2\pi)^6} \int \frac{d^4p}{(2\pi)^4} 
   \label{w_mu_nu_3} \\
&& \!\!\!\!\!\!\!\!\!\!\!\!\!\!\!\!\!\!\!\!\!  
2\pi \kd^+ ((\fq - \fp)^2)   \Bigg[ 
\tr \Big\{ \hat{T}_{\bar{q}} (p;p_1,p_2) \g^\mu (\f q - \f p) \g^{\nu} 
\Big\} \nn \\
&& \!\!\!\!\!\!\!\!\!\!\!\!\!\!\!\!\!\!\!\!\! \mbx + 
\tr \Big\{ \hat{T}_q (p;p_1,p_2) \g^\nu (\f q - \f p) \g^{\mu} \Big\}  
\Bigg],\nn 
\eea
where the parton-double-hadron overlap matrices are given
similarly as 
%the case of 
for single fragmentation functions as
\bea
\left[ \hat{T}_{\bar{q}}(p;p_1,p_2) \right]^{\alpha\beta} &=&  
\int d^4 x e^{i\fp \x \fx} \sum_{S - 2} \nn \\
%\Bigg( \prod_{f \in S-2} 
%\frac{d^3p_f}{2E_f (2\pi)^3} \Bigg) 
%
& & \hspace{-1.0 in}  
\lc 0 | \bar{\psi}_q^\beta (x) | p_1 p_2 S-2 \rc   
\lc p_1 p_2 S-2 | \psi_q^\alpha (0) | 0 \rc \label{t_aq}
\eea 
and
\bea
\left[ \hat{T}_q(p;p_1,p_2) \right]^{\alpha\beta} 
&=&  \int d^4 x e^{i\fp \x \fx} \sum_{S - 2} \nn \\
%\Bigg( \prod_{f \in S-2} 
%\frac{d^3p_f}{2E_f (2\pi)^3} \Bigg) 
%
& & \hspace{-1.0 in}
\lc 0 | \psi_q^\alpha (x) | p_1 p_2 S-2 \rc  
\lc p_1 p_2 S-2 | \bar{\psi}_q^\beta (0) | 0 \rc \label{t_q},
\eea
where the two quark field operators have the same
color index and average over colors is explicitly implied.

Up to this point the derivation of the factorized double 
inclusive fragmentation function has followed a path not entirely 
dissimilar to that of a single fragmentation function. At this point 
one may introduce the light-cone variables and their ratios, 
\[
z_1 = p_1^+ / p^+  \mbox{  ,  }  z_2 = p_2^+ / p^+  \mbox{  ,  }
\]
and $z = z_1 + z_2 = (p_1^+ + p_2^+)/p^+ \equiv p_h^+/p^+ $. Where 
$p_h$ is the total momentum of the pair of hadrons that are 
identified. We now take the collinear approximation
that the hadron momenta $p_1,p_2,p_h$ are 
almost collinear with respect to the quark (antiquark) 
momentum at high energies and are thus dominated 
by their $+$ components for light-cone vector $\fn$
chosen in the direction of the outgoing quark or antiquark.
Essentially, $p_1^+ >> p_1^-,{p_1}_\perp $ 
and the same is true for $p_2$ and $p_h$. The overlap 
matrices $\hat{T}_{\bar{q}}, \hat{T}_q$ have a Dirac matrix 
structure and hence can be decomposed in a basis of products 
of $\g$ matrices ($1,\g^\mu,\si^{\mu\nu},\g^5\g^\mu,\g^5$). 
The only term of this basis to survive is 
$\g^\mu$: even combinations are set to zero in the 
trace and combinations containing $\g^5$ vanish under a 
spin sum. As $\hat{T}_{\bar{q}}, \hat{T}_q$  are scalers and 
only depend on the three almost collinear momenta, we 
obtain at leading twist the following decomposition of the overlap 
matrices:
\bea
\hat{T}_q(p;p_1,p_2) &=& \frac{\f p_h}{2} T_q (p;p_1,p_2) \;\;{\rm or} \nn \\
T_q(p;p_1,p_2) &=& \frac{ \tr [ \f n \hat{T}_q(p;p_1,p_2) ]}{2 \fn \x \fp_h},
\eea
where $T_q(p;p_1,p_2)$ is a scalar function.  The entire 
Dirac structure has been extracted into the $\g$ matrix.

With the aid of $\kd-$functions, we can introduce the
definition of the fractional momenta $z_1$ and $z_2$
into the hadronic tensor:
\bea
W^{\mu \nu} &=& N_c\int_0^1 d z_1 d z_2 \h(1-z_1-z_2) 
\int \frac{d^3p_1d^3 p_2}{4E_1E_2 (2\pi)^6}   \label{w_mu_nu_4}\\
&& \hspace{-0.3 in} 2\pi \kd^+ (\fq^2 - 2 \fq \x \frac{\fp_h}{z} )  
\sum_q e_q^2 \Bigg[ \tr\Big\{ \frac{\f p_h }{2} \g^\mu 
\big( \f q - \frac{ \f p_h }{z}  \big) \g^\nu \Big\} \nn \\
&\times& \int \frac{d^4 p}{(2\pi)^4} T_{\bar{q}} (p;,p_1,p_2) 
\kd (z_1 - \frac{p_1^+ }{p^+}) \kd (z_2 - \frac{p_2^+ }{p^+}) \nn \\
&+& \tr\Big\{ \frac{\f p_h }{2} \g^\nu 
\big( \f q - \frac{ \f p_h }{z}  \big) \g^\mu \Big\} 
\int \frac{d^4 p}{(2\pi)^4} T_{q} (p;,p_1,p_2)  \nn \\
&\times& \kd (z_1 - \frac{p_1^+ }{p^+}) \kd (z_2 - \frac{p_2^+ }{p^+})  
\Bigg]. \nn
\eea

%In the above equation, after introducing the $z_1,z_2$ integrations, 
%we invoke the collinear approximation on
In collinear approximation, we expand the hard part
\bea
\mathcal{H'}^{\mu \nu } = 
\tr \Big[ \frac{\f p_h}{2} \g^\mu \big( \f q - \f p \big) \g^\nu \Big] 
\kd^+ ( (\fq - \fp)^2 ). \label{hardpart}
\eea
in the transverse momentum of the hadrons $p_{h\perp}$ and take
only the leading term $\mathcal{H'}^{\mu \nu } (p^+) \simeq  
\mathcal{H'}^{\mu \nu } (p^+ = p_h^+/z) + ... $.
This approximation allows us to factor out the hard part 
from the $d^4 p$ integral. This is the first step in the 
factorization of the double hadron inclusive cross section.
%of the expressions into a hard part and the 
%fragmentation functions which encode the soft physics. 
Based on the collinear approximation we have also 
dropped the term $ \fp^2$ or $\fp_h^2 / z^2 $ from the argument of the 
$\delta^+$ function (\ie $ \fq^2 - 2\fq \x \fp >> \fp^2 $). Given the 
light-cone structure of the four-vector $\fp$ a further 
simplication of the argument of the $\kd-$function may be obtained: 
\bea
\fq^2 - 2\fq \x \fp &=& q^+q^- 
- ( q^+ p^- + q^- p^+ - q_\perp \x p_\perp ) \nn \\ 
&=& Q^2 - \left( \frac{Q}{2} p^- + Q p^+ \right) , 
\mbox{    as  $q_\perp = 0$} \nn \\
\mbx = Q^2 - Q p^+ &=& Q^2 - Q \frac{p_h^+}{z},
\mbox{    as  $p^- << p^+ = p_h^+/z $ } \nn \\
&\simeq & \fq^2 - 2 \fq \x \frac{\fp_h}{z}
\eea
 
With the above simplifications and reorganization of
arguments in the two internal $\delta$ functions,
%it may be noted that 
%$\mhp^{\mu \nu} = \mhp^{\nu \mu}$. In the remaining, the 
%discussion will focus on just the quark fragmentation. 
%The antiquark piece may be obtained by a mere replacement of 
%the overlap matrices (Eqs.~(\ref{t_aq},\ref{t_q})). The quark 
%sector of the 
the hadronic tensor can be written as
\bea
W_q^{\mu \nu} &=& N_c \int_0^1 d z_1 d z_2 \h(1-z_1-z_2) 
\int d\T p_1 d\T p_2   \label{w_mu_nu_5}\\
&& 2\pi \kd^+ (Q^2 - 2 Q\frac{p_h^+}{z} )  
\mh^{\mu \nu} \nn \\
&\times& \sum_q e_q^2 \int \frac{d^4 p}{(2\pi)^4} 
\bigg[ T_q (p;,p_1,p_2)+T_{\bar{q}}(p;,p_1,p_2)\bigg] \nn \\
&\times& \frac{{p^+}^2}{z_1z_2} 
\kd (p^+ - \frac{p_1^+ }{z_1}) \kd (p^+ - \frac{p_2^+ }{z_2}), \nn
\eea
where we have introduced the shorthand notation 
$d \T p = d^3 p/(2\pi)^3 2 E_p$. The variable 
$p^+$ is overdetermined and thus one of the 
$\delta$ functions acts really on the integrations 
external to $d^4p$. These may be extracted and 
further reorganized as follows, 
\bea 
&& ... \frac{1}{z_1z_2} \int \frac{d^4 p}{(2\pi)^4} {p^+}^2 
\kd (p^+ - \frac{p_1^+ }{z_1}) \kd (p^+ - \frac{p_2^+ }{z_2}) ... \nn \\ 
&=& ...\frac{1}{z_1z_2} \kd (\frac{p_1^+ }{z_1} - \frac{p_2^+ }{z_2})
\int \frac{d^4 p}{(2\pi)^4} {p^+}^2 \kd (p^+ - \frac{p_1^+ }{z_1}) ... \nn \\
&=& ... \frac{1}{z_1z_2} \kd (\frac{p_1^+ }{z_1} - \frac{p_h^+ - p_1^+ }{z_2})
\int \frac{d^4 p}{(2\pi)^4} {p^+}^2 \kd (p^+ - \frac{p_1^+ }{z_1}) ... \nn \\
&=& ... \frac{1}{z_1 + z_2} \kd ( p_1^+ - \frac{p_h^+ z_1}{z_1 + z_2}) 
\int \frac{d^4 p}{(2\pi)^4} {p^+}^2 \kd (p^+ - \frac{p_1^+ }{z_1}) ... \nn \\
&=& ... \frac{1}{z} \kd ( p_1^+ - \frac{(p_1^+ + p_2^+ )z_1}{z})
\int \frac{d^4 p}{(2\pi)^4} {p^+}^2 \kd (p^+ - \frac{p_h^+ }{z}) ... \nn \\
&=& ... p_h^+ \kd ( z_2 p_1^+ - z_1 p_2^+) 
\int \frac{d^4 p}{(2\pi)^4} \kd ( z - \frac{p_h^+ }{ p^+}) ... ,
\eea
where $z = z_1 + z_2 $. Substitution of these expressions 
into the hadronic tensor $W^{\mu \nu}$ followed by a 
substitution of $W^{\mu \nu}$ back into Eq.~(\ref{s_w}) 
leads to the following expression for the total cross 
section for \epem annihilation,
\bea
\si &=& e^4N_c\sum_q e_q^2\int dz_1 dz_2 \h(1-z_1-z_2) 
\frac{\ml_{\mu \nu} \mh^{\mu \nu}(Q) }{8 Q^6} \frac{z}{2} \nn \\
\times & & \!\!\!\!\!\!\!\!\!\!\! 
\int d\T p_1 d\T p_2 2 \pi \frac{z}{2 Q} 
\kd^+ (p_h - z Q/2) p_h^+ \kd (z_2 p_1^+ - z_1 p_2^+) \nn \\
\times & & \!\!\!\!\!\!\!\!\!\!\!
\int \frac{d^4p}{(2\pi)^4} 
\kd (z - \frac{p_h^+}{p^+}) \bigg[ T_q(p;p_1,p_2) 
+ T_{\bar{q}} (p;p_1,p_2) \bigg],
\label{sigma_2}
\eea 
where the external $\kd-$function has been used to set
\[
\mhp^{\mu \nu} (p_h, Q) \equiv \frac{z}{2Q}
\delta^+(p_h-zQ/2) \mh^{\mu \nu} (Q).
\]
The dependence of the hard part on hadronic variables is replaced 
with the appropriate partonic variables.
%the external factor of $z/2$ is obvious. 
Differentiating the above equation with respect 
to $z_1,z_2$ leads to the double differential 
cross section outlined in Eq.~(\ref{LO_Dz1z2})). 
Before the extraction of the double inclusive 
fragmentation function, some simplification of
the above equation is in order.
In contrast to the definition of the
single fragmentation function, there are double hadronic 
integrals $d^3 p_1 , d^3 p_2$ and two sets of 
$\kd-$functions as opposed to one.
The cautious reader will note that the overlap matrices 
$\hat{T}_q (p;p_1,p_2) , \hat{T}_{\bar{q}} (p;,p_1,p_2)$ 
[see Eqs.~(\ref{t_aq},\ref{t_q})]  are 
dimensionally different from the overlap matrices in the 
definition of the single inclusive functions (see Ref.~\cite{osb03}).

%One also notes that the 
%hard part is now expressed solely in terms of the
%partonic variables. There also remains the problem 
%of energy dimensions of 
%The resolution of all these problems 

\begin{figure}[htb!]
%\begin{center}
%  \epsfxsize 80mm
%\hspace{0cm}
  \resizebox{3in}{3in}{\includegraphics[0.5in,1in][9.5in,10in]{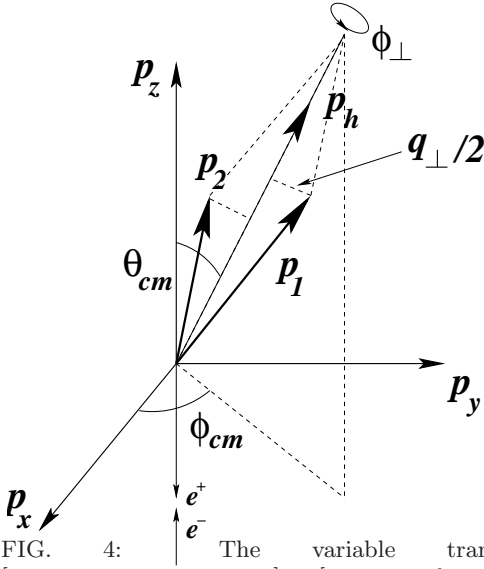}} 
%\vspace{0.25cm}
    \caption{The variable transform from 
    $[{p_1}_x,{p_1}_y,{p_1}_z,{p_2}_x,{p_2}_y,{p_2}_z]$ to 
    $[p_1,p_2,q_\perp,\h_{cm},\phi_{cm},\phi_\perp]$.}
    \label{fig4}
%  \end{center}
\end{figure}

To simplify, we begin with a 
variable transformation. One essentially changes from the 
set $[{p_1}_x,{p_1}_y,{p_1}_z,{p_2}_x,{p_2}_y,{p_2}_z]$ 
to the set $[p_1,p_2,q_\perp,\h_{cm},\phi_{cm},\phi_\perp]$
%The new set of variables are 
as illustrated in Fig.~\ref{fig4}.
This choice is not entirely arbitrary. The discussion 
of the NLO in the previous section required us to incorporate 
transverse momenta $q_\perp$ up to a semihard scale $\mu_\perp$
into the bare fragmentation function. This particular variable 
transformation allows us to isolate the $q_\perp$ integration. 

The new vector $\vp_h = \vp_1 + \vp_2 $ has the 
three components of mostly massless four-vectors. The 
requirement that $p_h^+ = p_1^+ + p_2^+$ is trivially fulfilled.
The new ``body-fixed'' variable $q_\perp$ quantifies 
the component of $\vp_1-\vp_2$ that lies on the plane 
perpendicular to $\vp_h$ and $\phi_\perp$ is the azimuthal
angle of $\vq_\perp$ on this plane. The angles $\h_{cm}$ and $\phi_{cm}$ 
quantify the direction of $\vp_h$ with respect to the \epem 
beam direction as the $z$ axis. The Jacobian for this tranformation 
is simply 

\bea
J = \frac{q_\perp}{4} (p_1 + p_2)^2, \nn
\eea

\nt
at leading twist. With these new 
variables one may easily relate the partonic variable in the 
hard part with the hadronic variable $\vp_h$ as $\vp/z$, \ie the 
sum of the 3-momenta of the detected hadrons is collinear with 
the 3-momenta of the fragmenting quark or antiquark. It may 
be demonstrated that the corrections to this statement 
contribute at higher twist. It should be pointed out, in 
passing, that this is a more accurate statement than the 
assumption of collinearity between the leading hadron and the
fragmenting quark (antiquark) in the case of single 
fragmentation, as in most cases a dominant part of the jet's momenta 
is contained in the momenta of the leading hadrons.

With these new variables we can evaluate the inner product 
of the leptonic tensor and the hard part of the hadronic tensor,
\bea
\ml_{\mu \nu} \mh^{\mu \nu} = 4 Q^4 (1 + \cos^2\h_{cm} )
\eea 
and obtain the double differential cross section as 
\bea 
\frac{d\si}{dz_1 dz_2} &=& \h(1-z_1-z_2) \sum_q \frac{e^4 e_q^2 N_c}{2s} 
\int \frac{dp_1 dp_2 }{(2\pi)^5 4 p_1 p_2} \nn \\ 
& & d \cos\h_{cm} d \phi_{cm} d q_{\perp} 
d \phi_\perp  \frac{q_\perp}{4} (p_1 + p_2)^2 \nn \\
&\times& \frac{z}{2Q} \kd^+ \left(p_h - z\frac{Q}{2}\right) 
\frac{p_h^+}{z_1 z_2} 
\kd\left(\frac{p_1^+}{z_1} - \frac{p_2^+}{z_2} \right) \frac{z}{2}  \nn \\
&\times& (1 + \cos^2\h_{cm})  
\int \frac{d^4 p}{(2\pi)^4} \kd \left( z - \frac{p_h^+}{p^+}  \right) \nn \\
&\times& \bigg[ T_{\bar{q}}(p;p_1,p_2) + T_q(p;p_1,p_2) \bigg] .
\label{sigma_3}
\eea
Assuming that the overlap matrices, $T_{\bar{q}}(p;p_1,p_2)$
and $T_q(p;p_1,p_2)$, are independent of the angles, $\theta_{cm}$,
$\phi_{cm}$ and $\phi_\perp$, one can carry out the integrations
over these variables.
%It is a trivial matter to perform the integrations 
%over $\phi_{cm}, \phi_\perp$
%as nothing else depends on them. The only term depending on $\h_{cm}$ is the 
%inner product of the two tensors; 
%integrating which lead to a simple factor of $\frac{8}{3}$.
It may once again be stipulated that the following 
factorization is being performed and expected to 
succeed only at high energies and momenta (large $Q^2$ limit). 
In this limit we note 
\[
p^+ \simeq p\left[ 1 - O\left( \frac{q_\perp^2}{ p^2} \right) \right]. 
\] 
Hence, in the collinear limit $p,p^+ >> q_\perp$, one may 
substitute $p \ra p^+$ in the entire integrand. Regardless 
of the presence of $\delta$ functions it may be demonstrated
that the correction to this approximation is suppressed by at least 
a power of $Q^2$. The two remaining 
$\kd-$functions external to the $d^4p$ integration, \ie 
\[
\kd^+ \left(p_1 + p_2 - z\frac{Q}{2}\right) 
\kd\left(\frac{p_1^+}{z_1} - \frac{p_2^+}{z_2} \right),
\]
may be used to evaluate the $p_1$ and $p_2$ integrals. 
Factoring out the LO total \epem annihilation cross 
section $\sigma_0^{q\bar{q}}$, we obtain the factorized 
double differential cross section in a form 
similar to the structure of Eq.~(\ref{LO_Dz1z2}) as,
\bea 
\frac{d\si}{dz_1 dz_2} &=& \sum_q \si_0^{q\bar{q}} \int 
\frac{ q_\perp dq_\perp}{4(2\pi)^2}  \frac{z^4}{4z_1z_2}
\int \frac{d^4 p}{(2\pi)^4}   \\
&\times & \bigg[ T_{\bar{q}}(p;p_1,p_2) + T_q(p;p_1,p_2) \bigg]
\kd \left( z - \frac{p_h^+}{p^+}  \right) . \nn 
\eea

We thus arrive at the definition of the leading order 
double inclusive fragmentation function as

\bea
D_q^{h_1,h_2}(z_1,z_2) &=& \int 
\frac{dq_\perp^2}{8(2\pi)^2}  \frac{z^4}{4z_1z_2}
\int \frac{d^4 p}{(2\pi)^4}   \label{dihad_def} \\
&\times & T_{q}(p;z_1p,z_2p) 
\kd \left( z - \frac{p_h^+}{p^+}  \right) .\nn
\eea

\begin{figure}[htb!]
%\begin{center}
%  \epsfxsize 80mm
%\hspace{0cm}
  \resizebox{4in}{4in}{\includegraphics[0in,0in][6.0in,6in]{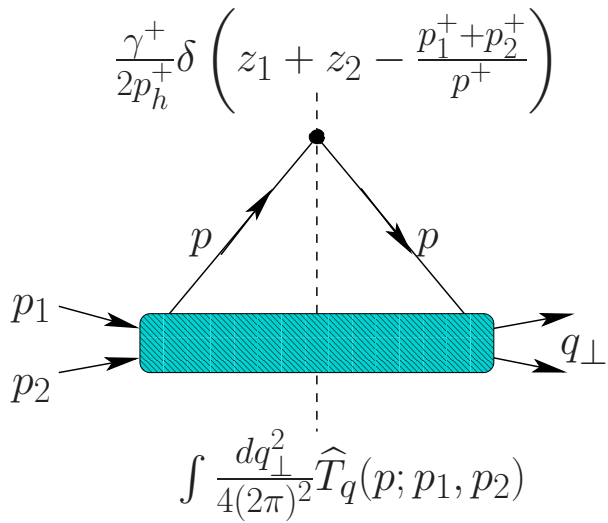}} 
%\vspace{0.25cm}
    \caption{The cut-vertex representation of the dihadron fragmentation
    function. }
    \label{cutvert1}
%  \end{center}
\end{figure}

\nt
In cut-vertex notation, the dihadron fragmentation function may also be 
expressed by the following equation 
\[
D_q^{h_1 h_2} (z_1, z_2) = \frac{z^4}{4z_1 z_2} \tilde{T}_q(z_1,z_2) ,
\]
where $\tilde{T}(z_1,z_2)$ is given by the 
diagram in Fig.~\ref{cutvert1}.
Note that the bare cut-vertex has undergone no change as 
compared to the single hadron fragmentation function, except it 
takes as input the sum of the fractional momenta $z = z_1 + z_2$. The soft 
hadronic sector is slightly modified by the exclusion of two 
hadronic momenta (instead of one) and the integration of 
the transverse momenta $q_\perp$.

Note that the definition of the dihadron fragmentation functions
in their factorized form as in Eq.~(\ref{LO_Dz1z2})
seems to depend on our choice of variable transformation 
(see. Fig.~\ref{fig4}). It may indeed be possible to use 
a different variable transformation and 
obtain a similar factorization. The sole constraint on the choice of 
transformation to be used is based on the factorization of the NLO 
expressions into a form similar to that of Eq.~(\ref{sigma_glue_mass}),
with the same definition of the double fragmentation functions. 
We will demonstrate this in the next section. A second 
motivation for this choice of variables is the ability to isolate the 
transverse momentum $q_\perp$, which is integrated over. In the operator
definition of the dihadron fragmentation functions, $q_\perp$
is generated non-perturbatively. So we will call it intrinsic.
The upper limit of integration of this intrinsic $q_\perp$ 
has not been specified in Eq.~(\ref{dihad_def}). From
our discussion in the previous section of the parton model 
and of the NLO processes in the next section, the upper limit may 
be set as $\mu_\perp$.
%({\bf comments by XNW})
We will assume that hadron pairs with the relative transverse
momentum $q_\perp>\mu_\perp$ are generated only perturbatively.

%%%%%%%%%%%%%%%%%%%%%%%%%%%%%%%%%%%
%%%%%%%%%%%%%%%%%%%%%%%%%%%%%%%%%%%
%%%%%%%%%%%%%%%%%%%%%%%%%%%%%%%%%%%
%%%%%%%%%%%%%%%%%%%%%%%%%%%%%%%%%%%
%%%%%%%%%%%%%%%%%%%%%%%%%%%%%%%%%%%

\section{Cross section at NLO and DGLAP evolution }

%%%%%%%%%%%%%%%%%%%%%%%%%%%%%%%%%%%
%%%%%%%%%%%%%%%%%%%%%%%%%%%%%%%%%%%
%%%%%%%%%%%%%%%%%%%%%%%%%%%%%%%%%%%
%%%%%%%%%%%%%%%%%%%%%%%%%%%%%%%%%%%%
%%%%%%%%%%%%%%%%%%%%%%%%%%%%%%%%%%%

%In the preceeding section the exact operator definition 
%of the dihadron fragmentation function was presented. 
%The final results were expressed in the cut-vertex formalism. 
%As has been pointed out this validates the factorized 
%cross section presented in Eq.~(\ref{LO_Dz1z2}). 
With the definition of the dihadron fragmentation functions in the operator formalism,
which is shown to factorize from the hard parton cross section in
the LO, we are ready to study the DGLAP evolution of the dihadron
fragmentation functions by computing the double inclusive cross 
section at next to leading order (NLO). The calculation will also
justify the factorized form of  Eq. ~(\ref{sigma_glue_mass}) and 
our LO definition of the dihadron fragmentation functions.
 
%The total cross section at next to leading order for the production
%of hadronic states from an \epem annihilation is 
%
%\bea
%\si &=& \frac{1}{2s} \sum_Z \prod_{f \in Z} \int \frac{d^3 p}{ (2 \pi )^3 2 E_f} \nn \\
%
%&\times& (2\pi)^4 \kd^4 ( \vk_1 + \vk_2 - \sum_f  \vp_f ) | \mat_{e^+ e^- \ra Z} |^2
%
%\label{sig_NLO}
%
%\eea
%\nt 

The NLO matrix element of \epem annihilation process can be obtained 
from the perturbative expansion of the $S$-matrix with an interaction 
Hamiltonian which includes an interaction potential corresponding 
to a quark of color $j$ interacting with an antiquark of color $i$ 
(or visa versa) and a gluon of color $a$:
\bea
\mh_s = i g t^a_{i,j} \bar{\psi}_i \g^\rho \psi_j  A_\rho^a .
\eea
The $t^a_{i,j}$'s represent the Gell-Mann matrices.

The DGLAP evolution arises from collinear gluon bremstrahlung
in the final state of two jet events. As in the case at LO in the 
previous section, we can again neglect the interference between
the two jets in either the partonic or the hadronic level in the leading
log and leading twist approximation. Therefore, the sum over all 
hadronic states is decomposed into the sum over two complete sets 
of hadronic states, each overlapping with one of the opposite-moving 
jets. Invoking the parton-hadron duality, the sum over the hadronic
states in the ``back-side'' jet can be replaced by a sum over 
partonic states. 
%moving in opposite directions. 
%Two particular hadrons ($h_1,h_2$) with momenta $\vp_1$ and $\vp_2$ in the 
%same jet are then identified. In the end, the focus of this section will 
%be on the NLO contribution to the dihadron fragmentation function 
%of the two hadrons ($h_1,h_2$).
Therefore, partonic processes within the ``back-side'' jet
will not contribute to the evolution of the fragmentation
function in the opposite side which has been defined 
within a rather strict collinear approximation. 
%In short a gluon emanating from 
%the quark-antiquark pair with momenta in a direction opposite 
%to that of the identified hadrons will not modify the fragmentation 
%of the hadrons, based on the approximation that the two jets 
%fragment almost independently of each other. Such processes 
%will thus not be considered and the sum over all hadronic states 
%moving in the direction opposite to the identified hadrons 
%will be replaced as before with 
%the sum over all momenta of an outgoing quark or antiquark.
As in the case of single fragmentation function,
we will also assume, in addition, that there is no interference
between the fragmentation of the leading parton and the radiated
gluon that has a minimum transverse momentum set by the factorization
scale $\mu$.

Hence, in NLO, the sum over all states may be expressed as
\bea
|S\rc = |k_{q(\bar{q})}\rc  \times | S -2, \fp_1 \fp_2 \rc .
\eea
%\nt 
%The arrows in subscript are purely a pnemonic to indicate that the two states 
%represent particles moving in opposite directions. 
Depending on how different operators are contracted with 
the outgoing hadronic state $|S-2\rc$,  we may identify three 
different cases:
%and the nature of the partonic 
%state that proceeds in the direction opposite to that of the 
%identified hadrons. 
%T%he six different out states are 
\bea
| S -2, \fp_1 \fp_2 \rc &=& | S -2, \fp_1 \fp_2 \rc
\times  | (p-l)_g \rc \nn \\
&+& | S -2, \fp_1 \fp_2 \rc
\times  | (p-l)_{\bar{q}(q)} \rc \nn \\
&+& | S -1, \fp_1 \rc
\times  | S-1, \fp_2 \rc . 
\label{diff_states}
\eea
These cases differ in the partonic operator 
that contracts with the hadronic state.
In the above, the parton with momentum $\fk$ which proceeds 
in a direction opposite to that of the identified hadrons 
alternates between a quark and an antiquark. 
In the rest of this paper, we will always take it as 
an antiquark in order to focus on the quark 
fragmentation function. The case for the antiquark 
fragmentation function will be formally identical to 
that of the quark. 
%The states $X,Y$ also represent a sum over all hadronic states with 
%the understanding is that the state $S = X + Y$. From each of the 
%states $X,Y$ we have identified one hadron. 

%In the following, contributions from the 
%second, the fourth and the sixth line of Eq.~(\ref{diff_states}) 

In the following subsections, we will evaluate contributions from these different
cases in detail. Roughly speaking, the first line 
of Eq.~(\ref{diff_states}) represents contributions where the 
fragmenting quark undergoes a split into a quark and a gluon and  
the two identified hadrons emanate from the quark offspring. The
second contribution represents the case where both identified 
hadrons emanate from the gluon. The last contribution 
represents the case where one hadron emanates from each of the 
quark and gluon offspring.

%%%%%%%%%%%%%%%%%%%%%%%%%%%%%%%%%%%%
%%%%%%%%%%%%%%%%%%%%%%%%%%%%%%%%%%%%
%%%%%%%%%%%%%%%%%%%%%%%%%%%%%%%%%%%%

\subsection{NLO contribution from quark fragmentation }

%%%%%%%%%%%%%%%%%%%%%%%%%%%%%%%%%%%%
%%%%%%%%%%%%%%%%%%%%%%%%%%%%%%%%%%%%
%%%%%%%%%%%%%%%%%%%%%%%%%%%%%%%%%%%%

Proceeding with the evaluation of the double 
differential cross section at next-to-leading order, 
the focus in this subsection will be on isolating contributions 
to the cross section that contain an explicit expression for a 
dihadron fragmentation function of a quark. 
The outstate in this subsection will solely be restricted to the 
first line of Eq.~(\ref{diff_states}). The instate is simply that 
of an incoming \epem pair. Insertion of the interaction operator 
density $ T[ \mh_{e^+ e^- \g} \mh_{q \bar{q} \g} \mh_{q \bar{q} g} ]$, 
followed by a contraction of the outgoing antiquark operator 
with the states $| k_{\bar{q}} \rc $
and gluon operator with  $|(p-l)_g \rc $, leads to the 
following matrix element:
\bea 
\mat^{i} &=& i \sum_q e_q  e^2 g t^a  
\bar{v}_{k_2} \g^\mu u_{k_1} \frac{g_{\mu \nu}}{\fq^2 + i\e} \nn \\
&\times & \lc \fp_1 \fp_2 S-2 | \bar{\psi}_q (0) | 0 \rc 
\Bigg\{ \frac{\g^\nu (-\f\fq + \f\fl) \g^\rho }{(\fq - \fl)^2 + i \e}  \nn \\
&+& \frac{\g^\rho (\f \fq - \f \fk ) \g^\nu }{(\fq - \fk)^2 + i \e } \Bigg\}
v^s(k) {\ve_a^\lambda}^*_\rho (p-l) \nn \\ 
&\times&  (2\pi)^4 \kd^4(\fq -\fk - \fp-\fl ) . \label{nlo_qrk_1}
\eea

\begin{figure}[htb!]
%\begin{center}
%  \epsfxsize 80mm
%\hspace{0cm}
  \resizebox{4in}{5in}{\includegraphics[0.0in,1.0in][5.0in,7.5in]{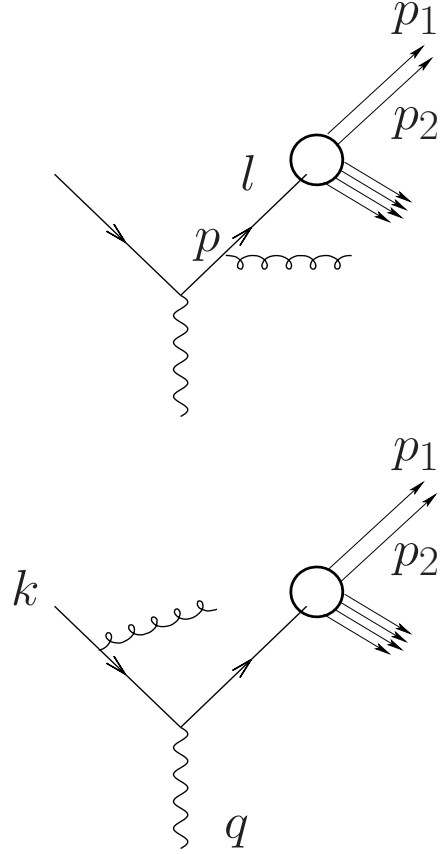}} 
%\vspace{0.25cm}
    \caption{The leading log contribution to the NLO modification of the 
    quark fragmentation function. }
    \label{nlo_qrk}
%  \end{center}
\end{figure}

In the above equation there are two terms with differernt momentum 
dependences within the  curly brackets. The reader will 
readily note that the second term is the Feynman rule for the 
process indicated in the upper panel of Fig.~\ref{nlo_qrk}, while 
the other term consists of the Feynman diagram 
where the gluon is radiated from the antiquark line, as 
shown in the lower panel of Fig.~\ref{nlo_qrk}.  

%The matrix element is squared and then inserted into the expression 
%for the cross section in Eq.~(\ref{sig_NLO}). 
In computing the NLO cross section, one may once again 
factorize the cross section into a leptonic and a hadronic 
piece [see Eq.~(\ref{s_w})]. 
%Averaging over the spins of the \epem pair gives us the familiar 
%leptonic tensor ${\ml}_{\A \nu} 
%= \sum \tr [u \bar{u} \g^\A v\bar{v} \g^\nu]$. 
%The hadronic tensor $W^{\A \nu}$ contains a sum over all final 
%states: both hadronic and partonic. 
Summing over all final states of the outgoing antiquark, gluon and 
hadrons from the fragmenting quark (besides $h_1$ and $h_2$) 
followed by an incorporation of  minor simplifications, the 
hadronic tensor may be expressed as 
\bea
W^{\A \nu} &=&  \int \frac{d^3 p_1 d^3 p_2}{(2\pi)^6 4 E_1 E_2} 
\int \frac{d^4 l}{(2\pi)^4}  \int \frac{d^4 p}{(2 \pi)^4} \label{W_nlo_1} \\ 
&\times&  2 \pi \kd^+ ( ( \fp - \fl )^2) 2 \pi \kd^+ ( (\fq - \fp )^2 ) 
g^2 N_c C_F  d_{\rho \si} (p-l) \nn \\
&\times& \tr \Bigg[ (\f \fq- \f \fp) \Bigg\{ 
\frac{ \g^\A \f \fp \g^\rho }{ \fp^2 - i \e } + 
\frac{ \g^\rho (\f \fl - \f \fq ) \g^\A}{ (\fl -\fq)^2 - i\e } \Bigg\}  \nn \\
&\times& \hat{T}_q(\fl;\fp_1,\fp_2) 
\Bigg\{ \frac{ \g^\si \f \fp \g^\nu }{ \fp^2 + i \e } + 
\frac{ \g^\nu (\f \fl - \f \fq ) \g^\si}{ (\fl -\fq)^2 + i\e } \Bigg\} \Bigg] .\nn 
\eea
In the above equation the color factor $N_c C_F$ comes from 
the factor $\tr[t^a t^b]\kd^{ab}$.  
%subsequent to performing the sum over the colors of the gluon.
For brevity, we have omitted in the above the sum over quark 
flavors weighted with fractional charge, $\sum_q e_q^2$.
The sum over polarizations of the gluon leads to the 
factor $d_{\rho \si} $. In light-cone gauge $d_{\rho \si}$  is given as 
\bea 
d_{\rho \si}(p-l) =  g_{\rho \si} - \frac{(p-l)_\rho n_\si 
+ (p-l)_\si n_\rho}{(\fp-\fl) \x \fn}
\eea
The overlap matrix element $\hat{T}_q(\fl;\fp_1,\fp_2)$ is the same as
defined in the previous section, with the final fragmenting quark momentum
reduced to $\fl$. The remaining portion of Eq.~(\ref{W_nlo_1}) 
is also easy to trace. The two $\delta$ functions essentially 
stipulate that the gluon and antiquark be released onshell. 
%The trace over the Dirac matrix structure finds its origin 
%in the sum over all final state spins. 
The first set of terms in the curly brackets 
represents the process indicated by upper panel of Fig. \ref{nlo_qrk}. 
The second set of terms indicates the case where the gluon is emitted
from the outgoing antiquark, with all other features remaining 
unchanged, \eg, the fragmenting quark has momentum $\fl$, the 
gluon still has a momentum of $\fp-\fl$.

Within the collinear approximation, we again assume 
that  $l^+ , p_h^+ >> l^-,p_h^-,l_\perp,{q}_\perp$. 
As a result, the overlap matrix 
element may be factorized via the following approximation:
\bea
\hat{T}_q(\fl;\fp_1,\fp_2) &=& \int d^4 x e^{i\fl \x \fx } 
\sum_{S-2} \nn \\ 
%\left[ \prod_{f \in S-2}\int \frac{d^3 p_f}{(2\pi)^3 2E_f} \right] 
%
& & \!\!\!\!\!\!\!\!\!\!\!\!\!\!\!\!\!\!\!\! \mbx \times \lc 0 | \psi_q(x) 
| \fp_1, \fp_2, S-2 \rc \lc \fp_1, \fp_2, S-2 | \psi_q(0) | 0 \rc  \nn \\
&\simeq& \frac{ \f \fp_h }{2} T_q(\fl,\fp_1,\fp_2).
\eea
Within the collinear approximation, we assume that the momentum 
of the final fragmenting quark is collinear with that of the 
hadrons emerging from the parton fragmentation. We thus define
a new fractional momentum, $z' = p_h^+/l^+$. This allows 
the partonic four-momentum vector $\fl$ to be replaced 
with the hadronic four-vector: \ie $\fl = \fp_h/z'$.
 
%This allows for a factorization of the term: 
%\[ 
%\tilde{T}_q(z',q_\perp) = \int \frac{d^4 l}{(2\pi)^4} \kd \left( z' - 
%
%\frac{p_h^+}{l^+} \right)
%
%\tr \Bigg[ \frac{\g^+}{2p_h^+} \hat{T}_q(\fl;\fp_1,\fp_2) \Bigg]
%\]
%\nt
%
%from the remaining expression and corresponds to the cut-vertex 
%structure in the lower right hand side of Fig. (\ref{nlo_cutvert_1}). 
%In the remaining expressions the replacement $\fl = \fp_h/z'$ has 
%been made. 
As a result the factor $(\fl -\fq)^2$  is approximated as:
\bea 
(\fl - \fq)^2 \simeq Q^2 - 2 Q l^+ \simeq Q^2 - 2 Q p_h^+/z' . \label{2nd_denom}
\eea

Unlike the case of the LO process, the 
transverse momentum of the quark which emanates from the 
electromagnetic vertex is non-vanishing, $p_\perp \neq 0$
(this is also the transverse momentum carried by the gluon). 
As a result, the negative longitudinal momentum $p^-$ is 
constrained by one of the $\delta^+$ functions as,
\bea
(\fp - \fp_h/z')^2 &\simeq&  2p^+p^- - p_\perp^2 - 2p^-p_h^+/z' = 0  \nn \\
&\im& p^- = \frac{p_\perp^2}{2(p^+ - p_h^+/z')} .
\eea

The hadron fractional forward light-cone momentum is still defined as 
$z = p_h^+/p^+$ as in the LO case.
At leading twist one may replace all occurences of $\fp_h$ with 
$z'\fl$. The corrections to this approximation are down by 
powers of $Q^2$.  Incorporating the above approximations into 
the expression for the hadronic tensor, we obtain: 
\nt
\bea
W^{\A \nu} &=&  \int d z_1 d z_2 dz' \int 
\frac{d^3 p_1 d^3 p_2}{(2\pi)^6 4 E_1 E_2} 
\int \frac{d^4 p}{(2 \pi)^4} \label{W_nlo_2} \\
&\times& g^2 N_c C_F  \kd ( z_1 - p_1^+/p^+ ) 
\kd ( z_2 - p_2^+/p^+ ) d_{\rho \si} (p-l)\nn \\ 
&\times&  2 \pi \kd^+ ( 2p^-(p^+ - l^+) - p_\perp^2  ) 
2 \pi \kd^+ ( Q^2 - 2Qp^+ ) \nn \\
&& \!\!\!\!\!\!\!\!\!\!\!\!\!\!\!\!\!\!\!\!\!\mbx 
\times \tr \Bigg[ (\f \fq- \f \fp) \Bigg\{ 
\frac{ \g^\A \f \fp \g^\rho }{ p_\perp^2 l^+/(p^+-l^+) } + 
\frac{ \g^\rho (\f \fl - \f \fq ) \g^\A}{ Q^2 - 2Q l^+ } \Bigg\}
\nn \\
&& \!\!\!\!\!\!\!\!\!\!\!\!\!\!\!\!\!\!\!\!\!\mbx \times
 \f \fl \frac{z'}{2} 
\Bigg\{ \frac{ \g^\si \f \fp \g^\nu }{p_\perp^2 l^+/(p^+-l^+) } + 
\frac{ \g^\nu (\f \fl - \f \fq ) \g^\si}{ Q^2 - 2Q l^+ } \Bigg\} 
\Bigg]_{\fl = \fp_h/z'} \nn \\
& & \hspace{-0.4 in}\times\int \frac{d^4 l}{(2\pi)^4} 
\kd \left( z' - \frac{p_h^+}{l^+} \right)
T_q(\fl;\fp_1,\fp_2) .\nn 
\eea
%\nt 
%In the above, every instance of the four-vector $\fl$ is meant to 
%indicate an occurance of the quantity $\fp_h/z'$. 

%In applying the leading twist approximation we are essentially 
%ignoring terms of the order of $q_\perp^2/l^+}^2$. 
A careful study of Eq.~(\ref{W_nlo_2}) reveals that the 
leading log contributions are dominated by the region 
where $p_\perp \ra 0$. As a result, the 
part of $W^{\A\nu}$ which represents the square of the 
process in the lower panel of Fig.~\ref{nlo_qrk}
has no leading log contribution. The leading log contribution 
comes from the square of the first term corresponding to 
the square of the process in the upper panel of Fig.~\ref{nlo_qrk}. 
In the light-cone gauge, the interference terms have no 
contribution at leading log as in the case of single fragmentation
functions (see chapter 3 of Ref.~\cite{fie95}).
This can be demonstrated to hold at leading twist merely 
by completing the trace as indicated in Eq.~(\ref{W_nlo_2}) and 
extracting the $p_\perp^2$ dependence from the numerators. 

%There it was 
%demonstrated for the case where the incoming \epem directions we 
%integrated over. Thus,
%similar to the case of single fragmentation functions, that 
%the leading logarithmic contribution emanates from diagrams that have the 
%structure of the upper panel of Fig. (\ref{nlo_qrk}). 

The remaining factorization into the hard and soft piece proceeds 
as in the LO case, leading to the hadronic tensor at NLO at leading 
log and leading twist, 
\bea
W^{\A \nu} &=&  \int d z_1 d z_2 dz'  
\int \frac{d^3 p_1 d^3 p_2}{(2\pi)^6 4 E_1 E_2} 
\int \frac{d^4 p}{(2 \pi)^4} \label{W_nlo_3}\\ 
&\times& g^2 N_c C_F \kd ( z_1 - p_1^+/p^+ ) 
\kd ( z_2 - p_2^+/p^+ )   \nn \\ 
&\times&  \frac{2 \pi \kd^+ \left( p^- - \frac{p_\perp^2}{2(p^+ - l^+)}  
\right)}{2(p^+ - l^+)} 
2 \pi \kd^+ \left( Q^2 - \frac{2Qp^+_h}{z} \right) \nn \\
&\times & \tr \Bigg[ d_{\rho \si}(p-l) (\f \fq- \f \fp)
\frac{\g^\A \f \fp \g^\rho \f \fl (z'/2) \g^\si \f \fp \g^\nu }
{[p_\perp^2 l^+/(p^+-l^+)]^2 }
\Bigg]_{\fl = \fp_h/z'} 
\nn \\
&\times&\int \frac{d^4 l}{(2\pi)^4} 
\kd \left( z' - \frac{p_h^+}{l^+} \right)
T_q(\fl;\fp_1,\fp_2) .\nn 
\eea

%In the above expression, the extraction of the hard piece is 
%the same as in the previous section. 
%The Dirac operator within the trace along with the gluon spin 
%sum $d_{\rho \si}$ is written symbolically as 

The trace of the Dirac matrices can be carried out by brute force.
The leading-twist part can be obtained in a more straightforward way
by rewriting the matrices within the trace symbolically as
\[
(\f \fq- \f \fp) \g^\A \hat{C} \g^\nu .
\]
The leading twist portion of the matrix $\hat{C}$,  may be written as 
\bea 
\hat{C} = \frac{\f \fp_h}{2} C , \nn
\eea
\nt 
with the factor $C$ expressed as a trace:
\bea 
C = \tr \Bigg[ d_{\rho \si} \frac{\g^+}{2 p_h^+} 
\frac{\f \fp}{\fp^2} \g^\rho \f \fl \frac{z'}{2} \g^\si 
\frac{\f \fp}{\fp^2}
\Bigg]_{\fl = \fp_h/z'},  \label{split-c}
\eea
where $\fp^2=p_\perp^2 l^+/(p^+-l^+)$.
This procedure is very similar to the one used to construct the collinear 
approximation to the overlap matrices $\hat{T}_q(p;p_1,p_2)$.
One can complete the above trace and obtain the regular splitting
function $(1+y^2)/(1-y)$, where $y = z'/z$. It represents the 
probability for the radiation of a gluon from a quark prior to 
its fragmentation.

\begin{figure}[htb!]
%\begin{center}
%  \epsfxsize 80mm
%\hspace{0cm}
  \resizebox{3.25in}{4.5in}{\includegraphics[0.5in,1in][6.5in,9in]{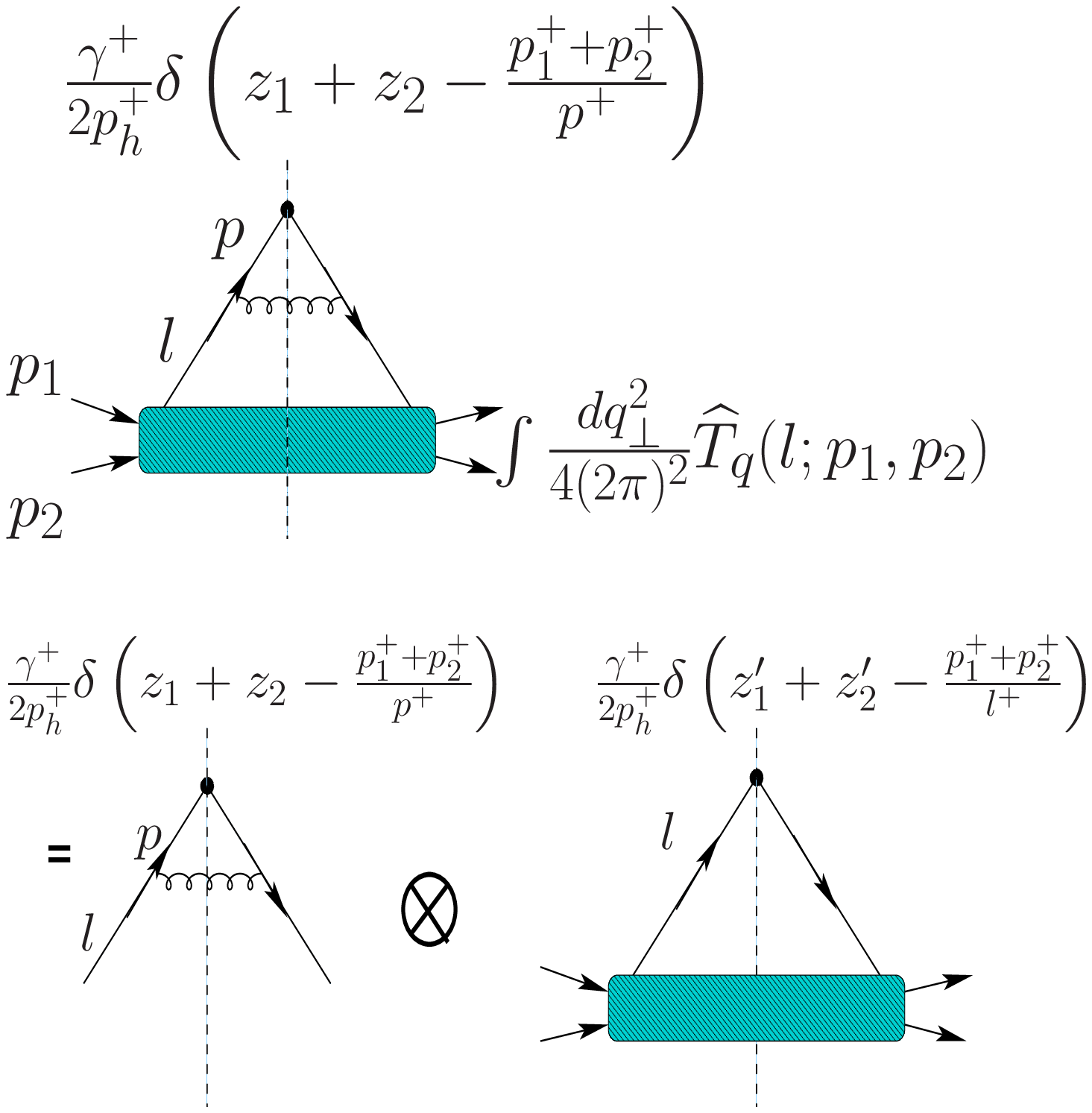}} 
%\vspace{0.25cm}
    \caption{Next-to-Leading order 
    cut-vertex for quark fragmentation and its factorization.}
    \label{nlo_cutvert_1}
%  \end{center}
\end{figure}

Using the above approximation one may extract the same 
hard part as in Eq.~(\ref{hardpart}),
\bea\
\mhp^{\A \nu} &=& \tr \Bigg[ (\f \fq - \f \fp) \g^\A  \frac{p_h^{\nu}}{2} \g^\nu  \Bigg] 
\kd ( (\fq - \fp_h/z)^2 ) \nn \\
&=& \mh^{\A \nu} \kd ( (\fq - \fp_h/z)^2 ).
\eea 
\nt
Extracting this hard part and comparing with 
Eqs.~(\ref{w_mu_nu_4},\ref{w_mu_nu_5}) and the resulting 
cut-vertex diagram in Fig.~\ref{cutvert1} we note that the soft 
part of Eq.~(\ref{W_nlo_3}) begins to display a 
structure as illustrated in Fig.~\ref{nlo_cutvert_1}.

To complete the calculation and obtain a factorized form
of the NLO contribution to the double inclusive cross section, 
the integration over the tranverse momentum of the identified 
hadrons will have to be factored into the fragmentation function. 
Similarly as in the case of LO calculation, one has to transform 
the basis of the momentum integrations of the two identified 
hadrons to the basis indicated in Fig.~\ref{fig4}.  

In the steps leading to Eq.~(\ref{W_nlo_3}), the 
approximation that the momenta of the hadrons and the 
fragmenting parton are collinear has been made. Particular
among these are the approximations that $l^+>>l_\perp$,$l^-$.
These essentially indicate that the invariant mass of the final 
fragmenting quark is negligible compared to its forward light-cone 
momentum. This is identical to the approximation made on the 
momentum $\fp$ in the LO calculation (see Fig.~\ref{fig3}).
Since the NLO process has a collinear divergence when
$p_\perp^2\rightarrow 0$, we will only consider the leading log
contribution.

We thus obtain the following factorized form for the hadronic tensor: 
\bea 
W^{\A \nu} &=& \hspace{0.2in}
\int dz_1 dz_2 \int \frac{d p_1 d p_2 }{ (2 \pi)^4 4 p_1 p_2} 
d \cos{\h_{cm}} d \phi_{cm} d \phi_{\perp} \nn \\
&\times&  g^2 N_c C_F\int_z^1 \frac{dy}{y^2} z \int \frac{dp^+ dp^- dp_\perp^2 }{(2\pi)^4} 
2 \pi \mh^{\A \nu} \kd((\fp -\fl^2) \nn \\
&\times&Cp^{+2} \kd(p_1^+ - z_1 p^+)\kd(p_2^+ - z_2 p^+)\frac{z}{2y} \nn \\
%
%&\times& \tr \Bigg[ \frac{\g^+}{2p_h^+} \frac{\f \fp}{\fp^2} 
%
%\g^\rho \f \fl \g^{\si} \frac{\f \fp}{\fp^2} \Bigg] 
%
%d_{\rho \si} \Bigg|_{\fl = y\fp_h/z} \nn \\
%
&\times& \int \frac{d q_{\perp}^2}{4 (2 \pi)^2} 
\int \frac{d^4 l }{(2\pi)^4} 
\kd (z' - \frac{p_h^+}{l^+})  T_q(\fl;\fp_1,\fp_2).
\eea

\nt

%Separating the $\kd-$function from the hard tensor:
%
%\bea\
%\mhp^{\A \nu} &=& \tr \Bigg[ (\f \fq - \f \fp) \g^\A  \frac{p_h^{\nu}}{2} \g^\nu  \Bigg] 
%
%\kd ( (\fq - \fp_h/z)^2 ) \nn \\
%
%&=& \mh^{\A \nu} \kd ( (\fq - \fp_h/z)^2 ) 
%\eea 

Further factorization of  $\mh^{\A \nu}$ from the $d^4 p$ 
integration leads to the factorized form for the NLO correction to
the dihadron fragmentation function as illustrated in 
Fig.~\ref{nlo_cutvert_1} in terms of cut vertices.  
\nt
%The factorization into the two pieces shown in the lower panel is 
%simply achieved by factoring out $\tilde{T}_{q}$, and expreessing it  
%as a trace. Performing the trace followed by a contraction of the 
%indices in the factor 
%\bea 
%C_F \tr \Bigg[ \frac{\g^+}{2p_h^+} \frac{\f \fp}{\fp^2} 
%
%\g^\rho \f \fl \g^{\si} \frac{\f \fp}{\fp^2} \Bigg] 
%
%d_{\rho \si} \Bigg|_{\fl = y\fp_h/z}
%
%\eea
%
%\nt
%we obtain the regular splitting function, 
\nt
Inserting the hadronic tensor into the expression 
for the double differential cross section with minor 
simplifications, we obtain the factorized NLO contribution
to the double inclusive cross section,
%expected expression for the regular DGLAP evolution 
%contribution from quark fragmentation. 
\nt
\bea 
\frac{d^2 \si}{d z_1 d z_2} &=&  \sum_q \frac{4 \pi}{3} 
\frac{\A^2 e_q^2 N_c}{Q^2}  \label{si_nlo_8} \\
&\times& \frac{\A_s}{2\pi} \int_{\mu^2}^{Q^2} 
\frac{d p_{\perp}^2 }{p_{\perp}^2} 
\int_{z}^1 \frac{dy}{y^2} C_F \frac{1+y^2}{1-y} \nn \\
&& \!\!\!\!\!\!\!\!\!\!\!\!\!\!\!\!\!\!\!\!\!\!\!\!\!\!\!\!\!\!\!\!
\frac{(z/y)^4}{4 (z_1/y) (z_2/y)} \int \frac{d q_\perp^2}{8 (2 \pi)^2} 
\int \frac{d^4 l}{(2 \pi)^4} \kd \left( \frac{z}{y} 
- \frac{p_h^+}{l^+} \right) T_q(l;p_1,p_2). \nn
\eea

The last line of the above equation may be easily 
identified as $D(z_1/y,z_2/y,\mu^2)$, the 
dihadron fragmentation functions scaled up by the momentum fraction $y$, 
carried by the 
quark emanating from the split. Physically this represents 
the contribution to the fragmentation 
functions at a higher order brought about by gluon 
radiation carrying away with it a momentum fraction 
$1-y$. By itself this process displays both an infrared 
divergence as $y \ra 1$ and a 
collinear divergence as $p_\perp \ra 0$. The infrared divergence 
will be canceled by the 
virtual diagram contribution as shown in the next section. 
The collinear divergent part will be 
combined with the collinear divergent part of gluon
fragmentation and absorbed into the renormalized 
fragmentation function.

%%%%%%%%%%%%%%%%%%%%%%%%%%%%%%%%%%%%%%%%%%%%
%%%%%%%%%%%%%%%%%%%%%%%%%%%%%%%%%%%%%%%%%%%%  
%%%%%%%%%%%%%%%%%%%%%%%%%%%%%%%%%%%%%%%%%%%%

\subsection{NLO contribution from gluon fragmentation}

%%%%%%%%%%%%%%%%%%%%%%%%%%%%%%%%%%%%%%%%%%%%
%%%%%%%%%%%%%%%%%%%%%%%%%%%%%%%%%%%%%%%%%%%%  
%%%%%%%%%%%%%%%%%%%%%%%%%%%%%%%%%%%%%%%%%%%%

%In the preceding subsection, we evaluated the 
%leading log and leading twist contribution to 
%the quark dihadron fragmentation function emanating 
%from quark fragmentation. 
%We expressed the result in the form of Feynman 
%diagrams containing the cut-vertices. 
We now proceed with the contribution from gluon 
fragmentation in the NLO processes. Essentially the out-going 
hadronic state is replaced with the 
second line of Eq.~(\ref{diff_states}). Insertion of the 
interaction operator density followed by the contraction 
of the quark and antiquark operator with the out-going  
states $|k_{\bar{q}} \rc$ and $| (p-l)_q \rc$ , leads to the 
following matrix element:

\bea 
\mat^{ii} &=& i\sum_q e_q e^2  g t^a \bar{v}_{k_2} 
\g^\mu u_{k_1} \frac{g_{\mu \nu}}{ q^2 + i\e}
\label{nlo_glue_1} \\
&\times& \lc p_1 p_2 S-2 | A^a_\rho (0) |0 \rc \bar{u}^r(p - l) 
\Bigg\{ \frac{\g^\rho \f \fp \g^\nu}{\fp^2 + i\e} \nn \\ 
&+& \frac{\g^{\nu} ( - \f \fk - \f \fl )}{(\fk + \fl)^2 + i\e} \Bigg\} v^s(k) 
(2\pi)^4 \kd (\fq - \fk -\fp -\fl). \nn
\eea

%On squaring the above matrix element followed by summing over all final 
%states, averaging over all initial states and factoring out the 
%electromagnetic tensor ($\ml^{\mu \nu}$) we obtain the hadronic tensor. 
As in the previous subsection, the approximation of 
very high energies is made, allowing the isolation of the leading 
twist and leading log contribution of the corresponding
hadronic tensor. This is obtained as 
\bea 
W^{\A \nu} &=& \int 
\frac{dp_1 dp_2 d\cos{\h_{cm}} d\phi_{cm} d\phi_\perp }{(2\pi)^4 4 p_1 p_2} 
\int \frac{d^4 p}{(2\pi)^4} \frac{d^4 l}{(2 \pi)^4} \nn \\ 
&& 2 \pi \kd ((\fq - \fp)^2) 2\pi \kd ((\fp - \fl)^2) g^2 N_c C_F  \hat{T}_g (l;p_1,p_2)\nn \\
&\times& d_{\rho \si} \tr \Bigg[ 
(\f \fp - \f \fl) \frac{\g^{\rho} \f \fl \g^{\A}}{\fl^2 + i\e} 
(\f \fq - \f \fp) \frac{\g^\nu \f \fl \g^{\si} }{\fl^2 - i \e} \Bigg], 
\label{W_nlo_g}
\eea
\nt
where the gluon overlap matrix element $T_g (l;p_1,p_2)$  is defined as
\bea 
\hat{T}_g (l;p_1,p_2) &=& \int d^4 x e^{i\fl\x \fx} \sum_{S-2} \nn \\
%\Bigg[ \prod_{f \in S-2}
%
%\int \frac{d^3 p_f}{(2\pi)^3 2 E_f} \Bigg]   
%
&& \int \frac{d q_{\perp}^2}{8 (2\pi)^2}
\times \lc 0 | A^a_\mu (x) | p_1,p_2,S-2 \rc \nn \\ 
&\times& \lc p_1,p_2,S-2 | A^b_\nu(0) | 0 \rc 
\frac{\kd^{ab} d^{\mu \nu}}{16} . \label{glue_matrix}
\eea

\begin{figure}[htb!]
%\begin{center}
%  \epsfxsize 80mm
%\hspace{0cm}
  \resizebox{4in}{3.5in}{\includegraphics[0in,0in][6in,5in]{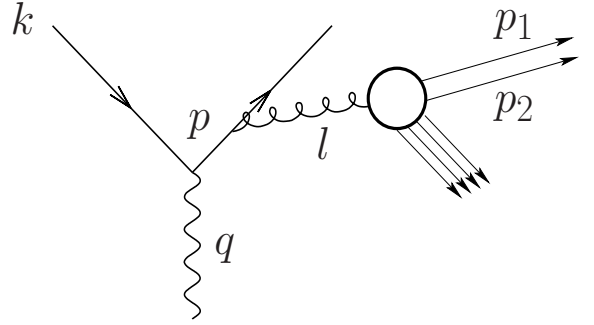}} 
%\vspace{0.25cm}
    \caption{The leading log gluon fragmentation contribution to the NLO 
    modification of the quark fragmentation function. }
    \label{nlo_glue}
%  \end{center}
\end{figure}

A diagrammatical representation of this gluon fragmentation process
can be illustrated as Fig.~\ref{nlo_glue}. The procedure leading to 
the extraction of the leading log and leading twist is rather 
similar to the case for the single fragmentation function and to 
the case of NLO process of quark fragmentation in the last subsection.
%Hence, they will not be further expanded upon here. 
Contracting the hadronic tensor with its leptonic counterpart 
we obtain the gluon contribution to the NLO double differential 
cross section:  
\bea 
\frac{d^2 \si}{d z_1 d z_2} &=& \sum_q \frac{4 \pi}{3} 
\frac{\A^2 e_q^2 N_c }{Q^2} \\
&& \frac{\A_s}{2\pi} \int_{\mu^2}^{Q^2} 
\frac{dp_{\perp}^2}{p_{\perp}^2} \int_z^1 
\frac{dy}{y^2} C_F \frac{1 + (1-y)^2}{y} \nn \\
&& \frac{(z/y)^3}{2 (z_1/y) (z_2/y) } \int \frac{d q_\perp^2}{8 (2 \pi)^2} 
\int \frac{d^4l}{(2\pi)^4}  \nn \\
&&\times \kd \left( \frac{z}{y} - \frac{p_h^+}{l^+} \right) 
\hat{T}_g (l;p_1,p_2) . \nn
\eea

\begin{figure}[htb!]
%\begin{center}
%  \epsfxsize 80mm
%\hspace{0cm}
 \resizebox{3.25in}{4.5in}{\includegraphics[0.5in,1in][6.5in,9in]{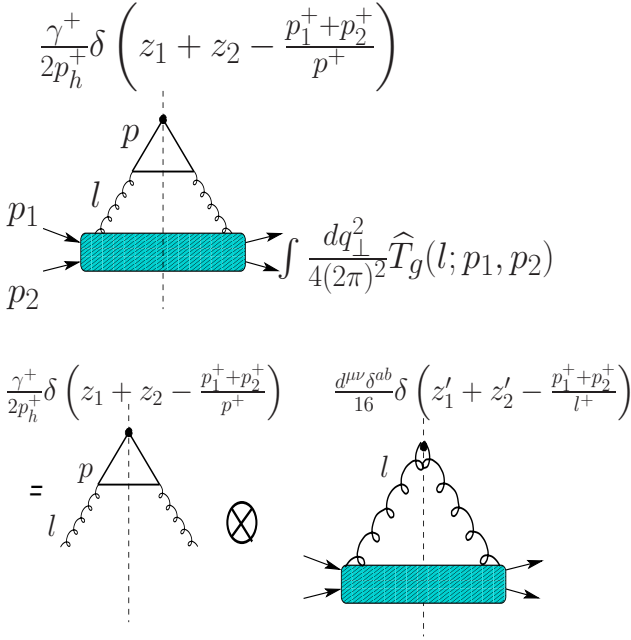}} 
%  \resizebox{3in}{4in}{\includegraphics[0in,0in][.0in,9in]{nlocutvert2.eps}} 
%\vspace{0.25cm}
    \caption{Next-to-Leading order cut-vertex representation of the
     gluon fragmentation contribution. }
    \label{cutvert2}
%  \end{center}
\end{figure}

It may come as no surprise that the above equation may also be 
derived from a set of Feynman rules involving cut-vertices. The 
cut-vertex diagrams are illustrated in Fig.~\ref{cutvert2}. The 
rules are indicated in the figure. As a result of this computation we 
may now present the cut-vertex expression for the gluon dihadron fragmentation
function (indicated in the lower right hand corner of Fig.~\ref{cutvert2}):

\bea 
D_g(z_1,z_2) &=& \frac{z^3}{2 z_1 z_2 }  \int \frac{dq_\perp^2}{8(2\pi)^2}
\int \frac{d^4l}{(2\pi)^4} \nn \\
&\times& \kd \left( z - \frac{p_h^+}{l^+} \right) \hat{T}_g (l;p_1,p_2),
\eea

\nt
where the factor $\hat{T}_g(l;p_1,p_2)$ is given in Eq.~(\ref{glue_matrix}). 

%{\bf move this discussion to Sec. V} 
%It should once again be pointed out that the $q_{\perp}$ integration in 
%Eq.~(\ref{glue_matrix}) is limited to a maximum value of $\mu$ the scale 
%at which the leading order fragmentation functions are defined. To 
%obtain the fragmentation functions at the scale $Q^2$ the transverse 
%momentum between the identified hadrons must also be allowed to rise. 
%This contribution is not indluded in this or the preceding subsection.
%This is evaluated perterbatively in the subsequent subsection.

%%%%%%%%%%%%%%%%%%%%%%%%%%%%%%%%%%%%%%%
%%%%%%%%%%%%%%%%%%%%%%%%%%%%%%%%%%%%%%%
%%%%%%%%%%%%%%%%%%%%%%%%%%%%%%%%%%%%%%%

\subsection{NLO contribution from quark and gluon single fragmentation}

%%%%%%%%%%%%%%%%%%%%%%%%%%%%%%%%%%%%%%%
%%%%%%%%%%%%%%%%%%%%%%%%%%%%%%%%%%%%%%%
%%%%%%%%%%%%%%%%%%%%%%%%%%%%%%%%%%%%%%%

\begin{figure}[htb!]
%\begin{center}
%  \epsfxsize 80mm
%\hspace{0cm}
  \resizebox{4in}{3.5in}{\includegraphics[0in,0in][6in,5in]{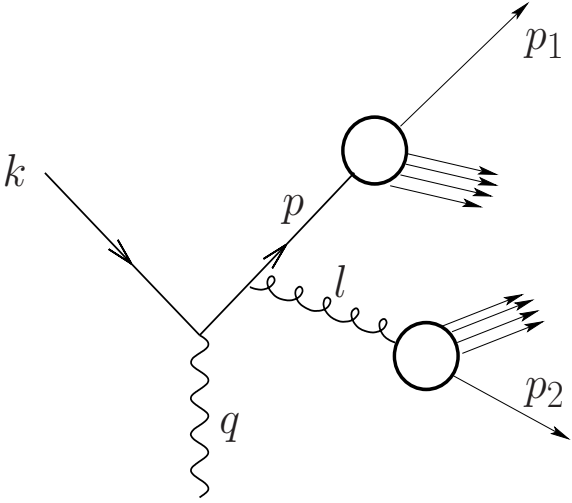}} 
%  \resizebox{3in}{3in}{\includegraphics[1in,3in][8.0in,8in]{nlofrag3.eps}} 
%\vspace{0.25cm}
    \caption{The leading log mixed contribution to the NLO modification of the quark dihadron 
fragmentation function.}
    \label{nlofrag3}
%  \end{center}
\end{figure}

In the previous subsections, we have evaluated two 
separate contributions to the NLO double 
fragmentation functions. In both of these 
we have assumed that the relative transverse momentum 
of the two detected hadrons is intrinsic and is limited
by a scale $\mu_\perp$. In the definition of the dihadron 
fragmentation functions, the hadrons are detected with given 
fractions of forward longitudinal momentum but the transverse 
momenta are integrated over. Thus, all allowed transverse momenta 
between the detected hadrons must be included. In this paper, we
will assume that all hadron pairs with relative transverse
momentum larger than $\mu \gtrsim \mu_\perp$ are generated perturbatively.
In the next-to-leading order, such hadron pairs can be produced 
from the independent fragmentation of the quark and gluon
after their split, as illustrated in Fig.~\ref{nlofrag3}.
Such a scenario has been considered in 
Refs. \cite{kon78,kon79a,kon79b} where the 
double inclusive cross section with two detected hadrons in 
\epem collisions with a fixed transverse momemtum between 
them was computed. The authors argued that in the case that
the transverse momentum lies in a  semihard region 
$\Lambda_{QCD} << q_\perp << Q$, the dominant contribution 
to the cross section comes from the process where 
the fragmenting parton undergoes a semihard split into 
two independent partons which then fragment independently.

%This being the scale at 
%which the dihadron fragmentation functions were defined at 
%leading order;
%the transverse momentum squared of the detected hadrons in that 
%case was restricted to values smaller than $\mu^2$. 
%However, the next-to-leading order evaluation is being 
%performed at a much larger scale $Q^2$. We would 
%naively expect all quantities to scale accordingly. As a 
%result the tranverse momenta between two detected hadrons 
%which belong in the same jet is now restricted to assume 
%values which only have to be much smaller than $Q^2$. 

%The final expression for the DGLAP evolution of the dihadron 
%fragmentation function must include both the change due 
%to independent parton radiation with undetected hadrons 
%(the two previous sections), as well as the possibility 
%of the detected hadrons to emerge with larger values 
%of transerse momenta. 

Under the condition that $q_\perp << Q$ these hadrons 
can still be considered to belong to the same jet. 
Moreover, when $\Lambda_{QCD} << q_\perp$, the 
higher order contributions from multiple  
gluon vertex corrections to the semihard vertex are 
non-leading and thus the fragmentation of the 
two partons emanating from the split may be considered 
as independent. In Ref.~\cite{kon79b}, the authors 
demonstrated that the Sudakov double logarithms 
from the higher order vertex corrections are 
absent in the region $\Lambda_{QCD} << q_\perp$. In the 
interest of completeness we will repeat this derivation in 
a slightly different language in the Appendix.

%In the definition of the dihadron fragmentation function 
%at leading order we had indicated that the scale at which 
%the fragmentation functions are to be defined must be such that
%$\mu \geq \mu_\perp >> \Lambda_{QCD}$. All values of 
%transverse momenta below $\mu$ are integrated and included 
%within the leading order definition of the fragmentation 
%function. The process indicated in Fig. (\ref{nlofrag3}) 
%will solely include hadrons with transverse momentum greater than 
%this scale.  Evaluation of the leading log and leading twist 
%secton of this contribution will be the main topic of this 
%subsection. 

We will evaluate the leading log and leading twist 
contribution in which the hadron pair comes from the 
independent fragmentation of the quark and gluon after a semi-hard
split. We start again with the matrix 
element for this process,

\bea
\mat^{iii} &=& i\sum_q e^2 e_q g t^a 
\frac{g_{\nu \mu}}{q^2 + i\e} \bar{v}_{k_1} \g^\mu u_{k_2} \nn \\
&\times& \lc p_1, S_1-1 | \bar{\psi}_q(0) | 0 \rc 
\lc p_2, S_2-1 | A^a_\rho (0) | 0 \rc \nn \\
&\times& \Bigg[ \frac{\g^\rho (\f \fq - \f \fk ) 
\g^\nu}{(\fq - \fk)^2 + i\e} + 
\frac{ \g^\nu (\f \fp_{S_1} - \f \fq) \g^\rho }
{(\fp_{S_1} - \fq)^2 + i\e } \bigg] v(k) \nn \\ 
&\times& (2\pi)^4 \kd^4 (\fq - \fp_{S_1} - \fp_{S_2} -\fk ).
\eea

\nt
The second term inside the square bracket corresponds to
gluon emission from the antiquark in the quark direction
and will not contribute in the leading log approximation.
The out-state in this case is chosen 
to be the last line of Eq.~(\ref{diff_states}). 
In this case, the sum over all 
final states of hadrons from the initial quark 
has been broken into two identical complete sets. In each of the 
sets, $S_1$ and $S_2$, a single hadron will be identified. Unlike in the 
previous two subsections, where one of the quark or gluon  
operators is contracted with a partonic state; 
both the quark and the gluon operators will be contracted with
hadronic states in this case.
The cross section constructed from this matrix element contains 
%a sum over all final states. This is decomposed as 
two separate sums over hadronic states; one of which has an 
overlap with a quark state, the other with a gluon state. 
The hadronic basis of states moving in the ``away-side'' jet 
will be replaced as before with the sum over all momentum states
of a single antiquark: 
%It may be noted, yet again that a matrix element such as the one above 
%is sensible perturbatively when there is a large enough transverse 
%momentum between the identified hadrons. Cases where this is not 
%have already been included previously.  

\bea
\sum_S =\frac{d^3 p_1}{ (2\pi)^3 2 E_1} \frac{d^3 p_2}{(2\pi)^3 2 E_2}
\frac{d^3 k}{ (2 \pi)^3 2 k} \sum_{S_1-1} \sum_{S_2-1} .
% 
%\prod_{f \in S-2}  \int \frac{ d^3 p_f }{ (2\pi)^3 2 E_f } \nn \\
%
%&\times& \frac{d^3 p_1}{(2\pi)^3 2 E_1} \frac{d^3 p_2}{(2\pi)^3 2 E_2} 
%
%\frac{d^3 k}{(2\pi)^3 2 k} \nn \\ 
%
%\prod_{f_1 \in X-1} \int \frac{d^3 p_{f_1}}{(2\pi)^3 2 E_{f_1}} \Bigg)
%
%\Bigg( 
%
%\prod_{f_2 \in Y-1} \int \frac{d^3 p_{f_2}}{(2\pi)^3 2 E_{f_2}} \Bigg) \nn \\
%
\eea

In the last two subsections, the quark attached to the 
electromagnetic vertex is assigned the momentum $\fp$, while 
the quark or gluon which materializes from the split is assigned the 
momentum $\fl$. In this subsection, the momentum of the fragmenting
gluon and quark will be set to be $\fl$ and $\fp$. The quark attached 
to the electromagnetic vertex will thus have a momentum of $\fp+\fl$.
This cosmetic reshuffling is solely to ease the extraction of the single 
fragmentation functions.

Again, we focus on the leading log portion of 
the matrix elements. This essentially restricts our 
attention to the square of the matrix element for the 
process depicted in Fig.~\ref{nlofrag3}. 
The integrated cross section for this process may be expressed as:

\bea 
\si^{iii} &=& \frac{1}{2s} \sum_q 
\int d^4 y e^{i\fy\x\fl} e^4 e_q^2 g^2  
\frac{ \ml_{\nu \B} }{4 (\fq^2)^2 } N_c C_F \label{si_nlo_10} \\
&& \!\!\!\!\!\!\!\!\!\!\!\!\!\!\!\! 
\int \frac{d^3 p_1d^3 p_2}{(2\pi)^6 4p_1p_2}  \frac{d^4 k d^4 l}{(2\pi)^8} 
\kd^+(\fk^2) \int d^4x e^{ \fx \x (\fq - \fl - \fk - \fp)} e^{i\fx\x\fp} \nn \\
&& \!\!\!\!\!\!\!\!\!\!\!\!\!\!\!\! 
\sum_{S_1-1} \sum_{S_2-1} \tr \Bigg[ \f \fk
\frac{\g^\B (\f \fq - \f \fk) \g^{\nu} }{(\fq - \fk)^2 - i \e} \nn \\
&&\!\!\!\!\!\!\!\!\!\!\!\!\!\!\!\! 
\times \lc 0 | \psi_q(x) | p_1, S_1-1 \rc 
\lc S_1-1 , p_1 | \bar{\psi}_q(0) |0 \rc
\frac{\g^\rho (\f \fq - \f \fk) \g^\nu }{(\fq-\fk)^2 + i \e} \Bigg] \nn \\
&& \!\!\!\!\!\!\!\!\!\!\!\!\!\!\!\! 
\times \frac{\kd^{a b}}{2}
\lc 0 | A_\si^b(x) | p_2 , S_2-1 \rc \lc S_2-1, p_2 | A_\rho^{a} | 0 \rc , \nn
\eea

\nt
where two identities of unity

\[
1 = \int d^4 l \int \frac{d^4 x}{(2\pi)^4} e^{i\fl \x \fx} ,
\]

\nt 
are inserted.
%Where, we have averaged over the spins of the incoming \epem pair and 
%summed over the colors and spins of the outgoing states. 
As in the preceding subsections the standard shift of partonic 
variables is introduced \ie $\fk \ra \fq - \fp - \fl$
and $\int d^4k\rightarrow \int d^4p$.
The squares of the overlap matrix elements between the 
partonic operators $\psi, A^a_\rho$ and the hadronic states will 
result in the single fragmentation functions. 
Absorbing the integrals over undetected hadron states and the Fourier 
integrals, the $\hat{T}$ matrix 
elements may be written as (see Sec. II or Ref.~\cite{osb03}):
\bea 
[\hat{T}_g ]^{ba}_{\si \rho} &=& \int d^4 y e^{\fy \x \fl} \sum_{S_2-1}  \nn \\
%\prod_{f_2 \in Y-1} \int 
%
%\frac{d^3 p_{f_2}}{(2\pi)^3 2 E_{f_2}} \nn \\
%
&& \!\!\!\!\!\!\!\!\!\!\!\!\!\!\! 
\lc 0 | A^b_\si (y) | S_2-1,p_2\rc \lc p_2 , S_2-1 | A^a_\rho(0) |0 \rc ,
\eea  

\nt
which leads to the definition of the gluon fragmentation 
function at leading twist. The same extraction 
may also be performed for the quark overlap matrix operator:

\bea 
[\hat{T}_q ]^{\alpha\gamma} &=& \int d^4 x e^{\fx \x \fp}\sum_{S_1-1} \nn \\
%\prod_{f_1 \in X-1} \int 
%
%\frac{d^3 p_{f_1}}{(2\pi)^3 2 E_{f_1}} \nn \\
%
&& \!\!\!\!\!\!\!\!\!\!\!\!\!\!\! 
\lc 0 | \psi_q^\alpha (x) | S_1-1,p_1\rc \lc p_1 , S_2-1 
| \psi_q^\gamma(0) |0 \rc ,
\eea

\nt 
which leads to the definition of the quark fragmentation 
function at leading order. 
%The reader will note that contrary to the previous subsections we are now 
%factoring out quantum overlap matrix elements which will lead to single fragmentation 
%functions as opposed to the double fragmentation functions. 
Within the collinear approximation applied to the above two
 matrix elements, they may be approximated at leading twist as 

\bea 
\mbox{} [\hat{T}_g ]^{ba}_{\si \rho} (l;p_2) &\simeq& 
\kd^{ab} d_{\si \rho}(l) T_g (l;p_2) , \\
%\eea
%\bea
\mbox{} \hat{T}_q (p;p_1) &\simeq& 
\frac{\f \fp_1}{2} T_q (p;p_1) .
\eea

\nt
With the definition of the single fragmentation functions
and the collinear approximation, the cross section for 
independent fragmentation may be expressed in a simplified form as

\bea
\si^{iii} &=& \frac{1}{2s} \sum_q \frac{e^4 e_q^2 N_c}{4Q^4}  \ml_{\B \nu} g^2 C_F 
\int \frac{d^3 p_1 d^3 p_2}{(2\pi)^6 4E_1 E_2} \nn \\
&& \!\!\!\!\!\!\!\!\!\!\!\!
\int \frac{d^4 l  d^4 p}{(2\pi)^8} 2\pi \kd^+( (\fq - \fl -\fp)^2 ) 
T_g(l;p_2) T_q(p;p_1) d_{\rho \si}(l) \nn \\ 
&& \!\!\!\!\!\!\!\!\!\!\!\!\!\!\!\!\!\!\!\!\!\!\!\!\!\!\!\! \mbx \times
\tr \Bigg[ (\f \fq - \f \fl - \f \fp) \g^\B \Bigg\{ 
\frac{\f \fl + \f \fp}{(\fl +\fp)^2} \g^\si \frac{\f \fp_1}{2} 
\g^\rho \frac{\f \fl + \f \fp}{(\fl +\fp)^2} \Bigg\} \g^\nu \Bigg]. 
\eea

\nt
This is the leading log contribution to the inclusive
cross section for the production of two identified hadrons at 
next-to-leading-order where each hadron emanates from the independent 
fragmentation of a parton. The overlap matrix elements which lead to the 
definition of the fragmentation functions, $T_g$ and $T_q$ have already been 
factored out. 
%The remaining expression contains a trace over eight 
%$\g$ matrices. The terms isolated inside the curly brackets along with 
%the factors $T_q $ and $T_g d_{\rho \si}$ 
This represents a NLO 
contribution to the double fragmentation of the quark emanating 
from the electromagnetic vertex.

Again, we use a collinear approximation to isolate the leading
twist part of the terms inside the curly brackets:

\bea 
\frac{\f \fl + \f \fp}{(\fl + \fp)^2} \g^{\si} \frac{\f \fp_1}{2}
\g^{\rho} \frac{\f \fl + \f \fp}{(\fl + \fp)^2}d_{\rho \si}(l) \nn \\
\simeq \frac{\f \fp_h}{2}
\tr \Bigg[ \frac{\g^+}{2 p_h^+}
\frac{\f \fl + \f \fp}{(\fl + \fp)^2} \g^{\si} \frac{\f \fp_1}{2}
\g^{\rho} \frac{\f \fl + \f \fp}{(\fl + \fp)^2}d_{\rho \si}(l)
 \Bigg].
\eea 

\nt
%At leading twist, and a summation over all final spins, the above 
%Dirac matrix may be approximated as 
%\bea 
%\hat{C} = \frac{\f \fp_h}{2} C = \frac{z}{2} (\f \fl + \f \fp) C,
%\eea
%\nt 
%where the leading twist portion under the collinear approximation has been 
%isolated in the second equality. The factor $z$ has not as yet been formally 
%introduced, for the moment it serves only as the ratio $p_h^+/(p^+ + l^+)$.
%The factor $C$ may be obtained as a trace itself, 
%
%\bea 
%C = \tr \Bigg[ \frac{\g^+}{2 p_h^+} \hat{C} \Bigg] .
%\eea

After introducing two momentum fractions $z'_1$ and $z'_2$ through
a multiplicative factor of unity,

\bea 
1 = \int_0^1 dz'_1 dz'_2 \kd \Bigg( z'_1 - \frac{p^+_1}{p^+} \Bigg) 
\kd \Bigg( z'_2 - \frac{p^+_2}{l^+} \Bigg),
\eea

\nt
and a rearrangement of the integrals we have the following
factorized from of the cross section at leading twist,

\bea
\si^{iii} &=& \frac{1}{2s} \sum_q \frac{e^4 e_q^2 N_c}{4Q^4}  
\ml_{\B \nu} g^2 C_F 
\int \frac{dz'_1 dz'_2 d^3 p_1 d^3 p_2}{(2\pi)^6 4E_1 E_2} \nn \\
&& \!\!\!\!\!\!\!\!\!\!\!\!\!\!\!\!\!\!\! \mbx \times
\tr \Bigg[ (\f \fq - \f \fl - \f \fp) \g^\B \frac{\f \fp_h}{2} \g^\nu \Bigg] 
2\pi \kd^+( ( \fq - \fl -\fp )^2 )  
\Bigg|_{ \fp = \fp_1/z'_1 }^{ \fl = \fp_2/z'_2 } \nn \\ 
&&  \!\!\!\!\!\!\!\!\!\!\!\!\!\!\!\!\!\!\! \mbx \times
\Bigg[ d_{\rho \si}(l) \tr \Bigg\{  \frac{\g^+}{2 p_h^+} 
\frac{\f \fl + \f \fp}{(\fl + \fp)^2} \g^{\si} \frac{\f \fp_1}{2}  
\g^{\rho} \frac{\f \fl + \f \fp}{(\fl + \fp)^2}
\Bigg\} \Bigg]_{ \fp = \fp_1/z'_1 }^{ \fl = \fp_2/z'_2 } \nn \\ 
&& \!\!\!\!\!\!\!\!\!\!\!\!\!\!\!\!\!\!\! \mbx \times
\int \frac{d^4 l }{(2\pi)^4}  \kd \Bigg( z'_2 - \frac{p^+_2}{l^+} \Bigg) T_g (l;p_2) \nn \\
&& \!\!\!\!\!\!\!\!\!\!\!\!\!\!\!\!\!\!\! \mbx \times
\int \frac{ d^4 p}{(2\pi)^4}  \kd \Bigg( z'_1 - \frac{p^+_1}{p^+} \Bigg) T_q (p;p_1) .
\label{si_nlo_12}
\eea

\begin{figure}[htb!]
%\begin{center}
%  \epsfxsize 80mm
%\hspace{0cm}
 \resizebox{3.25in}{5in}{\includegraphics[0.5in,1in][6.5in,10in]{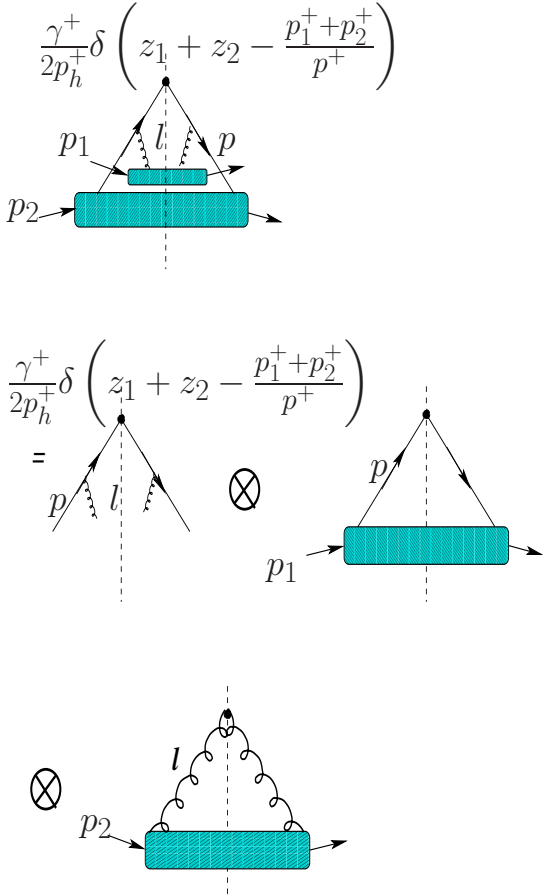}} 
%  \resizebox{3in}{5in}{\includegraphics[1in,1in][8.0in,10in]{nlocutvert3.eps}} 
%\vspace{0.25cm}
    \caption{Next-to-Leading order cut-vertex representation of the
     mixed contribution. }
    \label{cutvert3}
%  \end{center}
\end{figure}

\nt
It is apparent that the second line of the above equation 
corresponds to the hard cross section of an \epem pair annihilating 
via a single virtual photon to a $q\bar{q}$ pair. The third line 
corresponds to the splitting of the quark into a quark and gluon. Note the 
absence of any $\kd-$function maintaining an on-shell condition. 
This indicates that neither the quark nor gluon is being cut.
The fourth and fifth line indicate the independent 
fragmentation of the quark and gluon into hadrons with the 
identification of a single hadron from each of these sources. 
The cut-vertex structure of this process resembles 
that of Fig.~\ref{cutvert3}. The trace over the Dirac matrix 
structure of the third line may be performed, followed by 
a contraction of the Lorentz indices to obtain:
\[
\frac{z'_1 8 (p^+_1/z'_1 + p^+_2/z'_2 ) }{4p_h^+ (\fp_1/z'_1 + \fp_2/z'_2 )^2}
\frac{1+y^2}{1-y} ,
\]

\nt
where the variable $y$ is introduced once again as the integral over a 
$\kd-$function 

\bea
\int_0^1 dy \kd \Bigg(  y - \frac{ p_1^+/z'_1 }{ p_1^+/z'_1 + p_2^+/z'_2 } \Bigg).
\label{oth_delta}
\eea

\nt
One notes that the variable $y$ is essentially the ratio of the 
forward light-cone momentum of the offspring quark to that of the 
parent. This leads to the same splitting function as that of a 
quark splitting to a quark and a gluon. 
We may have chosen $y$ to represent the ratio of the energy of the 
gluon to that of the parent quark. This would have resulted in a splitting 
function similar to that of the preceding subsection. 
Changing the order of integration between the various 
ratios $y,z'_1,z'_2$, we define the quantities $z_1 = z'_1/y$ and 
$z_2 = z'_2/(1-y)$. Note that the double differential 
cross section $d^2 \si/dz_1 dz_2$ involves the ratios 
$z_1, z_2$ of the hadronic forward light-cone momentum to that of the 
parent quark emanating from the electromagnetic vertex. 
Following this we may again switch the order of integration, 

\bea 
&& \int_0^1 dz'_1 \int_0^1  dz'_2 \int_0^1 dy \nn \\
&=& \int_0^1 dy \int_0^1  dz'_1 \int_0^1 dz'_2 \nn \\
&=& \int_0^1 dy \int_0^{1/y}  \frac{dz_1}{y} \int_0^{1/(1-y)} \frac{dz_2}{(1-y)}  \nn \\
&=& \int_0^1 dz_1 \int_0^1 dz_2 \int_{z_1}^{1-z_2} \frac{dy}{y(1-y)} .
\eea

The integration over the hadronic momenta may now be subjected to the same 
variable transformation as demonstrated in Fig.~\ref{fig4}. 
Within the collinear approximation it may be easily demonstrated that

\bea 
(\fl + \fp)^2 &=& \frac{q_\perp^2 {p_h^+}^2 y(1-y)}{4p_1^+ p_2^+ z_1 z_2}.
\eea

\nt
The $\kd-$function introduced in Eq.~(\ref{oth_delta}) may also 
be similarly simplified to obtain,

\bea 
\kd \Bigg( y - \frac{ p_1^+/z'_1 }{ p_1^+/z'_1 + p_2^+/z'_2 } \Bigg) 
&=& \kd \Bigg[ \frac {y(1-y)}{p^+ + l^+} 
\Bigg( \frac{p_2^+}{z_2} - \frac{p_1^+}{z_1} \Bigg) \Bigg] \nn \\
&=&  \frac{p^+ + l^+}{y(1-y)} \kd \Bigg( \frac{p_2^+}{z_2} - \frac{p_1^+}{z_1} \Bigg),
\eea

\nt
where the quantities $p^+$ and $l^+$ are subjected to the condition of constraint 
introduced in Eq.~(\ref{si_nlo_12}).  The $\kd-$function is now 
similar to the second $\kd-$function in 
Eq.~(\ref{sigma_3}), and may be used to extract the hard cross 
section $\si_0^{q\bar{q}}$ [see Eq.~(\ref{si_0})].  

%({\bf Again, the argument is too lengthy})
In the last two subsections, there is an integration 
over the transverse momentum of the 
parent quark emanating from the EM vertex. 
%Indeed, the 
%assumption of a vanishing value by this 
%quantity is the source of the uncancelled 
This integration has a collinear divergence 
that must be absorbed into the renormalized 
fragmentation functions. 
%The method of derivation 
%consisted in aligning t
The light-like 
null vector $\fn$ was alligned in a direction such that its three-components 
remained opposite to those 
of $\fp_h$, the sum of the momenta of the detected hadrons. 
By definition $\fn$ has no 
transverse component. As both hadrons originated from the same 
parton, the preceding 
condition along with the assumption of collinearity of the 
final hadrons with the fragmenting 
parton, constrained the transverse component of the 
fragmenting parton to be near vanishing. 
In this case we have again chosen the three components 
of $\fn$ to be opposite to those of 
$\fp_h$. However, the two detected hadrons in this 
subsection emerge from two different 
partons which come from a split of the 
parent quark. As a result, whereas 
both fragmenting partons have a transverse 
component ($l_\perp,p_\perp$) 
proportional to the transverse component of the 
detected hadrons $q_\perp$, the 
parent quark is restricted to near vanishing values of a 
transverse component. 
In conformity with the previous subsections and with 
the DGLAP evolution of the 
single fragmentation function, the integration over a 
transverse component will remain that
of a partonic momentum. The hadronic transverse component 
$q_\perp$ may be trivially related to its partonic counterpart thus:
\nt
\bea 
q_\perp = 2 l_\perp \frac{z_1 z_2}{ y(1-y) (z_1 + z_2) }. \label{q_to_l}
\eea

With these approximations, it may be easily demonstrated that 
the double differential cross section of independent fragmentation assumes the form as 
sketched in the concluding lines of Eq.~(\ref{sigma_glue_mass}), i.e.,
\nt
\bea 
\frac{d^2 \si ^{iii}}{ d z_1 d z_2 } &=& \sum_q \si_0^{q\bar{q}} 
\frac{ \A_s }{ 2\pi } \int \frac{dl_\perp^2}{l_\perp^2} 
\int_{ z_1 }^{ 1 - z_2 } \frac{ dy }{ y(1-y) } \nn \\ 
& & C_F \frac{ 1 + y^2 }{ 1 - y } D_q \left( \frac{z_1}{y} \right)  
D_g \left( \frac{z_2}{1-y} \right).
\label{si_nlo_14}
\eea
\nt
Where $D_q$ and $D_g$ are the single fragmentation functions 
of a quark and a gluon \cite{col89} 
(see also Sect. II).
 
The careful reader will have noted that the order 
of integration between $y$ and $l_\perp$ has 
been switched; whereas from Eq.~(\ref{q_to_l}) we 
see that if $q_\perp$ is independent of 
$y$ then $l_\perp$ is a function of $y$. There are 
two ways to resolve this. We may ascribe 
a multiplicative $y$ dependence to the maximum value of $q_\perp$. 
Indeed this would be an artificial dependence, and 
would physically mean the inclusion of an 
extra piece of phase space in the $l_\perp$ integration. 
Any such multiplicative factor 
would have no influence on the DGLAP evolution equations 
as these would involve a differentiation 
with respect to $Q^2$ \ie

\bea 
\frac{\prt}{\prt Q^2} \int^{\kd^2Q^2} \frac{dl_\perp^2}{l_\perp^2} &=& 
\frac{\prt}{\prt Q^2} \int^{Q^2} \frac{dl_\perp^2}{l_\perp^2}.
\eea  

\nt
Yet another approach may be to view the order of 
integrals in Eq.~(\ref{si_nlo_14}) as merely
symbolic, the differentiation with respect to $Q^2$ 
leads to the same evolution equations.

There remains one last contribution to the inclusive cross 
section for the production of two hadrons from a parent quark:
this is obtained trivially by switching $z_1$ and $z_2$ 
in Eq.~(\ref{si_nlo_14}), \ie the hadron with momentum 
fraction $z_1$ originates from the fragmentation of the 
gluon rather than the quark.   
With the addition of the above mentioned contribution, 
we complete the discussion of the real radiative corrections to 
the inclusive cross section of same-side dihadron 
production. The contributions discussed in this 
subsection possess no infrared divergence 
as the $y$ integration is terminated at $1-z_2$ or $1-z_1$,
contrary to dihadron fragmentation from the single quark
whose infrared divergence is cancelled by the virtual correction.

%The contributions of the 
%first subsection displays an infrared divergence. 
%In the case of single fragmentation 
%functions, such a divergence is cancelled by a virtual correction. 
%Such is also the case here. In the remaining we will focus 
%our attention solely on the non-singlet fragmentation functions. 
%This will be carried out solely in the interest of 
%simplicity. Thus we defer the discussion of the 
%virtual correction to the next section. In this section we have completed 
%our program of demonstrating that the parton model like picture used to 
%derive the DGLAP evolution equations in Sec. II are indeed the dominant 
%behaviour demonstrated by the NLO cross sections. This justifies the 
%derivation of the evolution equations in Sec. II. 

%%%%%%%%%%%%%%%%%%%%%%%%%%%%%%%
%%%%%%%%%%%%%%%%%%%%%%%%%%%%%%
%%%%%%%%%%%%%%%%%%%%%%%%%%%%%%
%%%%%%%%%%%%%%%%%%%%%%%%%%%%%%

\section{Renormalized fragmentation functions and DGLAP evolution}

%%%%%%%%%%%%%%%%%%%%%%%%%%%%%%%
%%%%%%%%%%%%%%%%%%%%%%%%%%%%%%
%%%%%%%%%%%%%%%%%%%%%%%%%%%%%%
%%%%%%%%%%%%%%%%%%%%%%%%%%%%%%

\begin{figure}[htb!]
%\begin{center}
%  \epsfxsize 80mm
%\hspace{0cm}
  \resizebox{4in}{3.5in}{\includegraphics[0in,0in][6in,5in]{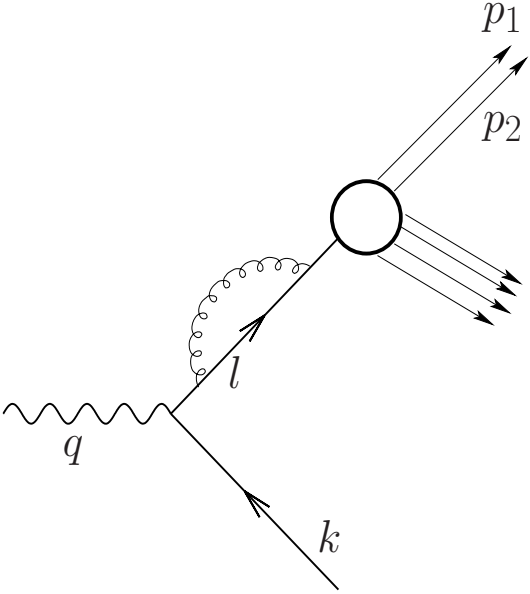}} 
%  \resizebox{3in}{3in}{\includegraphics[1in,3in][8.0in,8in]{nlofrag4.eps}} 
%\vspace{0.25cm}
    \caption{The leading log self-energy contribution to the NLO 
    modification of the quark fragmentation function. }
    \label{nlo_self}
%  \end{center}
\end{figure}

So far we have calculated the real radiative corrections to the dihadron
fragmentation functions. One notices that the contribution from
quark fragmentation after a gluon radiation in Eq.~(\ref{si_nlo_8})
contains an infrared divergence. Such an infrared divergence will be canceled 
by virtual contributions from interference diagrams such as those of 
Fig.~\ref{nlo_self}. It is well known that in gauge 
theories, in light-cone gauge, the leading log contribution 
is contained solely within the self-energy diagram and not 
shared between the self-energy
and the vertex correction, as is the case in the Feynman gauge \cite{col89}. 
The hadronic tensor for such a virtual correction is 
\bea 
W^{\mu \nu} &=& \sum_{S-2} 
%\Bigg( \prod_{f  \in S-2}   \int \frac{d^3 p_f}{(2\pi)^3 2 E_f}  \Bigg)
%
\int \frac{d^3 p_1 d^3 p_2 }{(2\pi)^6 4 E_1 E_2} \frac{d^4 p}{(2\pi)^4} 
2\pi \kd ( (\fq - \fp)^2  )
\nn \\
&& \hspace{-0.5 in} \times \tr \Bigg[ \widehat{T}_q(p;p_1,p_2)  \{ -i \Sigma (p) \}  
\frac{i}{\f \fp} \g^{\nu} ( \f \fq - \f \fp ) \g^{\mu} \Bigg],
\eea
\nt 
where, $-i\Sigma(p)$  respresents the one-loop quark self-energy 
of a quark with four-momentum $\fp$.  

The hadronic tensor may again be factorized at leading twist 
following much the same procedure as those of the last section. 
There remains the integration over the internal gluon momentum 
$\fl$. The leading behaviour of this integral, as in the case of 
single fragmentation, in the part of phase space that includes 
a pinch singularity on the $l^-$ contour, and endpoint 
singularities on the $l_\perp $ and $l^+$ contours.  The pinch 
singlularity may occur only in the region where 
\[
0< l^+ <p^+ \mbox{\hspace{1cm} and \hspace{1cm}}
0 < l_\perp^2 < \fp^2 \leq Q^2.
\] 
The derivation of the leading behaviour, which mostly mirrors the 
calculation for single fragmentation functions, will not be 
presented in full detail here. We refer the reader to Ref.~\cite{col89b}, 
for details. The final result of the self-energy 
correction to the partial double differential cross section is 
\bea 
\frac{d^2 \si}{ d z_1 d z_2} &=& - \sum_q \si_0^{q\bar{q}} \frac{\A_s}{2\pi} 
\int^{Q^2}\frac{d l_\perp^2}{ l_\perp^2} \nn \\
&\times& \int_0^1 \frac{1+y^2}{1-y}  D_q^{h_1 h_2} (z_1,z_2). \label{si_nlo_20}
\eea

\nt
Note that the transverse integration is over $l_\perp$ and not the 
parent quark momenta. The variable $y$ is defined as $y = l^+/p^+$. 
The overall negative sign should not be a cause for alarm, as this is 
only a part of the total cross section. Combining the above 
equation with Eq.~(\ref{si_nlo_8}) leads to the cancellation of 
the infrared singularity as $y \ra 1$. One can effectively
combine the virtual and real corrections with a ``$+$''-function,

\bea
P_{q \ra qg} (y) &=& C_f \bigg[ \frac{1+y^2}{(1-y)} \bigg]_+ \nn \\ 
&=& C_f\bigg[ \frac{1+y^2}{(1-y)_+}+\frac{3}{2}\delta(1-y)\bigg],
\eea

\nt
where the ``$+$''-function is defined as

\bea
\int_0^1dy \frac{F(y)}{(1-y)_+}\equiv \int_0^1 dy \frac{F(y)-F(1)}{1-y}
\eea
with $F(y)$ being any function that is sufficiently smooth at $y=1$.

%Calculationally this leads to the formulation of the splitting `+' function:
%\bea 
%P_{q \ra qg} (y)_+ &=& C_F \Bigg( \frac{1 + y^2}{1 - y} \Bigg)_+  \nn \\
%%
%&=& \lim_{\B \ra 0} \frac{1 + y^2}{1-y} \h (1 - y - \B) \nn \\ 
%
%&-&   \kd(1 - y - \B) \int_0^{1-\B} dx \frac{1 + x^2}{1-x}
%\eea
%The reader will note that the second piece on the right hand side 
%of the above equation
%is nothing but the contribution form the virtual diagram of Fig. (\ref{nlo_self}) and 
%Eq.~(\ref{si_nlo_20}). This modification effects only the first splitting 
%function of the 
%dihadron fragmentation function evolution equation (Eq.~(\ref{ns_dglap})).  The second 
%splitting function which leads to independent fragmentation of the quark and the gluon 
%does not possess an infrared divergence (as the $y$ integration is bounded by $1-z_2$) 
%and also has no obvious virtual correction. With this final modification the 
%evolution of the non-singlet quark fragmentation function with the energy of the 
%\epem annihilation may be computed. 

In the remaining, we will focus on the non-singlet (NS) fragmentation
functions for simplicity. In this case, the contribution from dihadron
gluon fragmentation drops out. 
Summing all three types of contributions 
from the last section, we obtain the NLO contribution to 
the NS dihadron fragmentation function,

\bea
D^{h_1,h_2}_{NS} (z_1,z_2,Q^2) &=& D^{h_1,h_2}_{NS} (z_1,z_2) \nn \\
& & \hspace{-1.1in} \mbx + \frac{\A_s}{2\pi} \int^{Q^2} 
\frac{d p_{\perp}^2 }{p_{\perp}^2} 
\int_{z}^1 \frac{dy}{y^2} C_F \Bigg( \frac{1+y^2}{1-y} \Bigg)_+
D^{h_1,h_2}_{NS}(\frac{z_1}{y};\frac{z_2}{y}) \nn \\
& & \hspace{-1.1in} \mbx + \frac{ \A_s }{ 2\pi } \int^{Q^2} \frac{dl_\perp^2}{l_\perp^2} 
\int_{ z_1 }^{ 1 - z_2 } \frac{ dy }{ y(1-y) } \nn \\ 
& & \hspace{-1.1in} \mbx \times C_F \frac{ 1 + y^2 }{ 1 - y } 
D_{NS}^{h_1} \left( \frac{z_1}{y} \right)  
D_g^{h_2} \left( \frac{z_2}{1-y} \right),
\eea

\nt
where the leading order fragmentation functions are defined as
matrix elements of field operators in Eq.~(\ref{dihad_def})
and (\ref{dz1}). This has exactly the same structure as we
have outlined in the parton model in Eq.~(\ref{dqq_coll}).
Therefore, the definition of renormalized fragmentation
functions and the derivation of the DGLAP evolution can be
similarly applied here.

The renormalized dihadron fragmentation function is defined as

\bea
D^{h_1,h_2}_{NS}(z_1,z_2,\mu^2) &\equiv& D^{h_1,h_2}_{NS}(z_1,z_2) \nn \\
& & \hspace{-1.0in} \mbx + \frac{\A_s}{2\pi} \int^{\mu^2} 
\frac{d p_{\perp}^2 }{p_{\perp}^2} 
P_{q\ra q g} * D_{NS}^{h_1 h_2}  \nn \\
& & \hspace{-1.0in} \mbx + \frac{ \A_s }{ 2\pi } \int^{\mu^2} 
\frac{dl_\perp^2}{l_\perp^2} 
\hat{P}_{q \ra q g} \bar{*} \bigg[D_{NS}^{h_1} D_g^{h_2}\bigg],
\eea

\nt
%where $m_g$ is fictitious gluon mass that is used to regulate the 
%collinear divergence.
We point out, once again, that the 
scale $\mu$ at which the renormalized functions are defined 
is chosen above the semihard scale $\mu_\perp$. At this 
scale, corrections to the renormalized quantities may be 
evaluated perturbatively.
In terms of the renormalized dihadron fragmentation function, 
the dihadron fragmentation function at NLO can be
expressed as

\bea
D^{h_1,h_2}_{NS} (z_1,z_2,Q^2) &\equiv& D^{h_1,h_2}_{NS}(z_1,z_2,\mu^2) \nn \\
& & \hspace{-1.0in} \mbx +\frac{\A_s}{2\pi} \log(\frac{Q^2}{\mu^2})
P_{q\ra q g} * D_{NS}^{h_1 h_2}(\mu^2)  \nn \\
& &\hspace{-1.0in} +\frac{ \A_s }{ 2\pi } \log(\frac{Q^2}{\mu^2})
\hat{P}_{q \ra q g} \bar{*} \bigg[D_{NS}^{h_1}(\mu^2) D_g^{h_2}(\mu^2)\bigg],
\label{final_nlo}
\eea

\nt
where we have also used the renormalized form of the single fragmentation
functions.

Note that we have introduced a minimum limit to the 
scale $\mu \gtrsim \mu_\perp$ in the definition of the renormalized dihadron 
fragmentation function. If one insisted in choosing $\mu < \mu_\perp$, then   
the contribution from independent
quark and gluon fragmentation after a semi-hard split would have to be factorized 
off at the scale $\mu_\perp$. In other words the minimum of the second 
logarithm in Eq.~(\ref{final_nlo}) would be set at $\mu_\perp^2 $  and 
would only contribute in the event that $Q^2$ is chosen greater than $\mu_\perp^2$.
In this way, we have assumed that only hadron pairs with relative transverse
momentum $q_\perp>\mu_\perp$ are generated perturbatively from
independent fragmentation of two separate partons in the
process of a perturbative cascade. For two partons whose relative
transverse momentum is smaller than $\mu_\perp$, nonperturbative
processes become important and their fragmentation cannot be
independent anymore. We include this part, which also contains 
a collinear divergence, in the renormalized dihadron fragmentation function.
This non-perturbative scale can also be considered as the intrinsic
relative transverse momentum of the dihadron fragmentation function 
and it should set the limit of the integration over $q_\perp$ in 
the matrix element definition of the dihadron fragmentation 
function in Eq.~(\ref{dihad_def}). If one wants to consider the
unintegrated (over $q_\perp$) dihadron fragmentation function, 
$\mu_\perp$ could set the initial condition for the $q_\perp$ 
distribution and can be used to study the evolution equation of the
angular distribution inside a jet. For now, this scale will
only set a limit of the physical scale $Q^2>>\mu^2>\mu_\perp^2>> \Lambda^2_{QCD}$ 
for the DGLAP evolution and will not enter the equation.
In the ensuing calculation of the evolution of the fragmentation functions 
we will not enter into such subtleties and always chose the starting scale 
$\mu \gtrsim \mu_\perp$.

To include the entire leading log modification, contributions from 
all the diagrams outlined in Fig.~\ref{insanity} have to be resummed 
into the scale dependent fragmentation functions. These are then 
differentiated to obtain the evolution equation which is given
exactly as in Eq.~(\ref{ns_dglap}) for NS dihadron fragmentation function.

%\bea 
%
%\frac{\prt D_{NS}^{h_1 h_2} (Q^2)}{\prt \log{Q^2}} &=& 
%
%\frac{\A}{2\pi} \Bigg[ P_{q\ra q g} * D_{NS}^{h_1 h_2} (Q^2)  \nn \\
%
%\mbx + && \!\!\!\!\!\!\!\!\!\! \hat{P}_{q \ra q g} 
%
%\bar{*} D_{NS}^{h_1} (Q^2) D_g^{h_2} (Q^2)  + 1 \ra 2 \Bigg] 
%
%\eea
%
%\nt
%The first splitting function $P_{q\ra q g} $ as derived 
%from Eq.~(\ref{si_nlo_8}) is merely,
%
%\bea 
%P_{q\ra q g} = C_F \frac{1+ y^2}{1-y},
%\eea
%
%\nt 

%%%%%%%%%%%%%%%%%%%%%%%%%%%%%%%
%%%%%%%%%%%%%%%%%%%%%%%%%%%%%%
%%%%%%%%%%%%%%%%%%%%%%%%%%%%%%

\section{Numerical results of Non-Singlet evolution }

%%%%%%%%%%%%%%%%%%%%%%%%%%%%%%
%%%%%%%%%%%%%%%%%%%%%%%%%%%%%%
%%%%%%%%%%%%%%%%%%%%%%%%%%%%%%

%In the preceding section, the NLO contribution to the double 
%inclusive cross section for the production of two hadrons 
%from an \epem annihilation 
%was computed. All the computations were, however, restricted 
%to the real contributions only. This allowed for an uncancelled 
%infrared divergence in Eq.~(\ref{si_nlo_8}) as $y \ra 1$. In the 
%case of single fragmentation functions this was cancelled 
%by the virtual contribution, and will be no different in this case.
%We will return to this topic momentarily. 
%
%The cross sections calculated in the preceding section will, as 
%mentioned in Sec. II be used to compute the correction to the 
%leading order definition of the fragmentation functions. 
%After factoring out the hard cross section
%$\si_0$, the remaining contribution will be included as a modification to 
%the fragmentation function. In the interest of simplicity we will 
%focus on the much simpler non-singlet fragmentation functions as 
%defined in Eq.~(\ref{non_singlet}).  The correction to these 
%functions at NLO emanates solely from the cross sections of 
%Eqs.~(\ref{si_nlo_8},\ref{si_nlo_14}), 

\begin{figure}[htb!]
%\begin{center}
%  \epsfxsize 80mm
%\hspace{0cm}
  \resizebox{3.2in}{4.1in}{\includegraphics[0.5in,1in][7.5in,10in]{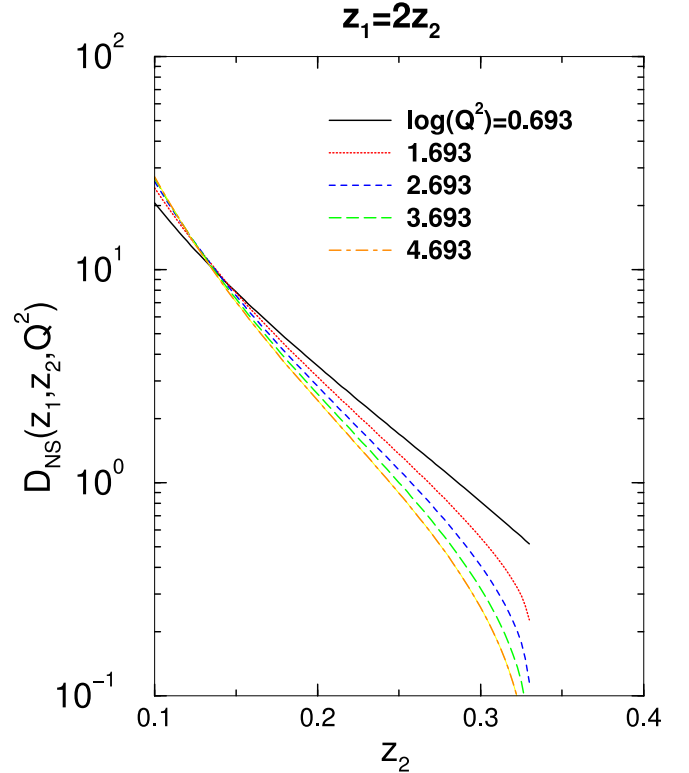}} 
%\vspace{0.25cm}
    \caption{Results of the evolution of the non-singlet quark dihadron 
    fragmention function $D_q^{h_1h_2}(z_1,z_2)$, where $z_1=2z_2$, from 
    $Q^2= 2 \gev^2 $ to $109 \gev^2$. See text for details.}   
     \label{res1}
%  \end{center}
\end{figure}

\begin{figure}[htb!]
%\begin{center}
%  \epsfxsize 80mm
%\hspace{0cm}
  \resizebox{3.2in}{4.1in}{\includegraphics[0.5in,1in][7.5in,10in]{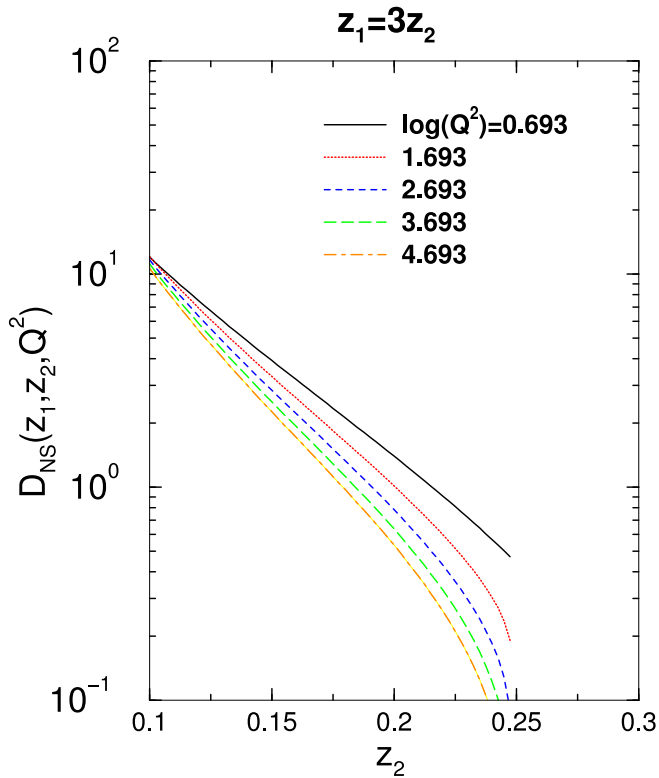}} 
%\vspace{0.25cm}
    \caption{Same as Fig.~\ref{res1} except $z_1=3z_2$. }
    \label{res2}
%  \end{center}
\end{figure}

In this section we will study numerically the DGLAP evolution of the
non-singlet dihadron fragmentation.
As in many other cases of DGLAP evolution, the solutions require
an initial condition of the fragmentation functions at an initial scale.
Such initial conditions, as in the case of single fragmentation functions,
are non-perturbative and are usually constructed from the 
experimental measurement of the single inclusive differential 
cross section $ d\si/dz$, according to Eq.~(\ref{LO_Dz}) at LO.
The evolution of the fragmentation functions with 
the energy scale of the reaction can then be calculated from the DGLAP 
equations.

The absence of any experimental data for two particle 
correlation in \epem annihilation forces us to 
formulate an ansatz of the initial condition. We simply use it as a toy model 
to illustrate the DGLAP evolution of the dihadron fragmentation
functions. We take the LO 
product of two single fragmentation functions as the initial condition 
for the evolution of the fragmentation function, \tie,

\bea 
D^{h_1 h_2}_q (z_1, z_2,\mu^2) &=& D^{h_1}_q (z_1,\mu^2) D^{h_2}_q (z_2,\mu^2)
\nn \\ 
& \times & \theta(1-z_1-z_2).
\eea

\begin{figure}[htb!]
%\begin{center}
%  \epsfxsize 80mm
%\hspace{0cm}
  \resizebox{3.2in}{4.1in}{\includegraphics[0.5in,1in][7.5in,10in]{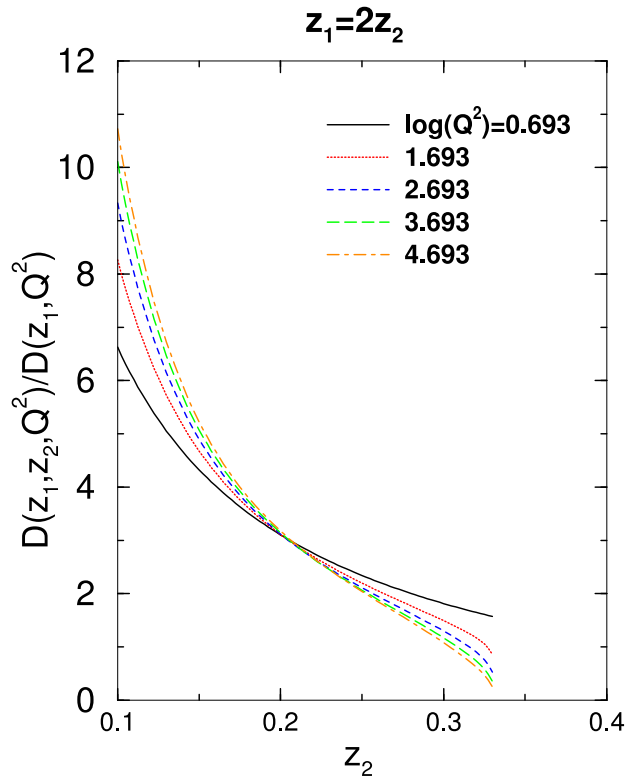}} 
%\vspace{0.25cm}
    \caption{Results of the ratio of the non-singlet quark dihadron 
    fragmention function $D_q^{h_1h_2}(z_1,z_2,Q^2)$ to the single 
    leading fragmentation function  $D_q^{h_1}(z_1,Q^2)$. In this 
    case $z_1=2z_2$ and
    $Q^2= 2 \gev^2 $ to $109 \gev^2$. See text for details.}
    \label{res3}
%  \end{center}
\end{figure}

\begin{figure}[htb!]
%\begin{center}
%  \epsfxsize 80mm
%\hspace{0cm}
  \resizebox{3.2in}{4.1in}{\includegraphics[0.5in,1in][7.5in,10in]{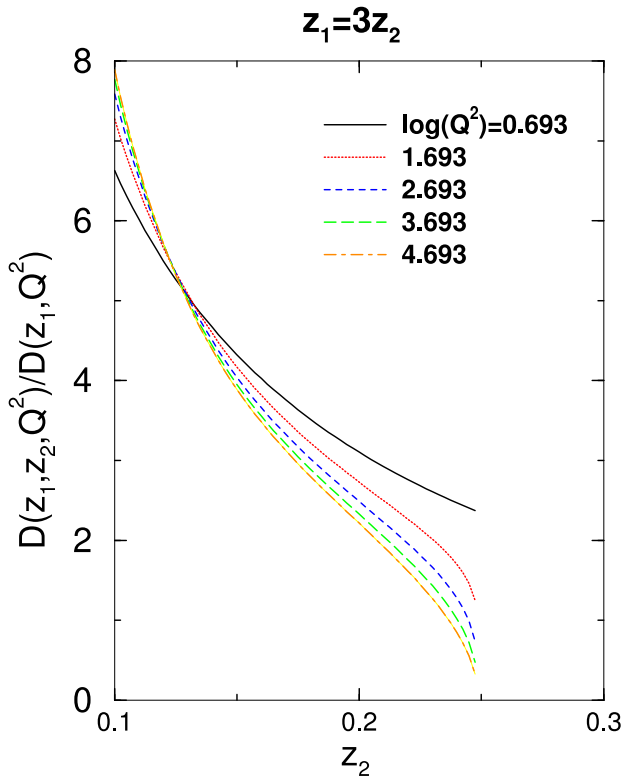}} 
%\vspace{0.25cm}
    \caption{Same as Fig.~\ref{res3} except $z_1=3z_2$.   }
    \label{res4}
%  \end{center}
\end{figure}

\begin{figure}[htb!]
%\begin{center}
%  \epsfxsize 80mm
%\hspace{0cm}
  \resizebox{3.2in}{4.1in}{\includegraphics[0.5in,1in][7.5in,10in]{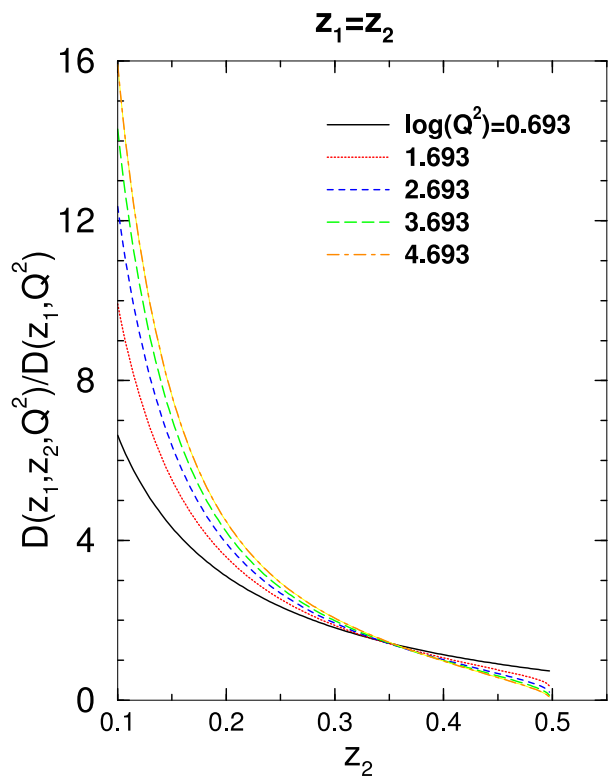}} 
%\vspace{0.25cm}
    \caption{Same as Fig.~\ref{res3} except $z_1=z_2$.  }
    \label{res5}
%  \end{center}
\end{figure}

\begin{figure}[htb!]
%\begin{center}
%  \epsfxsize 80mm
%\hspace{0cm}
  \resizebox{3.2in}{4.1in}{\includegraphics[0.5in,1in][7.5in,10in]{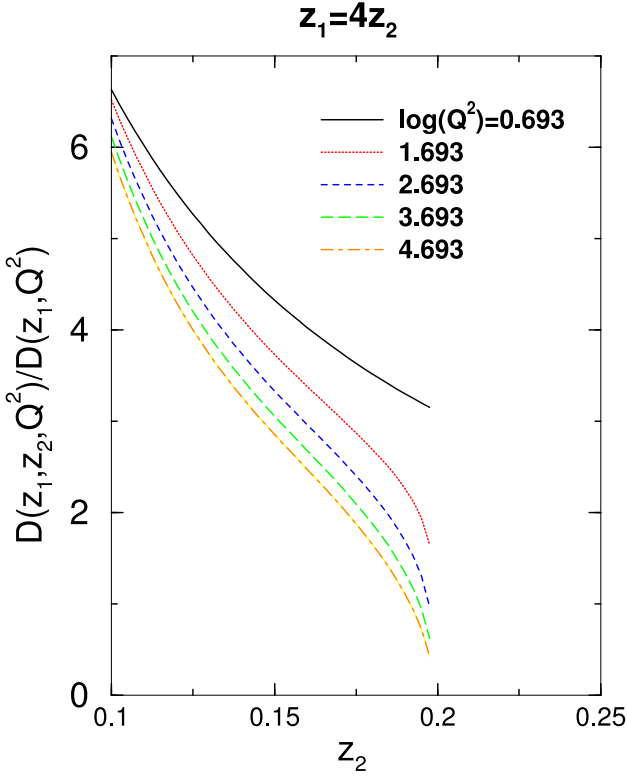}} 
%\vspace{0.25cm}
    \caption{ Same as Fig.~\ref{res3} except $z_1=4z_2$. }
    \label{res6}
%  \end{center}
\end{figure}

\begin{figure}[htb!]
%\begin{center}
%  \epsfxsize 80mm
%\hspace{0cm}
  \resizebox{3.2in}{4.1in}{\includegraphics[0.5in,1in][7.5in,10in]{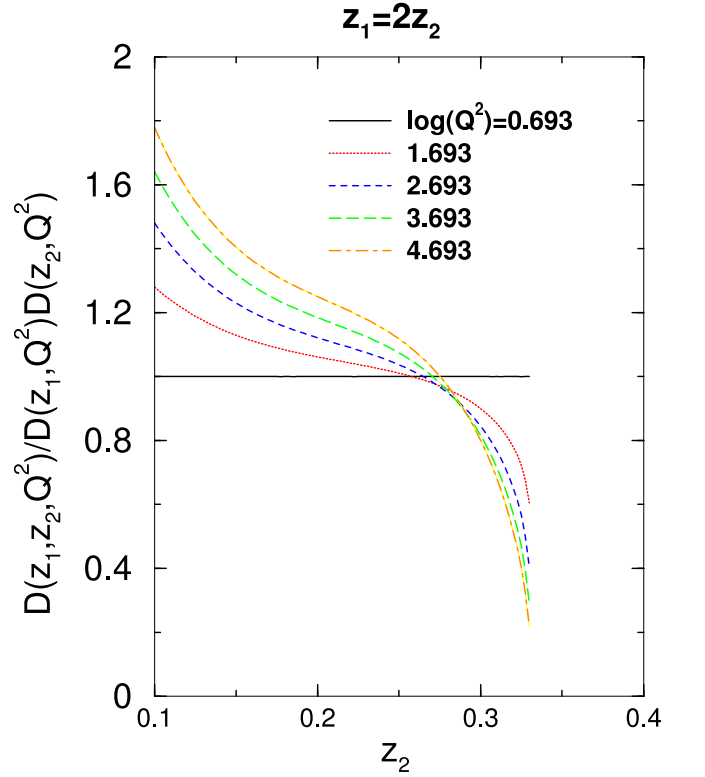}} 
%\vspace{0.25cm}
    \caption{Results of the ratio of the non-singlet quark dihadron 
    fragmention function $D_q^{h_1h_2}(z_1,z_2,Q^2)$ to the product of the single 
    fragmentation functions  $D_q^{h_1}(z_1,Q^2) D_q^{h_1} (z_2,Q^2)$. In this 
    case $z_1=2z_2$ and $Q^2= 2 \gev^2 $ to $109 \gev^2$. See text for details. }
    \label{res7}
%  \end{center}
\end{figure}

\begin{figure}[htb!]
%\begin{center}
%  \epsfxsize 80mm
%\hspace{0cm}
  \resizebox{3.2in}{4.1in}{\includegraphics[0.5in,1in][7.5in,10in]{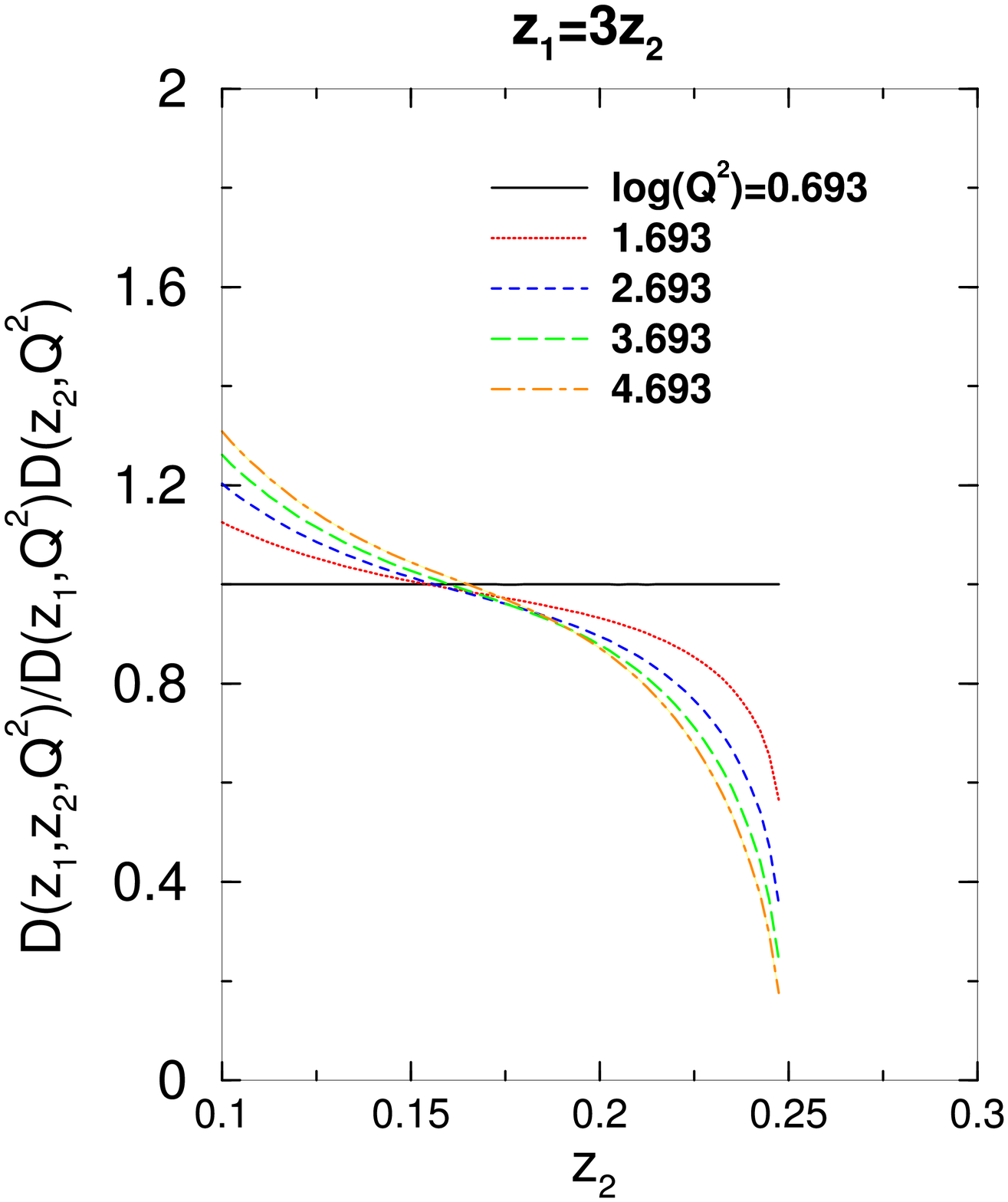}} 
%\vspace{0.25cm}
    \caption{ Same as Fig.~\ref{res7} except $z_1=3z_2$  }
    \label{res8}
%  \end{center}
\end{figure}

We set the initial 
condition at $Q^2 = 2 $GeV$^2$. This corresponds to $\log{Q^2} = 0.693$. 
While it may be argued that the initial energy is somewhat low for the 
applicability of pQCD in \epem annihilation, we consider it as just
the scale of the momentum transfer while the actual jet energy could be
sufficiently high.  
The differential equation corresponding to Eq.~(\ref{ns_dglap}) is then 
solved by the simple methods of a second order Runge-Kutta numerical 
estimation. Results are presented in Figs. \ref{res1}-\ref{res8} 
at intervals of $\Delta \log{Q^2} = 1.0$ 
%and correspond to the various lines as indicated in the plots. 
The initial 
condition is represented by the solid black line in all plots.
We stop the evolution at 
$\log{Q^2} = 4.693$, which corresponds to $Q^2 \simeq 109 $GeV$^2$.  

We present results where the leading particle possesses a
multiple of the momentum fraction of the next-to-leading 
particle. 
%We present results where the ratio is held fixed at integral values. 
We begin with plots of just the evolution of the non-singlet dihadron fragmentation 
function  at $z_1=2z_2$ in Fig.~\ref{res1} and $z_1=3z_2$ in Fig.~\ref{res2}.
In these and all other plots the results are always presented as a 
function of $z_2$.  
The results of the evolution are not qualitatively different from those 
of the single fragmentation function. Since the sum of the momentum fractions are 
constrained to unity \ie $z_1 + z_2 \leq 1$, we find that the fragmentation 
functions terminate at $z_2 = 1/(1+r)$ as appropriate. The intial condition, 
which is merely the product of two single fragmentation functions and is
not subjected to this constraint, does not show this behaviour. It is
imposed by hand, in the initial condition, that they vanish at and above this value.

In experiments, one can first identify a hadron as the leading hadron inside
a jet and use it as a trigger with given momentum $z_1$. Then, the associated
or the ``next-leading'' hadron distribution inside the same jet 
corresponds to the ratio of dihadron and single hadron fragmentation
functions, $D^{h_1,h_2}(z_1,z_2)/D^{h_1}(z_1)$.
We present results for this ratio at $z_1 = 2 z_2$ in Fig.~\ref{res3}, at 
$z_1= 3z_2$ in Fig.~\ref{res4} and in the extreme cases of $z_1=z_2$ 
in  Fig.~\ref{res5} and $z_1 = 4 z_2$ in Fig.~\ref{res6}. It should be 
pointed out that the $y$-axis in all these plots is linear and not 
logarithmic. Thus, one concludes that the ratio demonstrates little qualitative 
change for a variation of $Q^2$ by almost two orders of magnitude. 
In making such an observation, one must ignore the difference between the
solid line (initial condition) and the remaining lines (evolved functions) 
especially at large $z_2$. This is due to the fact that the initial 
condition is not subjected to the kinematic constraint as $z_2 \ra 1/(1+r)$.
However, the 
four plots are visually quite different from each other: the ratios of the 
evolved functions display a steady drop as compared to the initial condition 
as we progress from $z_1 = z_2$  to $z_1 = 4z_2$. The reader will also note 
that the maximum of the $y$-axis drops 
%as we plot cases 
with increasing 
$r$. 
%Thus it is not true that the evolution shows a more pronounced affect for 
%large values of $r$. 

As the energy of the \epem annihilation is raised the multiplicity 
must also increase. The cause for this is nothing other than the 
excess energy available for particle production. Following the 
physical picture of fragmentation proposed in Ref.~\cite{dok89}, 
one notes that at higher energies it becomes 
more probable that the two hadrons emanate from two causally disconnected 
sectors of the fragmenting jet. If this were the case, then at very high energies, 
the ratio of the double fragmentation function to the product of the single 
fragmentation functions 
($D_q^{h_1h_2}(z_1,z_2,Q^2)/D_q^{h_1}(z_1,Q^2) D_q^{h_1} (z_2,Q^2)$) should 
reach unity, especially for small values of
$z_1$, $z_2$. This has not turned out to be the case as evidenced by the plots of 
this ratio in Fig.~\ref{res7} for $z_1 = 2z_2$ and in Fig.~\ref{res8} for 
$z_1 = 3 z_2$. In these plots the ratio deviates from unity at small 
momentum fractions. Whether this is a facet of the choice of our initial 
conditions is as yet unclear.

%%%%%%%%%%%%%%%%%%%%%%%%%%%%%%%
%%%%%%%%%%%%%%%%%%%%%%%%%%%%%%
%%%%%%%%%%%%%%%%%%%%%%%%%%%%%%

\section{Discussions and Conclusions }

%%%%%%%%%%%%%%%%%%%%%%%%%%%%%%
%%%%%%%%%%%%%%%%%%%%%%%%%%%%%%
%%%%%%%%%%%%%%%%%%%%%%%%%%%%%%

In this paper, we have studied dihadron fragmentation functions 
within the framework of collinear factorization in the high-energy
$e^+e^-$ annihilation processes, starting with the operator definition.
Using the cut-vertex technique, we also derived the DGLAP evolution 
equation for the non-singlet dihadron quark fragmentation function. 
We solved the DGLAP evolution equation numerically, with a
simple ansatz for the initial condition.

Both the definition in operator formalism and the resultant DGLAP
evolution equations for the dihadron fragmentation functions have
remarkable similarities to the single fragmentation functions.
The first type of contribution to the evolution equations, that from gluon 
radiation before the fragmentation of the offspring quark (or gluon)
into a pair of hadrons, is very similar to the corresponding process
in single fragmentation functions. The second type, unique to the
dihadron fragmentation functions, comes from independent fragmentation
of the two offspring partons into two single hadrons of the observed
pair. Since this piece represents the incoherent fragmentation of 
the quark and the gluon, it is well defined only when the transverse 
momentum between the detected hadrons is large.

The relative transverse momentum between the two hadrons is
integrated over in the definition of the dihadron fragmentation
function. Hence, we have assumed that its non-perturbative contribution,
which resides in the operator definition, is restricted to
an intrinsic transverse momentum scale, $q_\perp<\mu_\perp$. Hadron
pairs with $q_\perp>\mu_\perp$ are assumed to be generated only
perturbatively. For this assumption to be justified, the semihard 
scale $\mu_\perp$ is chosen to be much larger than $\lambda_{QCD}$.
One has then to assume that the energy scales of 
the processes in question \ie $Q^2$ be much larger than the semihard 
scale $\mu_\perp$, such that a hierarchy of scales,
$\Lambda^2_{QCD} << \mu^2_\perp << Q^2$, is satisfied.

This study is motivated by the observation \cite{adl03} that the
same side correlations of two high $p_T$ hadrons in central
$Au+Au$ collisions remain approximately unchanged as compared
with that in $p+p$ and $d+Au$ collisions. Specifically in this
experiment, one measures the distribution, in azimuthal angle, 
of the secondary (or associated) 
hadron $\frac{1}{N_{trig}} \frac{d N}{d \phi}$ with respect
to the triggered high $p_T$ hadron. Neglecting the differences
in production cross section and fragmentation functions for
different parton species, the integrated yield of the correlation
around the peak at $\phi=0$ should be the ratio of dihadron and
single hadron fragmentation functions, 
$D_a^{h_1h_2}(z_1,z_2,Q^2) / D_a^{h_1}(z_1,Q^2)$, with $z_1$ and
$z_2$ being the momentum fractions of the triggered hadron
and associated hadrons, respectively. To understand the
observation in the framework of jet quenching, one has to study
the medium modification to a dihadron fragmentation function
due to parton energy loss. Since it has been shown in the case 
of single fragmentation functions that medium modification due 
to multiple scattering and induced gluon radiation closely 
resemble that of radiative corrections due to evolution in 
vacuum \cite{guowang}, the DGLAP evolution is expected to 
yield clues regarding  the medium modification of the dihadron 
fragmentation functions.

Our numerical results indeed show little change
of the ratio $D_q^{h_1h_2}(z_1,z_2,Q^2) / D_q^{h_1}(z_1,Q^2)$
as $Q^2$ is varied in a wide range of values.
The evolution is shown to be strongly dependent, however, 
on the ratio of the momentum fractions of the two hadrons ($r = z_1/z_2$).
In the results of Ref.~\cite{adl03} the ratio $r=z_1/z_2$ is 
essentially integrated over all values $\geq 1$.
In order to relate to the results in this paper one essentially 
must average the effects of evolution shown
in Figs. (\ref{res3}-\ref{res6}). One will immediately note 
that summing over different values of the ratio $z_1/z_2$ will 
lead to the observation of minimal change in the ratio of the 
fragmentation functions as a function of the $Q^2$ of the reaction.

No doubt, this study is but the first step in this effort. 
In the interest of simplicity, results for the computationally 
simpler non-singlet fragmentation functions were presented. 
The results for the evolution of the more physically relavant singlet 
fragmentation functions will be presented in a future effort. 
The DGLAP evolution equations for such functions 
will involve, in addition, the splitting of one gluon to 
two gluons and the coupled differential equations. 

In the above, we have demonstrated the factorization of the 
double differential cross section into a LO hard part and a
soft piece which encoded the nonperturbative information of 
converting partons into hadrons. A complete proof of factorization 
requires the extension of the calculation to all orders. There 
also remains the evaluation of the medium modifications to the dihadron 
fragmentation functions. We will address each of these issues, 
in turn, in future publications.

\begin{acknowledgments}
The authors wish to thank V. Koch, J. Qiu and G. Sterman
 for helpful discussions. 
This work was supported in part by the Natural Sciences and 
Engineering Research
Council of Canada, and in part 
by the Director, Office of Science, Office of High Energy and Nuclear Physics, 
Division of Nuclear Physics, and by the Office of Basic Energy
Sciences, Division of Nuclear Sciences, of the U.S. Department of Energy 
under Contract No. DE-AC03-76SF00098. 
\end{acknowledgments}

\appendix
\section{Sudakov double logarithms in light-cone gauge}

\begin{figure}[htb!]
%\begin{center}
%  \epsfxsize 80mm
%\hspace{0cm}
  \resizebox{4in}{3.5in}{\includegraphics[0in,0in][6in,5in]{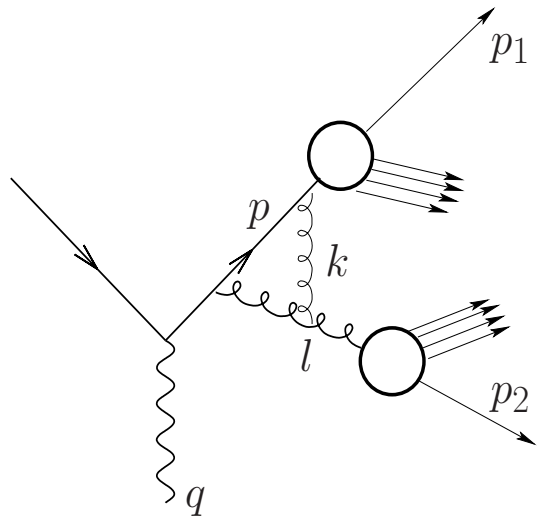}} 
%  \resizebox{3in}{3in}{\includegraphics[1in,3in][8.0in,8in]{nlofrag5.eps}} 
%\vspace{0.25cm}
    \caption{A next-to-next-to-leading order correction.}
    \label{jq1}
%  \end{center}
\end{figure}

In Subsect. c of Sect. IV we argued that one of the NLO contributions to 
the dihadron fragmentation functions would result from the convolution 
of two single fragmentation functions. The justification for the inclusion of 
this process rests on the assumption that the higher 
order diagrams 
%corrections to the split which consist of 
that have gluon lines 
%which 
connecting the outgoing quark and gluon 
can be ignored in a leading log analysis. 
The simplest 
higher order correction of this type emanates from the presence of a 
single gluon 
line connecting the outgoing quark and gluon as illustrated in 
Fig.~\ref{jq1}.

The presence of a Sudakov double logarithm in this 
diagram signals a leading log contribution 
%from this diagram 
to the 
infrared and collinear divergences of diagrams of the same order. 
Such a contribution will call into 
question our derivation of the second piece of the evolution equation 
[Eq.~(\ref{si_nlo_14})]. 
Derivation of this piece of the evolution equations required 
the identification of the leading log contributions from infrared and 
collinear divergences followed by factorization and a resummation of these 
into the single fragmention functions. Such a procedure may only be 
carried out if the leading log portions in higher order diagrams are 
contained solely in the selfenergy corrections of the outgoing 
gluon and quark lines and in real gluon or quark emmissions off these 
lines. A leading log contribution from the infrared or collinear 
sector of the diagram of Fig.~\ref{jq1} will call this procedure into 
question. In this appendix, we demonstrate that in the collinear 
sector of this diagram, evaluated  in light-cone gauge, the 
Sudakov double log is absent.

In this endevour we follow the techniques outlined in Ref.~\cite{col89} for 
the evaluation of the photon vertex. The leading contributions 
from the diagram of Fig.~\ref{jq1} may be evaluated as in \cite{col89} 
by the solution of the Landau equations. This leads to the presence 
of two regions of phase space where a double logarithm may arise:

\bea
p^\mu + k^\mu = \A_1 k^\mu &,&  \fk^2 = \fp\x\fk \nn \\ 
l^\mu - k^\mu = \A_2 k^\mu &,&  \fk^2 = \fp\x\fk  \label{col_pinch}
\eea

\nt
In the above, $\A_1,\A_2$ are mere real numbers. 
%and are not related to the 
%argument of the splitting functions described in the previous sections. 
These conditions essentially outline the cases where the 
gluon $k^\mu$ achieves collinearity with the outgoing 
quark with momentum $p^\mu$ or with the outgoing 
gluon with momentum $l^\mu$. Either case produces 
a double log contribution in the Feynman gauge. 

The expression for a quark with momentum $\fp + \fl$ splitting 
into an outgoing onshell quark with momentum $\fp$ and gluon 
with momentum $\fl$ with the splitting vertex corrected by  
a single gluon with momentum $\fk$ may be expressed as 

\bea 
& \bar{u}_i(p) &  \!\!\!
{ \Gamma ^a_{\ro} }_{i,j}  u_j (p+l) \ve_a^\ro(l) =  \bar{u}_i(p) 
\int \frac{d^4 k}{(2\pi)^4} i t^d g \g^\nu \nn \\
&\times& \!\!\! \frac{i (\f \fp + \f \fk ) }{(\fp+\fk)^2 + i \e} i t^c g \g^\mu 
\frac{i d_{\A \mu} (\fl - \fk)}{(\fl + \fk)^2 + i\e } g f^{acd} \nn \\
&\times& \!\!\! \bigg[ g^{\ro \A} (k - 2l)^\B + g^{\A \B} (l - 2k)^\ro + 
g^{\B \ro } (k + l)^\A \bigg] \nn \\
&\times& \!\!\! \frac{i d_{\B \nu} (k) }{\fk^2 + i\e} {\ve^a_\ro}^* (l) . \label{glue_vert}
\eea

It may be argued (see Ref.~\cite{col89}), that the double 
logarithm behaviour in vertex diagrams originates on the
collinear pinch surfaces outlined above in Eq.~(\ref{col_pinch}). 
The behaviour of this above vertex correction in the region where the 
internal gluon momentum $\fk$ becomes collinear to the the 
outgoing gluon momentum $\fl$ will now be analysed.
This corresponds to the 
second condition in Eq.~(\ref{col_pinch}). 
The $d^4k$ integration is decomposed into the light-cone 
variables $dk^+dk^-d^2k_\perp$. The focus is on the pinch 
sigularity which results from the two denominators 

\[
\frac{1}{\fk^2 + i\e} , \frac{1}{(\fl-\fk)^2 + i\e } .
\] 

The pinch between the two denominators arises in the $k^-$ integration 
solely in the region  $0<k^+<l^+$.  Evaluating the pole at 
$k^- = k_\perp^2/2k^+$. We obtain a pinch from the 
\[
\frac{1}{(\fl-\fk)^2 + i\e }
\] 
propagator in the region where 
$\vk_\perp \x \vl_\perp = k_\perp l_\perp$, \ie $\fk$ becomes 
collinear with $\fl$. To regulate the collinear divergence we introduce 
the variables $x,\kd \vk_\perp$:

\bea 
k^+ &=& x l^+  \nn \\
\vk_\perp &=& x \vl_{\perp} + \kd \vk_\perp . \label{subs}
\eea

Evaluating the integrand of Eq.~(\ref{glue_vert}) at the residue of the
pole $k^- = k_\perp^2/2k^+$ followed by the 
substitutions outlined in Eq.~(\ref{subs}) we obtain the split vertex correction
as

\bea 
& \bar{u}_i(p) &  \!\!\!
{ \Gamma ^a_{\ro} }_{i,j}  u_j (p+l) \ve_a^\ro(l) = S^a \int  \frac{dx}{x} 
\int  \frac{d^2 \kd k_\perp}{\kd k^2_\perp} \nn \\
&\times& \bar{u}(p) \g^\nu 
\frac{( \f \fp + x \f \fl ) p^+ l^+}{l_\perp^2 ( p^+ + l^+ )^2 } \g^\mu u(p+l) \nn \\
&\times& \bigg[ -g_{\A \mu} + \frac{(1-x) (l_\A n_\mu + l_\mu n_\A) }{(1-x) \fl\x \fn} \bigg] \nn \\
&\times& \bigg[ g^{\ro \A} (x-2)l^\B + g^{\A \B} (1-2x) l^\ro + g^{\B \ro} (x+1) l^\A \bigg] \nn \\
&\times& d_{\B \nu} (l) {\ve^a_\ro}^*(l).  \label{glue_vert_approx}
\eea 

\nt
In the above equation all factors of color together with multiplicative constants have been 
absorbed into the factor $S^\A$. The sum over polarizations of the gluon with momentum $\fl - \fk$
[\ie the factor $d_{\A \mu} (\fl - \fk) $] and the expression for the glue vertex have been simplified. 
In these and in the rest of the numerator, factors of $\kd \vk_\perp$ have been neglected, as 
the focus is on the region where $\kd k_\perp \ra 0$. 
If the numerator, under this approximation, turned out to be vanishing, then this would indicate
the leading contribution to be proportional to  $\kd k_\perp \ra 0$ and 
as a result no double logarithm
and no leading contribution from this loop would result. Under the relations afforded by the 
polarizations of the light-cone gauge:

\[
l^\B \x d_{\mu \B} = 0,  \mbox{\hspace{1cm}}  l^\ro \x {\ve^a_\ro}^*(l) = 0,
\]

\nt
it may be easily demonstrated that a contraction of the Lorentz indices $\A, \ro, \B$ lead to the 
numerator of the \emph{r.h.s.} of 
Eq.~(\ref{glue_vert_approx}) to become vanishing. This result is a property of the 
light-cone gauge. 

It may also be demonstrated, following similar methods, that the double logarithm 
emanating from the region of phase space where the gluon $k^\mu$ becomes collinear with the 
outgoing quark line is also suppressed due the vanishing of its coefficient as above. We 
leave the proof of this property to the reader.
The above arguments demonstrate the vanishing of the leading contributions form higher 
order corrections  to the split vertex. This property validates our picture of independent 
fragmentation.

%\bea
%\frac{ T }{ 2 \pi i } \oint_{ C_1 }  d q^0 f(q^0) \af \B \tanh 
% \left( \af \B ( q^0 - \mu ) \right) 
%
%&=& \frac{1}{ 2 \pi i }  \int_{ {i \ini - \e}_{C_3^b} }^{ -i \ini - \e } 
%
%d q^0 f( q^0 ) \left( -\af  + \frac{1}{ e^{ \B ( \mu - q^0 )  }  +  1 }  
%\right)  \nn \\
%
%\mbox{} + \frac{1}{ 2 \pi i }  \left(  \int_{ -i \ini + \e }^{ i \ini + \e } + 
%
%\int_{ i \ini + \mu - \e }^{ -i \ini + \mu - \e } + 
% \int_{ - i \ini + \mu + \e }^{ i \ini + \mu + \e }
%
%\right)_{C_3^a} & &  \!\!\!\!\!\!\! d q^0 f( q^0 ) 
%\left( \af  - \frac{1}{ e^{ \B ( q^0 - \mu )  }  +  1 } 
%\right)  .
%
%\eea 

%\vspace{2cm}

\end{document}